\documentclass{article}

\usepackage[preprint]{neurips_2026}

\usepackage[utf8]{inputenc}
\usepackage[T1]{fontenc}
\usepackage{amsmath,amssymb,amsfonts,mathtools}
\usepackage{booktabs}
\usepackage{graphicx}
\usepackage{float}
\usepackage{algorithm}
\usepackage{algpseudocode}
\usepackage{placeins}
\usepackage{microtype}
\usepackage{nicefrac}
\usepackage{xcolor}
\usepackage{xspace}
\usepackage{url}
\usepackage{hyperref}
\usepackage{multirow}
\usepackage{array}
\usepackage{tabularx}
\usepackage{tikz}
\usetikzlibrary{arrows.meta,fit,positioning,shapes.geometric}

\definecolor{MethodBlue}{HTML}{3B6FB6}
\definecolor{MethodTeal}{HTML}{2A9D8F}
\definecolor{MethodOrange}{HTML}{E69F00}
\definecolor{MethodRed}{HTML}{D55E00}
\definecolor{MethodGray}{HTML}{6B7280}

\hypersetup{
  colorlinks=true,
  linkcolor=MethodBlue,
  citecolor=MethodTeal,
  urlcolor=MethodBlue
}

\newcolumntype{Y}{>{\raggedright\arraybackslash}X}
\newcolumntype{C}[1]{>{\centering\arraybackslash}p{#1}}

\newcommand{\method}{\textsc{MOPD}\xspace}
\newcommand{\minirl}{\textsc{Agentic-miniRL}\xspace}
\newcommand{\labeler}{\textsc{SWE Labeler}\xspace}

\newcommand{\ssg}{\ensuremath{\mathrm{SSG}}\xspace}
\newcommand{\E}{\mathbb{E}}
\newcommand{\ind}{\mathbb{I}}

\newcommand{\up}{\ensuremath{\uparrow}}

\newcommand{\BasePro}{\ensuremath{52.64{\scriptstyle\,\pm\,0.28}}}
\newcommand{\BaseProA}{\ensuremath{51.73{\scriptstyle\,\pm\,0.43}}}
\newcommand{\BaseProB}{\ensuremath{54.39{\scriptstyle\,\pm\,1.17}}}
\newcommand{\BaseProC}{\ensuremath{51.87{\scriptstyle\,\pm\,1.27}}}

\newcommand{\UnifiedPro}{\ensuremath{55.50{\scriptstyle\,\pm\,0.46}}}
\newcommand{\UnifiedProA}{\ensuremath{54.75{\scriptstyle\,\pm\,1.33}}}
\newcommand{\UnifiedProB}{\ensuremath{56.72{\scriptstyle\,\pm\,1.07}}}
\newcommand{\UnifiedProC}{\ensuremath{55.10{\scriptstyle\,\pm\,1.44}}}

\newcommand{\OracleProA}{\ensuremath{59.58{\scriptstyle\,\pm\,1.40}}}
\newcommand{\OracleProB}{\ensuremath{58.87{\scriptstyle\,\pm\,1.31}}}
\newcommand{\OracleProC}{\ensuremath{57.48{\scriptstyle\,\pm\,0.64}}}

\newcommand{\CategoryMOPDPro}{\ensuremath{58.04{\scriptstyle\,\pm\,0.20}}}
\newcommand{\CategoryMOPDProA}{\ensuremath{58.07{\scriptstyle\,\pm\,1.13}}}
\newcommand{\CategoryMOPDProB}{\ensuremath{59.37{\scriptstyle\,\pm\,1.02}}}
\newcommand{\CategoryMOPDProC}{\ensuremath{56.63{\scriptstyle\,\pm\,1.10}}}

\newcommand{\BalancedPro}{\ensuremath{55.34{\scriptstyle\,\pm\,0.92}}}
\newcommand{\BalancedProA}{\ensuremath{54.45{\scriptstyle\,\pm\,0.85}}}
\newcommand{\BalancedProB}{\ensuremath{57.05{\scriptstyle\,\pm\,0.94}}}
\newcommand{\BalancedProC}{\ensuremath{54.59{\scriptstyle\,\pm\,1.25}}}

\newcommand{\ExpBaseProVTwoMean}{81.62}

\newcommand{\ExpBaseMultiMean}{67.89}

\newcommand{\ExpPooledMultiMean}{69.44}

\newcommand{\ExpBalancedMultiMean}{69.00}

\newcommand{\ExpMOPDProVTwoMean}{86.19}

\newcommand{\ExpMOPDProVTwoGainBase}{4.57}
\newcommand{\ExpMOPDHardMean}{42.48}

\newcommand{\ExpMOPDHardGainBase}{5.23}
\newcommand{\ExpMOPDMultiMean}{71.78}

\newcommand{\ExpMOPDMultiGainBase}{3.89}
\newcommand{\ExpMOPDProFullMean}{56.41}

\newcommand{\ExpMOPDLCBMean}{84.57}

\newcommand{\ExpBaseMoEHardMean}{28.10}

\newcommand{\ExpMOPDMoEProVTwoGainBase}{3.69}
\newcommand{\ExpMOPDMoEHardMean}{33.99}

\newcommand{\ExpMOPDMoEHardGainBase}{5.88}

\newcommand{\ExpMOPDMoEMultiGainBase}{3.00}

\newcommand{\ExpMOPDMoEProFullGainBase}{2.37}

\newcommand{\ExpMOPDMoELCBGainBase}{1.14}
\newcommand{\ExpMOPDProVTwoGainPooled}{4.62}
\newcommand{\ExpMOPDProVTwoGainBalanced}{3.84}

\newcommand{\ExpMOPDHardGainBalanced}{6.54}
\newcommand{\ExpMOPDMultiGainPooled}{2.33}
\newcommand{\ExpMOPDMultiGainBalanced}{2.78}

\newcommand{\ExpMultiBaseCILow}{1.44}
\newcommand{\ExpMultiBaseCIHigh}{6.56}

\title{One to More, More to One:\\
Category-Aware Iterative Expert Training\\
for Software Engineering Agents}

\newcommand{\equalcorresponding}{\textsuperscript{*\textdagger}}
\makeatletter
\g@addto@macro\@thanks{%
  \footnotetext[1]{Equal contribution.}%
  \footnotetext[2]{Corresponding authors: \texttt{aixiao.zj@alibaba-inc.com}, \texttt{jiangziyu.jzy@alibaba-inc.com}.}%
}
\makeatother

\author{%
  Jie Zhao\equalcorresponding\quad
  Ziyu Jiang\equalcorresponding\quad
  Suhang Zheng\\
  \textbf{Minghui Shan\quad Xiaoxiao Xu\quad Lin Qu}\\[4pt]
  Alibaba Group\\[6pt]
  {\small\href{https://huggingface.co/Logics-MLLM/Logics-SWE-Qwen3.6-27B}{\raisebox{-0.25em}{\includegraphics[height=1.4em]{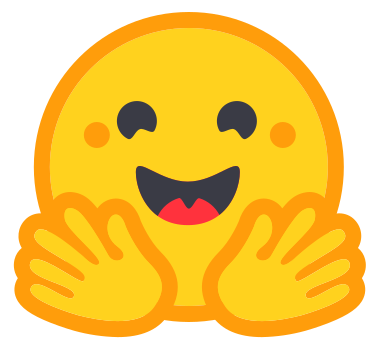}}\enspace\textcolor{orange!75!black}{\textbf{Model}\enspace$\cdot$\enspace\underline{Logics-SWE-Qwen3.6-27B}}}}
}

\begin{document}

\raggedbottom

\maketitle

\begin{abstract}
Repository-level software engineering (SWE) comprises heterogeneous task
categories, whose progress under pooled agentic reinforcement learning can
be uneven: gains in some categories coincide with regressions in others,
while aggregate resolution obscures these changes.  Motivated by this
category see-saw, we develop a category-aware expert-training and
policy-integration framework.  Executable task construction and SWE Labeler,
an evidence-grounded multi-axis labeling system, organize the training pools.
Initial category-specific RL improves average training success while leaving
uneven instance-level progress, motivating explicit consolidation of successful
behavior and policy-adaptive task selection. Same-origin category experts alternate long-horizon Agentic-miniRL with
Refresh--Repair--Expand (RRE): the updated policy refreshes instance mastery,
reuses its own verified successful trajectories for Repair SFT, and reselects
tasks for further RL.  Label-routed multi-teacher on-policy distillation
(MOPD) consolidates the experts into one deployable student, with ReLU-gated
reward extrapolation keeping only each teacher's improving direction over
the reference.  Expert training and policy integration require no external model
to provide solution trajectories or action targets.  On a fixed, audit-filtered Pro-618 subset, we connect Pooled RL and Balanced RL,
expert development, and single-model integration through overall and
per-category resolution and expert-gain recovery. In final-model
evaluations, MOPD achieves mean resolution of \ExpMOPDProVTwoMean\% on
SWE-bench Pro V2 and \ExpMOPDMultiMean\% on SWE-bench Multilingual,
exceeding the base model by \ExpMOPDProVTwoGainBase{} and
\ExpMOPDMultiGainBase{} percentage points, respectively.
The same three 27B experts also improve a Qwen3.6-35B-A3B student
across the evaluated task sets.
\end{abstract}

\section{Introduction}
\label{sec:introduction}

Software engineering agents must navigate repositories, edit code, invoke
tools, inspect execution feedback, and revise a patch across long, stateful
trajectories.  Benchmarks and agent-computer interfaces have made this process
executable and measurable~\citep{jimenez2024swebenchlanguagemodelsresolve,yang2024sweagentagentcomputerinterfacesenable,wang2025openhandsopenplatformai}.

The breadth of software engineering itself makes this training problem
structurally rich.  Repository-level tasks span distinct engineering
contexts: service and data-layer bug fixes, user-facing interface adjustments,
infrastructure and tooling changes, and performance or security patches each
demand different evidence sources, tool-interaction patterns, and verification
approaches.  Recent benchmarks make this heterogeneity concrete.  SWE-bench
Verified curates a human-validated subset of 500 Python instances with clear,
solvable specifications~\citep{openai2024swebenchverified}; SWE-bench Pro
moves beyond the bug-fix-dominated setting of Verified to more varied
repository contexts, maintenance intents, and modification
scales~\citep{deng2025swebenchpro}; and SWE-bench Multilingual extends the
evaluation to 300 tasks across 42 repositories and nine programming
languages~\citep{khandpur2025swebenchmultilingual}.  Although all remain
within repository-level SWE, they expose an agent to materially different
task demands under a shared executable-verdict interface.

Existing post-training strategies can be grouped into three families, each
leaving a distinct gap.

\paragraph{Joint training on the pooled mixture.}
Existing systems such as SWE-RL, SWE-Gym, and SWE-Master improve a shared SWE
policy using pooled task or trajectory collections, with training recipes
ranging from supervised fine-tuning to reinforcement
learning~\citep{wei2025swerladvancingllmreasoning,pan2025trainingsoftwareengineeringagents,song2026swemasterunleashingpotentialsoftware}.
This treatment is convenient but suppresses structure that matters to
optimization: a local logic fix, a cross-module compatibility change, a build
repair, and a security patch can differ in trajectory length, reward variance,
and data frequency.  A joint update may therefore help one task category and
hurt another even when both depend on overlapping underlying skills, yet a
single aggregate metric cannot reveal this internal redistribution.

\paragraph{Pipeline decomposition into subtasks.}
A second family decomposes issue resolution into functional stages such as
localization and repair.  Agentless reduces the
problem to hierarchical fault localization followed by
repair~\citep{xia2024agentless}; SWE-Fixer trains dedicated retrieval and
editing modules~\citep{xie2025swefixer}; AutoCodeRover orchestrates
structured subtask workflows~\citep{zhang2024autocoderover}.  These approaches
decompose the issue-resolution workflow into functional stages.  Our
decomposition is complementary: we partition the training task distribution
into categories and iteratively develop category-specific experts before
integrating them into a single policy.

\paragraph{Cross-domain expert splitting and fusion.}
A third line splits training by domain or capability and then consolidates
experts.  Branch-Train-Merge trains independent domain experts and merges
them~\citep{li2022branchtrainmerge,gururangan2024branchtrainmix}; MOPD routes
multiple teachers during on-policy
distillation~\citep{ma2026mopdmultiteacheronpolicydistillation}; and ExOPD
extrapolates beyond the teacher along a reference-anchored
direction~\citep{yang2026learningbeyondteacher}.  However, these methods
operate across conventional domains such as mathematics, code generation, and
instruction following, where the expert partition is given by domain
boundaries.  We investigate this specialization-and-integration approach
within repository-level SWE, using observable task categories to organize
expert training.  Our motivation is that when aggregate
scores are decomposed, opposing category-level changes can hide beneath
stable overall metrics, a phenomenon we call the \emph{category see-saw}.

This observation raises two questions: can balancing category exposure improve
coordinated learning within a shared policy, and can category-specific training
develop stronger specialists whose gains can be integrated into one model?
Category separation alone does not guarantee stronger experts. In our initial
expert RL runs, average training-instance success rates improve, but some
instances regress while others improve. These paired observations motivate
explicit consolidation of successful behavior and repeated reassessment of
which tasks provide useful training signal. We use Refresh--Repair--Expand
(RRE) to develop the category experts before evaluating their integration.

\begin{figure*}[t]
  \centering
  \includegraphics[width=\textwidth]{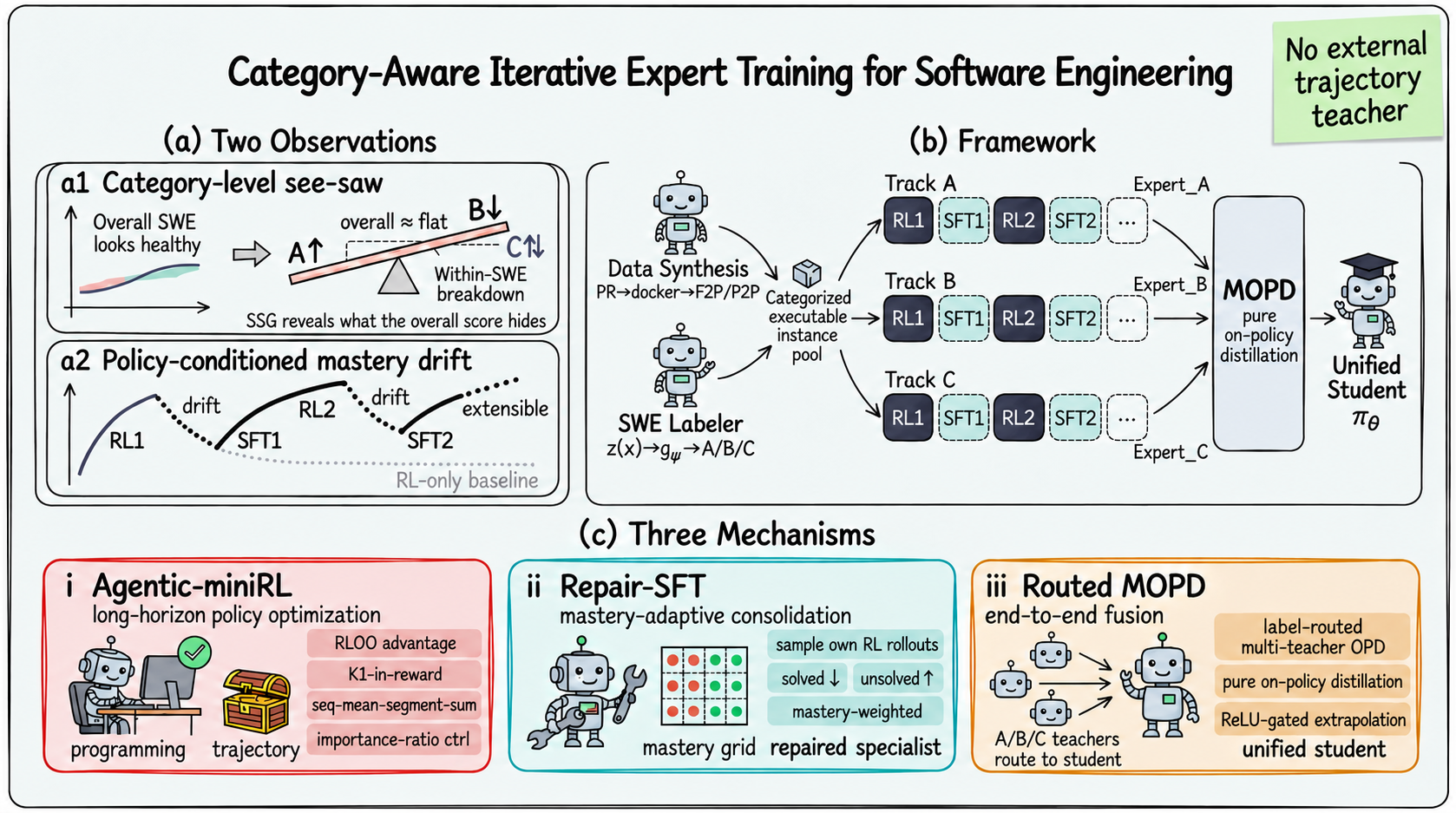}
  \caption{\textbf{Category-aware agentic RL for software engineering.}
  (a) Two observations motivate the framework: aggregate scores can conceal
  opposing category-level changes during joint RL, while a fixed task pool
  becomes stale as instance mastery drifts over long-horizon training.
  (b) \labeler organizes executable tasks into categorized training pools;
  three same-origin training tracks alternate \minirl with Repair SFT before their experts are consolidated into one student through
  pure label-routed MOPD.
  (c) The three mechanisms provide long-horizon policy improvement,
  mastery-aware repair, and multi-teacher on-policy distillation without an external
  trajectory teacher.}
  \label{fig:overview}
\end{figure*}

We take a single software engineering domain, organize its heterogeneous
tasks into categories, and develop category experts through iterative
Refresh--Repair--Expand training. We then bring these experts together through
multi-teacher on-policy distillation, aiming to retain their capabilities in
a single deployable agent (Figure~\ref{fig:overview}).
Cross-domain capability integration starts from
domains that are separable by construction, so an expert partition is given;
inside one conventional domain no such partition exists, and we instead
establish category granularity through explicit rules over observable task
labels. A hierarchical labeler makes these categories auditable, and
each category line trains a same-origin expert by alternating stable
long-horizon agentic RL with mastery refresh and Repair SFT.
During label-routed multi-teacher on-policy distillation (MOPD), reward
extrapolation augments imitation with a reference-anchored learning signal.
Sparse environment reward defines executable correctness during expert
training; routed teachers provide the dense policy signal during pure
on-policy distillation.

Categories provide a common task-side structure for organizing training and
measuring its outcomes. The central evaluation asks whether expert development
and subsequent integration
produce a single model with stronger overall and per-category resolution
than the Pooled RL and Balanced RL baselines.
We retain Pro-618 as a fixed set for training-process and category-level
analysis, and evaluate the final models on SWE-bench Pro V2 and
SWE-bench Multilingual. Extended results also
cover the original 731-task Pro population, code generation, and reuse of
the same 27B experts with a 35B-A3B student.

Our contributions are:
\begin{itemize}
  \setlength{\itemsep}{4pt}
  \setlength{\parsep}{0pt}
  \item \textbf{Category-aware specialization and integration within SWE.}
  To our knowledge, this is the first study to organize repository-level
  SWE tasks into semantic repository-domain categories, train
  category-specific RL experts, and consolidate them into a single policy
  through label-routed multi-teacher on-policy distillation (MOPD).
  Category-level evaluation connects the
  motivating see-saw observations to the practical objective: improving
  overall and per-category resolution in one deployable policy.
  \item \textbf{Self-improving experts through Refresh--Repair--Expand.}
  Instance-level diagnostics reveal that average initial-RL gains coexist
  with observed regressions. RRE couples policy development with data selection: each updated expert
  refreshes instance mastery, reuses verified successes from its preceding
  RL trajectories for Repair SFT, and selects a new training frontier from
  the broader pool.  \minirl supplies the long-horizon RL recipe.  Expert
  development and policy integration use no external model to provide
  solution trajectories or action targets.
  \item \textbf{SWE Labeler: an evidence-grounded, multi-axis labeling system.}
  Source-grounded label definitions and explicit decision rules support both
  issue--change instances and agent interaction trajectories.  Two hierarchical
  semantic axes and three ordinal scale axes provide a common foundation for
  category-level analysis, training-pool construction, and teacher routing.
  We apply the system to executable training data and multiple SWE benchmarks
  to characterize their task distributions.
\end{itemize}

\section{Related work}
\label{sec:related}

\paragraph{Executable data and training environments for SWE agents.}
SWE-bench established repository-level issue resolution as an executable task,
and SWE-Bench Pro extends this setting to broader, longer-horizon software
changes~\citep{jimenez2024swebenchlanguagemodelsresolve,deng2025swebenchpro}.
Subsequent work has expanded the training substrate through collected repository
tasks, procedurally generated instances, hybrid verifiers, and continuously
refreshed evaluation: representative systems include SWE-Gym, R2E-Gym,
SWE-smith, Skywork-SWE, and SWE-rebench
V2~\citep{pan2025trainingsoftwareengineeringagents,jain2025r2egymproceduralenvironmentshybrid,yang2025swesmithscalingdatasoftware,zeng2025skyworksweunveilingdatascaling,badertdinov2026swerebenchv2languageagnosticswe}.
The resulting executable feedback is scalable but not automatically reliable;
recent audits identify specification, environment, grading, and reward-hacking
failure modes, including incorrect patches that still receive positive verifier
outcomes~\citep{wang2026automatedbenchmarkauditing,rajan2026rewardhackability,zhao2026specbench}.
We use executable repository tasks as the training substrate for category
experts, with task audits and runtime anti-hacking controls preserving the
integrity of the outcome reward used by agentic RL.

\paragraph{Post-training software engineering agents.}
SWE-agent post-training has progressed beyond a single recipe.  Trajectory
supervision trains repository navigation, localization, and editing behavior, as
illustrated by SWE-Fixer and SWE-Dev~\citep{xie2025swefixer,wang2025swedev}.
Recent SFT-centered systems further scale validated trajectories, difficulty
curricula, and execution-backed refinement: SWE-Lego studies a strong SFT-only
recipe, whereas SWE-ZERO to SWE-HERO moves from execution-free supervision to a
smaller execution-validated stage~\citep{tao2026swelego,ludwig2026swehero}.
SWE-RL studies reinforcement learning with patch-similarity
rewards~\citep{wei2025swerladvancingllmreasoning}, while long-context multi-turn
RL uses execution feedback for repository-level
tasks~\citep{golubev2025traininglongcontextmultiturnsoftware}.  SWE-Master and
SWE-Prot\'eg\'e combine data curation, supervised fine-tuning, reinforcement
learning, or selective expert assistance in broader post-training
pipelines~\citep{song2026swemasterunleashingpotentialsoftware,kon2026sweprotege}.
Rather than treating SWE as one homogeneous post-training distribution, we work
at a finer granularity within this broad domain: evidence-grounded categories
define separate expert-training streams, and each expert is progressively
refined by agentic RL and Repair SFT drawn from its own successful rollouts.

\paragraph{Reinforcement learning for long-horizon agents.}
Multi-turn agents interleave language-model actions with stateful tool and
environment observations, yet often receive only a sparse task-level outcome.
This regime has motivated algorithms and empirical studies beyond single-turn
RLVR.  DAPO develops large-scale policy-optimization practices, ARPO targets
credit attribution and exploration around tool interactions, and recent
multi-turn studies analyze how environment complexity, reward sparsity, policy
gradient estimators, and horizon length affect learning
stability~\citep{yu2025dapo,dong2025arpo,wang2025practitionersguide,kim2026longhorizon}.
For repository-level SWE, long-context multi-turn RL demonstrates the importance
of optimizing directly against environment feedback
~\citep{golubev2025traininglongcontextmultiturnsoftware}.  MiniRL further relates
stable LLM RL to train--inference discrepancy and policy
staleness~\citep{zheng2025stabilizingreinforcementlearning}.  Agentic-miniRL
instantiates these principles for long-horizon SWE trajectories through
behavior-policy correction, RLOO advantages, the K1 reference penalty in the
reward path, and turn-aware loss reduction over assistant actions separated by
tool observations.

\paragraph{Heterogeneous SWE tasks and category-aware specialization.}
Repository-level SWE contains materially different problem structures even
when all instances share the same agent interface and executable success
criterion.  Recent benchmarks make this variation concrete: visual software
issues require multimodal grounding and interface-state reasoning, security
tasks require vulnerability reproduction and patch validation, and performance
engineering requires bottleneck localization and optimization under workload
constraints~\citep{yang2025swebenchmultimodal,lee2025secbench,shetty2025gso}.
Related post-training studies show that gains learned in one domain need not
transfer uniformly to others, and that sample interactions or optimization
conflicts can shape multi-domain outcomes
~\citep{hu2026breakingbarriers,liang2025evic,liang2026cgpo,ming2026ideal}.
These works mainly study differences among broad domains or capabilities.  We
instead examine heterogeneity \emph{within} SWE: a hierarchical labeling system
maps observable task evidence to fine-grained annotations, from which we derive
a small number of trainable categories.  This structure exposes category
see-saw hidden by aggregate scores and provides the routing basis for training
and evaluating category-specific experts.

\paragraph{Model merging and multi-teacher on-policy distillation.}
Expert models can be integrated either in parameter space or through teacher
supervision.  Task arithmetic and branch--train--merge combine independently
specialized parameter updates without teacher inference
~\citep{ilharco2022taskarithmetic,li2022branchtrainmerge}.  Knowledge
distillation instead transfers output distributions; GKD and MiniLLM evaluate
teachers on student-generated sequences, reducing the distribution mismatch of
static teacher trajectories, while on-policy context distillation conditions a
teacher on additional context
~\citep{agarwal2023gkd,gu2023minillm,ye2026opcd}.  MOPD extends on-policy
distillation to routed domain teachers, and ExOPD generalizes the objective with
reward extrapolation; recent analysis also identifies optimization-budget
imbalance as a source of incomplete multi-teacher integration
~\citep{ma2026mopdmultiteacheronpolicydistillation,yang2026learningbeyondteacher,gao2026openmopd}.
Our setting couples these ideas to intra-domain specialization: all experts
originate from the same SWE base model, their granularity and routing are
defined by observable category labels, and their behaviors are integrated on
the student's own long-horizon trajectories using label-routed MOPD with reward
extrapolation.

\section{Problem setup and empirical motivation}
\label{sec:problem}

We first formalize executable repository-level SWE tasks and the long-horizon
agents that interact with them.  We then show how aggregate joint-RL
performance can conceal opposing movements across heterogeneous SWE
categories.  Finally, a structured multi-axis SWE label space, mapped into
operational categories by deterministic rules, makes this heterogeneity
measurable and provides the basis for category-aware specialization and
integration.

\subsection{Repository-level SWE and joint agentic RL}
\label{sec:swe_task}

Following executable issue-resolution benchmarks such as SWE-bench, we model a
software-engineering instance as
\begin{equation}
  x=(\mathcal{R},b,q,\mathcal{E},\mathcal{V}).
  \label{eq:swe_instance}
\end{equation}
Here $\mathcal{R}$ is a repository, $b$ is the base revision at which the issue
is reproduced, $q$ is a natural-language problem statement, $\mathcal{E}$ is a
reproducible execution environment, and $\mathcal{V}$ is an executable
verifier~\citep{jimenez2024swebenchlanguagemodelsresolve}.  The environment
specifies the source tree, dependencies, runtime, and permitted tools.  The
verifier contains task-specific tests and non-regression checks.  The evaluated
agent observes $q$, the repository at $b$, and feedback from tools it invokes;
it does not observe the reference patch or hidden verifier outcomes.

Starting from $\mathcal{R}_0=\operatorname{checkout}(\mathcal{R},b)$, an agent
produces a terminal repository state $\mathcal{R}_T$ and hence a patch
$p=\operatorname{diff}(\mathcal{R}_0,\mathcal{R}_T)$.  In the binary setting
used throughout this paper, executable success is
\begin{equation}
  r^{\mathrm{env}}(x,p)=
  \ind\!\left[
    \operatorname{apply}(p)
    \wedge \mathcal{V}_{\mathrm{F2P}}(\mathcal{R}_T)
    \wedge \mathcal{V}_{\mathrm{P2P}}(\mathcal{R}_T)
  \right],
  \label{eq:swe_reward}
\end{equation}
where fail-to-pass tests encode the requested repair and pass-to-pass tests
guard against regressions.  Environment construction failures, timeouts, and
invalid patches are retained as separate audit outcomes rather than silently
interpreted as evidence about a task category.  This definition distinguishes
repository-level SWE from isolated code generation: success depends on a
stateful edit that remains compatible with the surrounding project.

\paragraph{Long-horizon tool-using policies.}
\label{sec:swe_agent}

A SWE agent couples a language-model policy with an agent--computer interface
that exposes repository navigation, file inspection and editing, shell
execution, and test feedback~\citep{yang2024sweagentagentcomputerinterfacesenable,wang2025openhandsopenplatformai}.
Let $s_t$ denote the environment state, including the current repository and
process state.  The interface reveals an observation $o_t=\Omega(s_t)$ and
executes a structured action $a_t\in\mathcal{A}$, inducing
\begin{equation}
  s_{t+1}\sim P(\,\cdot\mid s_t,a_t),
  \qquad
  \tau=(o_1,a_1,o_2,a_2,\ldots,o_T,a_T).
  \label{eq:swe_trajectory}
\end{equation}
Actions include search and inspection, patch-producing edits, commands or
tests, and a terminal submission.  Tool output, compiler or test errors, and
the accumulated repository changes become subsequent observations.  Because
the verifier is partly hidden and the complete repository state is not placed
in every model context, the policy acts on the interaction history
$h_t=(q,o_{1:t},a_{1:t-1})$:
\begin{equation}
  a_t\sim\pi_\theta(\,\cdot\mid h_t),
  \qquad
  J_{\mathrm{RL}}(\theta)=
  \E_{x\sim\mathcal{D},\,\tau\sim\pi_\theta}
  \bigl[r^{\mathrm{env}}(x,p(\tau))\bigr].
  \label{eq:joint_rl_objective}
\end{equation}
The scalar objective is appropriate for executable correctness, but it
aggregates over tasks with different maintenance intents and repository
contexts.  Consequently, two checkpoints with similar $J_{\mathrm{RL}}$ can
have materially different distributions of solved tasks.

\subsection{Empirical motivation: the category see-saw}
\label{sec:seesaw_motivation}

The Pooled RL trajectories motivating this study show uneven category-level
progress: improvements on some task groups coincide with regressions on
others, while aggregate resolution obscures these differences.  We call this
observed pattern the \emph{category see-saw}.

The underlying concern is broader than any one benchmark partition.
Repository-level SWE is not a homogeneous problem distribution.  Recent
benchmarks make this heterogeneity concrete: visual JavaScript
issues add multimodal grounding, asynchronous programming, and DOM/state
manipulation~\citep{yang2025swebenchmultimodal}; security engineering includes
vulnerability reproduction, proof-of-concept generation, and
patching~\citep{lee2025secbench}; performance optimization centers on bottleneck
localization and low-level code~\citep{shetty2025gso}; and research-repository
tasks emphasize environment setup, configuration, and
execution~\citep{bogin2024super}.  Although all remain within repository-level
SWE, these task families expose an agent to different evidence sources,
interaction patterns, and executable success criteria.  Joint RL nevertheless
updates one shared policy from a pooled stream under finite sampling and update
budgets.  Learning can therefore progress unevenly across these task-demand
mixtures: experience that benefits one category may transfer only partially to
others.  Closely related trade-offs have been observed in multi-domain LLM
fine-tuning and multi-domain RL, where sample interactions evolve throughout
training and gains in one domain can come at the expense of another
~\citep{liang2025evic,liang2026cgpo}.  The categories used here are observable
operational groupings of tasks; they do not identify isolated latent
capabilities or a particular conflicting gradient.

We study this within-SWE heterogeneity on Pro-618, a fixed, audit-filtered
subset of SWE-bench Pro.  The repository-domain label system in
Section~\ref{sec:labeler} is defined over general SWE instances.  Applying its
A/B/C grouping to Pro-618 yields three mutually exclusive evaluation groups:
service/data/security (\emph{Pro-A}), user-facing applications (\emph{Pro-B}),
and systems, tooling, and runtimes (\emph{Pro-C}), containing 221/201/196 tasks,
respectively.  We evaluate on 618 of the 731 SWE-bench Pro instances.
Community reports have repeatedly flagged individual SWE-bench Pro tasks for
faulty environments, broken container images, or unreliable evaluation logic.
Rather than quietly editing the evaluation set to improve our scores, we
exclude these instances through an explicit filter grounded in those public
findings.  Section~\ref{sec:pro618_construction} explains the benchmark
choice and audit filter; Appendix~\ref{app:pro618_audit} records the
construction protocol.

Let $S_c(t)$ be the success rate of checkpoint $t$ on category $c$, with $t=0$
denoting the base model, and let $\Delta_c(t)=S_c(t)-S_c(0)$.  The corresponding
overall gain is $\Delta_{\mathrm{all}}(t)=\sum_{c=1}^{K}w_c\Delta_c(t)$, where
$w_c$ is the category share used by the official micro average.  We quantify
simultaneous progress across categories by
\begin{equation}
  G_{\mathrm{sim}}(t)=\min_c\Delta_c(t),
  \qquad
  G_{\mathrm{sim}}^{\star}=\max_{t\in\mathcal{T}}G_{\mathrm{sim}}(t),
  \label{eq:simultaneous_gain}
\end{equation}
where $G_{\mathrm{sim}}(t)>0$ means that every category improves over the base
model at checkpoint $t$, and $G_{\mathrm{sim}}^{\star}$ is the largest gain
retained jointly by a single checkpoint in the evaluated trajectory
$\mathcal{T}$.  To measure how far the weakest category lags behind the overall
result, we define the
\emph{see-saw gap} (\ssg):
\begin{equation}
  \ssg(t)=\Delta_{\mathrm{all}}(t)-G_{\mathrm{sim}}(t).
  \label{eq:ssg}
\end{equation}
The \ssg is zero when all category gains are equal.  It grows when aggregate
improvement is concentrated in only part of the benchmark or when the overall
score hides a category regression.  At one checkpoint, \ssg measures how far
the least-improved category trails the aggregate gain; $G_{\mathrm{sim}}$
measures how much improvement is shared by every category.  Across checkpoints,
opposed category transitions characterize the temporal see-saw.
These summaries satisfy $\Delta_{\mathrm{all}}=G_{\mathrm{sim}}+\ssg$.
Table~\ref{tab:gain_interpretation} interprets representative changes relative
to our objective. We seek higher overall and minimum category gains, with a
narrower gap indicating better alignment between them. Gap reduction through
aggregate decline alone does not meet this objective; category-wise score
comparisons remain necessary to establish improvement in every category.

\begin{figure*}[t]
  \centering
  \includegraphics[width=\textwidth]{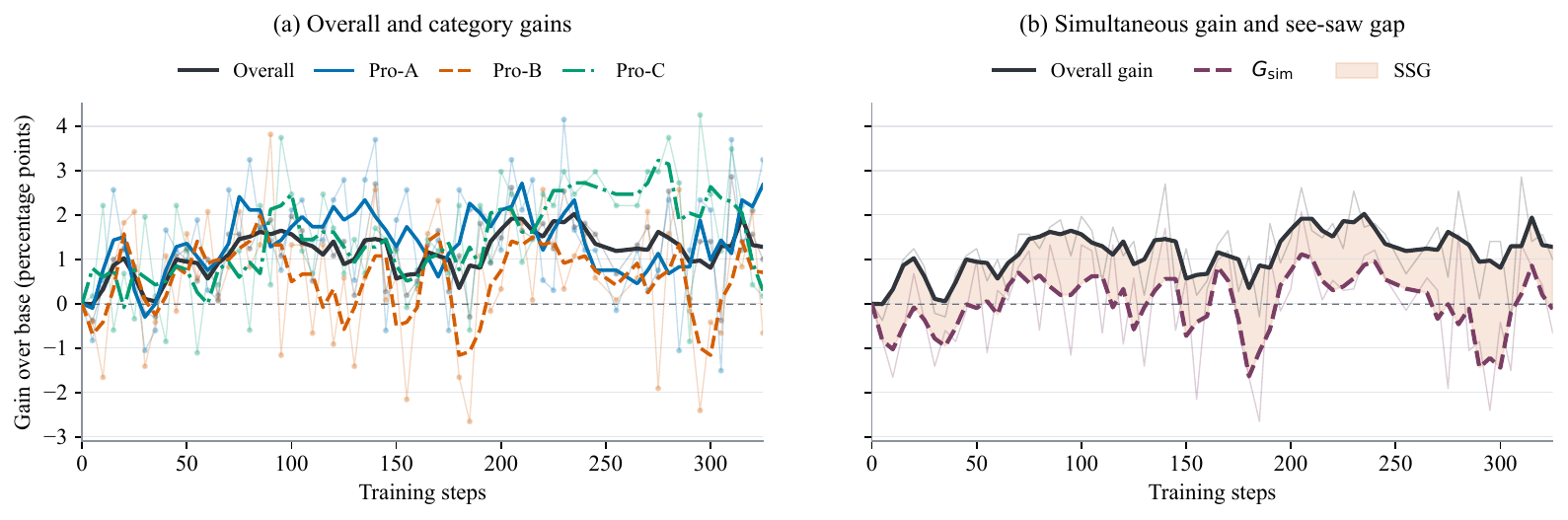}
  \caption{\textbf{Category dynamics under Pooled RL.}
  A single policy is trained on the full 6{,}723-task mixture and evaluated on
  the fixed Pro-618 split (Pro-A/B/C: 221/201/196 tasks).  Faint traces show the
  mean at each checkpoint; bold traces show centered
  three-checkpoint moving averages.
  Step 0 anchors the base model.  Left: overall and
  category gains relative to their respective base-model rates.  Right: overall
  gain and minimum category gain $G_{\mathrm{sim}}$; their
  shaded difference between the smoothed curves is the smoothed \ssg.}
  \label{fig:seesaw_motivation}
\end{figure*}

Figure~\ref{fig:seesaw_motivation} shows that under Pooled RL, gains in some
categories coincide with regressions in others (panel~a). Positive overall
gains can coexist with a negative minimum category gain (panel~b), showing
that aggregate improvement alone does not guarantee progress in every category.

\begin{table}[htbp]
  \centering
  \caption{\textbf{Interpreting representative changes in overall and category gains.}
  $G_{\mathrm{sim}}$ is the minimum category gain over the base; \ssg{} is
  the gap between overall and minimum category gains. Arrows compare two
  checkpoints or training settings using the same base and category weights.}
  \label{tab:gain_interpretation}
  \small
  \setlength{\tabcolsep}{5pt}
  \renewcommand{\arraystretch}{1.35}
  \begin{tabularx}{\linewidth}{@{}cccX@{}}
    \toprule
    Full gain & $G_{\mathrm{sim}}$ & \ssg{} & Interpretation relative to our objective \\
    \midrule
    $\uparrow$ & $\uparrow$ & $\downarrow$ & \textbf{Aligned improvement:} overall and minimum gains rise while the gap narrows. \\
    $\uparrow$ & $\uparrow$ & $\rightarrow$ & \textbf{Shared improvement with a stable gap:} overall and minimum gains rise together, with unchanged separation. \\
    $\uparrow$ & $\uparrow$ & $\uparrow$ & \textbf{Shared improvement with a wider gap:} both gains rise, but overall gain increases faster than minimum gain. \\
    $\uparrow$ & $\rightarrow$ / $\downarrow$ & $\uparrow$ & \textbf{Aggregate improvement without minimum-gain improvement:} overall gain rises while minimum gain stays unchanged or declines. \\
    $\downarrow$ & $\rightarrow$ / $\downarrow$ & $\downarrow$ & \textbf{Gap reduction through aggregate decline:} the gap narrows without improving the minimum gain; this does not meet our improvement objective. \\
    $\downarrow$ & $\uparrow$ & $\downarrow$ & \textbf{Minimum--aggregate trade-off:} the minimum gain improves and the gap narrows at the cost of overall gain. \\
    \bottomrule
  \end{tabularx}
  \par\smallskip
  \begin{minipage}{\linewidth}
    \footnotesize
    Full gain $=G_{\mathrm{sim}}+\ssg{}$; $\rightarrow$ denotes no change.
    Improvement in every category must still be checked against the A/B/C scores.
  \end{minipage}
\end{table}

These observations motivate two approaches to category-aware post-training.
The first adjusts the data composition of a shared training stream.  Training
mixture proportions affect multi-capability performance
\citep{ming2026ideal}, making category balancing a natural intervention to
evaluate.  We instantiate this approach with a subset that matches category
counts and base-model pass-rate strata.  The second changes the organization
of policy learning: train experts on category-specific pools, then integrate
their behaviors into a single model. In particular, does category-specific RL alone produce
experts that outperform Pooled RL on their respective categories?
Our experiments first examine Balanced RL, then assess the initial
experts and their subsequent development before evaluating single-model
integration.

\subsection{From category-aware diagnosis to specialization and integration}
\label{sec:labels_setup}

The see-saw above exposes a limitation of aggregate evaluation: a single
resolution score cannot show where joint RL improves, stagnates, or regresses
within the heterogeneous SWE task distribution.  We therefore use a structured
label space that is expressive enough to describe diverse SWE instances and
organized enough to support consistent category-level aggregation.  It provides
a common unit for measuring category-wise progress, constructing balanced
controls, and forming specialized training pools.  We write
\begin{equation}
  \mathbf{z}(x)=L_\phi(e(x)),
  \qquad
  c(x)=g_\psi(\mathbf{z}(x))\in\mathcal{C}\cup\{\varnothing\},
  \label{eq:category_map}
\end{equation}
where $e(x)$ is offline task evidence, $L_\phi$ is a frozen labeler,
$g_\psi$ is a deterministic category rule, and $\varnothing$ denotes an unrouted
task.  The fine-grained vector $\mathbf{z}(x)$ retains instance properties,
whereas $g_\psi$ maps them into operational categories.  At checkpoint $t$,
this grouping expands the scalar overall result into the category-performance
vector $(S_1(t),\ldots,S_K(t))$, revealing how gains and regressions are
distributed across the SWE task space.  The same map is applied to training-pool
construction, expert routing, and category-aware evaluation.

\labeler, described in Section~\ref{sec:labeler}, instantiates this structured
label space.  Its multiple axes describe observable instance properties such as
repository domain, maintenance intent, and modification scope.  The main
experiments derive Pro-A/B/C from a frozen Domain L1 rule; the remaining axes
support within-category stratification and diagnosis.  Here \emph{repository
domain} denotes the software ecosystem represented by a repository, while all
labels remain within the overarching domain of repository-level SWE.  Pro-A/B/C
are therefore one coarse operational grouping of a richer annotation space.
The same label space can organize training and held-out tasks independently of
whether they originate from Pro-618.

A \emph{label} is an observable instance annotation, a \emph{category} is an
operational grouping of labels, and a model \emph{capability} is latent.  One
category can require several capabilities, and one capability can support
several categories.  Our analysis therefore concerns category-wise performance,
specialization, and integration rather than a decomposition of latent
capabilities.

Our learning problem is therefore to produce a single deployable policy with
stronger aggregate resolution and category-level performance than Pooled RL and
Balanced RL.  We evaluate the overall score and score differences across
all evaluation categories.  The \ssg separately summarizes how unevenly those
gains are distributed.
Balancing adjusts category exposure within one policy's training stream.
Our expert-based approach instead allocates a separate training stream to
each category and then uses label-routed
multi-teacher on-policy distillation (\textbf{MOPD}) to consolidate the
experts' behaviors into one deployed policy.  The categories control training-pool construction and
teacher routing, while the deployed agent receives only ordinary task and tool
observations.  Section~\ref{sec:method} develops this label-grounded training
framework.

\FloatBarrier

\section{Method}
\label{sec:method}

Our framework organizes SWE post-training around category experts and their
integration into one deployed policy.  Executable task construction provides
the instances and verifiers, while SWE Labeler supplies the task categories.
Each category expert starts from a common base and follows the
Refresh--Repair--Expand (RRE) loop, using Agentic-miniRL for its RL phases
and its own verified trajectories for Repair SFT.  Label-routed multi-teacher
on-policy distillation then consolidates the experts into a single student.

\subsection{Executable SWE task construction}
\label{sec:data}

We construct the executable task pool from merged repository PRs and linked
issues. Each instance contains a base revision, an issue-style problem
statement, an environment specification, and executable tests. Fix patches
are separated from test patches, and an LLM rewrites the PR and issue content
into the problem statement. Leakage checks reject statements exposing
source or test paths or added patch code, with boilerplate exempted;
deduplication and contamination audits further filter the pool.
Environment construction follows RepoLaunch~\citep{zhang2025swebenchgoeslive},
producing a Docker image with rebuild and test commands.

We validate each instance by running its tests before and after the gold
patch, retaining fail-to-pass and pass-to-pass tests. Repeated gold-patch
checks filter flaky or time-decayed verifiers. Retained instances record
provenance, verifier status, labeler version, and split membership.
The resulting pool contains approximately 32K executable candidates;
Section~\ref{sec:experiments} describes the subsets used for training and
RRE expansion. These tasks provide the executable training substrate, while
the labeling system below supplies the categories used for expert training
and teacher routing.

\subsection{SWE Labeler: an evidence-grounded hierarchical labeling system}
\label{sec:labeler}

Repository-level SWE instances differ in maintenance intent, software ecosystem,
and task scale.  To make this heterogeneity actionable, we introduce \labeler,
an evidence-grounded hierarchical labeling system.  Crucially, the taxonomy is
constructed from established software-engineering concepts rather than a flat,
benchmark-specific inventory: each fine-grained label links an authoritative
source to an operational SWE rule, observable evidence signals, and explicit
boundaries with neighboring labels.  The resulting label space is interpretable,
auditable, and reusable for data construction, expert routing, and category-aware
evaluation.

\paragraph{Label space.}
Each instance receives two semantic axes---\emph{Task Type} and \emph{Repository
Domain}---and three orthogonal scale axes: modification scope, cognitive
complexity, and estimated resolution time.  Task Type captures the maintenance
intent of a change; Repository Domain captures the software ecosystem of the
repository; and the scale axes characterize the extent of modification, the
reasoning complexity, and the anticipated resolution time.  Given an instance
$x$, the Labeler produces the semantic and scale annotation
\begin{equation}
  \mathbf{z}(x)=\left(
  z^{\mathrm{task}}_{\mathrm{L1}},
  Z^{\mathrm{task}}_{\mathrm{L2}},
  z^{\mathrm{domain}}_{\mathrm{L1}},
  Z^{\mathrm{domain}}_{\mathrm{L2}},
  z^{\mathrm{scope}},
  z^{\mathrm{cognitive}},
  z^{\mathrm{time}}
  \right).
  \label{eq:label_vector}
\end{equation}
The final taxonomy contains 26 Task Type L1 families with 119 L2 labels and 21
Repository Domain L1 families with 108 L2 labels.  Each semantic axis has one
primary L1 label; its L2 field retains one to three applicable children, or
\texttt{unspecified} when no concrete child is supported by the evidence.  The
three scale axes each have four ordered levels, for 12 scale levels in total.
The Labeler additionally records programming language, specification type, and
dependency context as auxiliary metadata.

The full inventory and representative source chains appear in
Appendix~\ref{app:taxonomy_grounding}; Figure~\ref{fig:taxonomy_atlas}
visualizes the breadth of the two semantic axes.

\paragraph{Taxonomy grounding and label contract.}
The taxonomy operationalizes established software-engineering concepts.  The
Task Type axis draws on the maintenance and quality vocabulary of ISO/IEC~25010 and ISO/IEC/IEEE~14764,
defect and security mechanisms from MITRE CWE, and behavior-preserving
restructuring from Fowler's refactoring catalog
\citep{iso25010_2023,iso14764_2022,mitre_cwe682,mitre_cwe697,fowler2018refactoring}.
The Repository Domain axis organizes the software ecosystem in which a change
occurs; its fine-grained definitions are grounded in primary framework,
platform, and project documentation.  The three scale axes combine repository
topology with empirical SWE task analysis and cognitive-load theory into
four-level operational definitions
\citep{brand2025swebenchskills,sweller1988cognitiveload}.

At the fine-grained layer, every concrete label is a decision record
\begin{equation}
  \mathcal{E}(\ell)=
  \bigl(a_\ell,d^{\mathrm{src}}_\ell,d^{\mathrm{op}}_\ell,
  s^{+}_\ell,b_\ell\bigr),
  \label{eq:label_evidence}
\end{equation}
where $a_\ell$ identifies an authoritative source,
$d^{\mathrm{src}}_\ell$ its source concept,
$d^{\mathrm{op}}_\ell$ the operational SWE rule, $s^{+}_\ell$ observable
positive signals, and $b_\ell$ boundaries to neighboring labels.  The current
inventory instantiates this contract for all 119 Task Type L2 labels, 108
Repository Domain L2 labels, and 12 scale levels.  For example,
\texttt{bug-fix.logic\_error} is grounded in CWE-682 and CWE-697: it covers an
incorrect condition, calculation, or comparison, while index-boundary mistakes
map to \texttt{off\_by\_one} and unintended mutation maps to
\texttt{state\_corruption}.  This source-to-rule-to-boundary structure gives
the Labeler explicit semantics at the point where neighboring classes are most
easily confused. Appendix~\ref{app:labeler} details the label inventory,
annotation procedure, structural audit, and category mapping.

\begin{figure*}[t]
  \centering
  \includegraphics[width=\textwidth]{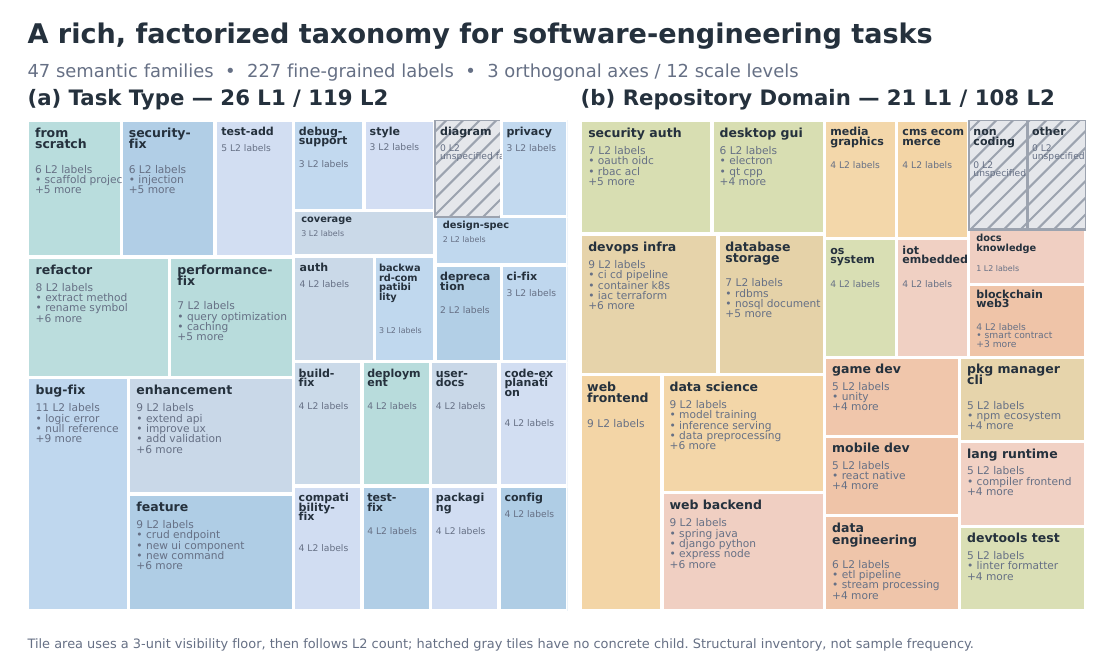}
  \caption{\textbf{Semantic breadth of the SWE Labeler taxonomy.}
  The two panels show the 26 Task Type and 21 Repository Domain L1 families.
  Tile area is proportional to the number of declared L2 labels, and tile text
  shows selected L2 examples.  The inventory represents the semantic resolution
  available to the Labeler.}
  \label{fig:taxonomy_atlas}
\end{figure*}

\paragraph{Task and trajectory evidence.}
Repository-level SWE is observed through two complementary evidence forms.
Benchmark and mined instances are typically static issue--change tasks (e.g.,
issue--PR pairs), pairing a task statement and repository state with a reference
change and executable tests.  Interactive settings instead yield complete
user--agent trajectories, which preserve the user request together with agent
actions, tool observations, edits, and verification outcomes.  Both forms
describe SWE work, but they expose different evidence for the same label
decision.  Two input adapters normalize them into a common instance
representation, after which the Labeler applies the same taxonomy and decision
rules.

For GitHub issue--change instances, the annotation input includes the repository
name and available GitHub metadata (description, topics, and primary language),
the problem statement, and bounded excerpts of the code patch and test patch.
Repository metadata establishes the project's purpose and software ecosystem;
the task statement and changes provide evidence of maintenance intent, affected
functionality, and change scale.  Both annotation stages receive this evidence.
If repository metadata is unavailable, the Labeler falls back to the repository
name and task/change content and marks this fallback in its domain rationale.

Annotation proceeds hierarchically.  In the first stage, the Labeler predicts
Task Type L1 and Repository Domain L1 from the broad record, together with the
auxiliary metadata.  Repository purpose and metadata provide the primary signal
for the Domain label, while change-specific evidence resolves multi-purpose
repositories.  In the second stage, the Labeler selects L2 labels only from the
children of the predicted L1 families and assigns the three scale axes.  This
conditional refinement concentrates the fine-grained decision on semantically
nearby alternatives; the output is checked against the declared L1--L2
hierarchy and admissible scale levels.  The trajectory adapter and the
issue--change adapter thus produce comparable labels across task records and
interaction trajectories.

\paragraph{Cross-benchmark label profiles.}
We applied the completed Labeler to 1,531 instances drawn from three SWE
benchmark corpora. The collection spans SWE-bench Verified, SWE-bench Pro,
and SWE-bench Multilingual, covering varied maintenance tasks and mirroring the
recent expansion of executable SWE data across repositories and language
ecosystems~\citep{deng2025swebenchpro,badertdinov2026swerebenchv2languageagnosticswe}.
Table~\ref{tab:labeler_benchmark_profiles} shows that the same label space
captures sharply different maintenance, ecosystem, and scale profiles.

\begin{table*}[t]
  \centering
  \caption{\textbf{Cross-benchmark profiles produced by SWE Labeler.}
  The three labeled corpora contain 1,531 instances. Percentages are computed
  within each corpus; the last column reports a representative high-mass scale
  label.}
  \label{tab:labeler_benchmark_profiles}
  \small
  \renewcommand{\arraystretch}{1.08}
  \begin{tabularx}{\textwidth}{@{}lrrY Y Y@{}}
    \toprule
    Labeling corpus & $N$ & Langs. & Leading Task Type & Leading Repository Domain & Scale signature \\
    \midrule
    SWE-bench Verified & 500 & 1 & \emph{bug-fix} (87.0\%) & \emph{web\_backend} (46.4\%) & 86.2\% single-file \\
    SWE-bench Pro & 731 & 3 & \emph{bug-fix} (47.3\%) & \emph{devops\_infra} (33.2\%) & 59.2\% cross-module \\
    SWE-bench Multilingual & 300 & 8 & \emph{bug-fix} (89.3\%) & \emph{lang\_runtime} (24.7\%) & 66.0\% single-file \\
    \bottomrule
  \end{tabularx}
\end{table*}

\paragraph{Routable category construction.}
The complete annotation $\mathbf{z}(x)$ is deliberately richer than a direct
expert router: assigning one expert to every Repository Domain L1/L2 label
would fragment the available RL examples and optimization budget.  We therefore
use Repository Domain L1 as the routing axis.  It identifies the software
ecosystem in which an agent must locate evidence, interact with tools, and
validate a change; Task Type may occur across these ecosystems, while Task
Type, L2 labels, and scale axes remain available for within-category
stratification and fine-grained analysis.

For the main configuration, we coarsen the 18 routable Domain L1 families into
three broad engineering contexts: service/data/security (A), user-facing
applications (B), and systems, tooling, and runtimes (C).  This three-route
configuration instantiates category-aware specialization and integration under
a fixed RL budget while retaining the full label vector for analysis.  We
construct the routes with a deterministic mapping from Repository Domain L1:
\begin{equation}
  c(x)=g_\psi\!\left(z^{\mathrm{domain}}_{\mathrm{L1}}(x)\right)
  \in \{A,B,C,\varnothing\}.
  \label{eq:category_mapping}
\end{equation}
The complete Domain L1 whitelist is given in
Appendix~\ref{app:routing}.  The same rule is applied to all labeled training
and evaluation records.  Applied to the Pro-618 longitudinal probe, its three
routes are denoted Pro-A/B/C.  The resulting partition controls joint-RL
sampling, category-expert training, and \textbf{MOPD}; the same labeled record
also supports category-aware evaluation.

\subsection{Agentic-miniRL}
\label{sec:minirl}

We build on the policy-gradient variant of
MiniRL~\citep{zheng2025stabilizingreinforcementlearning} and instantiate it for
long-horizon, multi-turn SWE trajectories.  Four optimization challenges
determine the objective.  First, executable verification normally provides a
sparse binary terminal reward.  Group-relative learning obtains a nonzero
executable-reward signal only from mixed-outcome rollout groups, which are
scarce for difficult instances; we use RLOO to avoid further attenuating these
signals through a self-inclusive baseline.  Second, repeated long-horizon
updates can move the policy away from a fixed reference; we regularize this
drift through K1 in the token-reward path.  Third, the rollout behavior policy
$\mu_b$, the pre-update learner $\pi_{\theta^-}$, and the current learner
$\pi_\theta$ need not coincide.  We therefore correct for behavior-probability
mismatch and bound policy updates at the token level.  Fourth, a trajectory
interleaves assistant action turns with tool observations, and the action turns
vary widely in length; we aggregate the token loss turn by turn.

For each instance, $\mu_b$ produces $G$ trajectories and their executable
returns.  The trainer stores the behavior log-probabilities, recomputes
log-probabilities under $\pi_{\theta^-}$, and evaluates a reference
$\pi_{\mathrm{ref}}$ that remains frozen within the RL phase.  MiniRL supplies
the behavior-corrected token-level policy surrogate.  We combine this
surrogate with instance-wise RLOO advantage estimation, K1 reward
regularization, and turn-aware reduction for Agentic SWE.  The resulting
computation follows the four operations above in order; only the final policy
loss is differentiated.

\paragraph{RLOO reward centering: preserving finite-group signal.}
Repository tasks differ substantially in their probability of being solved,
while executable verification provides only sparse terminal feedback.
Comparing rollouts of the same instance conditions the learning signal on the
task, but a learned critic for long, sparse-reward trajectories introduces an
additional estimation problem.  We instead adopt the leave-one-out
(RLOO)~\citep{ahmadian2024basicsrevisitingreinforcestyle} baseline.  For $G$
rollouts of the same instance with executable returns $R_1,\ldots,R_G$, RLOO
excludes each rollout from its own baseline:
\begin{equation}
  \bar R_i
  = R_i-\frac{1}{G-1}\sum_{j\neq i}R_j
  = \frac{G}{G-1}\left(R_i-\frac{1}{G}\sum_{j=1}^{G}R_j\right).
  \label{eq:rloo}
\end{equation}
Under i.i.d. on-policy sampling, the leave-one-out baseline for trajectory $i$
is independent of that trajectory conditional on the SWE instance.  Its
score-function contribution therefore has zero expectation,
$\mathbb{E}[b_{-i}\nabla_\theta\log\pi_\theta(\tau_i\mid x)]=0$, so subtracting
$b_{-i}$ preserves the expected REINFORCE policy gradient while providing a
parameter-free, instance-conditioned control variate for variance reduction.
This property is particularly useful for Agentic SWE: executable feedback
arrives only after a long tool-interaction trajectory, value estimation would
span heterogeneous intermediate tool states, and each sandbox rollout is
expensive.  RLOO reuses every trajectory both as an update sample and as
baseline evidence for the other group members, without training a critic.

Like mean-centered GRPO, RLOO supplies a within-instance learning signal; its
specific distinction is excluding the current rollout from its own baseline,
whose finite-group effect we quantify below.  For binary executable rewards,
suppose that $k$ of the $G$ rollouts succeed.
Equation~\ref{eq:rloo} assigns
\begin{equation}
  \bar R^{+}=\frac{G-k}{G-1},
  \qquad
  \bar R^{-}=-\frac{k}{G-1}
  \label{eq:rloo_binary}
\end{equation}
to every successful and failed rollout, respectively.  In a non-degenerate
group containing both successful and failed rollouts, a rare successful
trajectory therefore receives a strong positive weight, while its failed
siblings provide instance-matched negative evidence.  An all-failure or
all-success group has no within-instance contrast, which motivates pairing
group-relative learning with the pass-rate-aware instance selection described
in Section~\ref{sec:rre}.

The rightmost form of Equation~\ref{eq:rloo} also makes the relation to our
mean-centered, non-standardized GRPO baseline explicit.  If
$A_i^{\mathrm{GRPO}}=R_i-G^{-1}\sum_jR_j$, then
\begin{equation}
  \bar R_i=\frac{G}{G-1}A_i^{\mathrm{GRPO}}.
  \label{eq:rloo_grpo_relation}
\end{equation}
For the centered executable reward, mixed-outcome groups are the only groups
with nonzero advantages.  Including a rollout in its own group mean would
shrink every such advantage by $(G-1)/G$.  RLOO restores this finite-group
scale without changing the sign or relative ordering of the advantages.  In
our $G=8$ configuration, for example, the sole successful rollout in a
one-success group receives advantage $1$ under RLOO, compared with $7/8$ under
inclusive mean centering.  The emphasis on scarce successful behavior still
comes from within-instance comparison; the purpose of RLOO is to avoid further
attenuating that already limited signal.  Equations~\ref{eq:rloo}--\ref{eq:rloo_grpo_relation}
match the centering operation used in our implementation.

\paragraph{K1 regularization before return-to-go.}
Long-horizon training can accumulate drift from a fixed reference policy.  We
therefore place the sampled K1 log-ratio in the reward path, after RLOO has
centered the executable return and before return-to-go is computed.  For each
valid assistant token,
\begin{equation}
  \begin{aligned}
  d^{\mathrm{K1}}_{i,t}
    &= \log\pi_{\theta^-}(a_{i,t}\mid h_{i,t})
       -\log\pi_{\mathrm{ref}}(a_{i,t}\mid h_{i,t}),\\
  \widetilde r_{i,t}
    &= \mathbf{1}[t=T_i]\,\bar R_i-\beta d^{\mathrm{K1}}_{i,t},
  &
  A_{i,t}
    &= \sum_{u=t}^{T_i}\gamma^{u-t}\widetilde r_{i,u},
  \end{aligned}
  \label{eq:k1_return}
\end{equation}
where $T_i$ is the terminal valid token and rewards are zero on masked
tool-observation tokens.  Equation~\ref{eq:k1_return} mirrors the implemented
order: terminal reward expansion, token-level K1 subtraction, and a reverse
discounted cumulative sum.  The resulting $A_{i,t}$, rather than a
single trajectory-level scalar, weights each optimized token.

The location of the KL estimator matters.  Under on-policy sampling, K1 and
the nonnegative K3 estimator
$\exp(-d^{\mathrm{K1}})-1+d^{\mathrm{K1}}$ estimate the same reverse-KL
value, but their optimization gradients are not interchangeable.  K1 used as
a stop-gradient reward yields the score-function gradient of the
KL-regularized policy objective, whereas directly differentiating K3 as an
auxiliary loss gives a biased gradient for that objective
\citep{liu2025rethinkingkl,shah2025comedyestimators}.  Controlled LLM-RL
experiments also report stronger and more stable optimization with
K1-in-reward than with K3-in-loss~\citep{shah2025comedyestimators}.  We compute
K1 from $\pi_{\theta^-}$ and $\pi_{\mathrm{ref}}$ without gradient and optimize
it through the reward path, while retaining K3 only as a diagnostic.  This
reference regularizer is separate from the category-expert distillation term
in Section~\ref{sec:mopd}.

\paragraph{MiniRL token update: behavior correction with a proximal safeguard.}
The third challenge above arises because the rollout engine records behavior
probabilities under $\mu_b$, whereas the trainer optimizes $\pi_\theta$ after
recomputing the pre-update policy $\pi_{\theta^-}$.  Cross-engine numerical
differences and asynchronous policy lag can therefore make these probabilities
different.  MiniRL corrects this mismatch with the bounded importance weight
$w_{i,t}$.  In addition, the proximal ratio
$\rho^{\mathrm{prox}}_{i,t}$ defines an asymmetric safeguard on the direction
of the policy update.  For a valid assistant token,
\begin{equation}
  \begin{aligned}
  \rho^{\mathrm{prox}}_{i,t}(\theta)
    &= \frac{\pi_\theta(a_{i,t}\mid h_{i,t})}
             {\pi_{\theta^-}(a_{i,t}\mid h_{i,t})},\\
  w_{i,t}(\theta)
    &= \operatorname{clip}\!\left(
       \frac{\pi_\theta(a_{i,t}\mid h_{i,t})}
            {\mu_b(a_{i,t}\mid h_{i,t})},c_{\min},c_{\max}\right),\\[-1mm]
  M_{i,t}(\theta)
    &= 1-\mathbf{1}\!\left[
       (A_{i,t}>0 \land \rho^{\mathrm{prox}}_{i,t}>1+\epsilon_{+})\right.\\[-1mm]
    &\hspace{27mm}\left.\lor
       (A_{i,t}<0 \land \rho^{\mathrm{prox}}_{i,t}<1-\epsilon_{-})
       \right].
  \end{aligned}
  \label{eq:minirl_ratios_mask}
\end{equation}
The sign-aware mask suppresses a token update only after the policy has moved
beyond the corresponding boundary in the direction encouraged by its
advantage.  Truncated importance sampling provides the primary correction for
the rollout--trainer probability mismatch, while the mask acts as a proximal
safeguard.  Section~\ref{sec:experiments} reports the rollout-group size, KL
coefficient, and clipping bounds.

\paragraph{Turn-aware reduction and the final objective.}
The fourth challenge arises from the alternating structure of assistant action
turns and tool observations in each SWE trajectory; action turns vary sharply
in both count and length.  Let
$\mathcal{S}_i$ be the set of contiguous valid response segments in
trajectory $i$, each corresponding to one assistant turn.  The complete
Agentic-miniRL loss directly combines the mask, behavior correction, token
advantage, and turn-aware reduction:
\begin{equation}
  \boxed{\begin{aligned}
  \mathcal{L}_{\mathrm{Agentic\text{-}MiniRL}}(\theta)
  ={}&-\frac{1}{B_{\mathrm{valid}}}\sum_i\frac{1}{|\mathcal{S}_i|}
    \sum_{s\in\mathcal{S}_i}\sum_{t\in s}
    M_{i,t}(\theta)\,\operatorname{sg}\!\left(w_{i,t}(\theta)\right)\\[-1mm]
  &\hspace{22mm}{}\cdot A_{i,t}
    \log\pi_\theta(a_{i,t}\mid h_{i,t}).
  \end{aligned}}
  \label{eq:minirl_objective}
\end{equation}
This is the minimized loss implemented by the trainer.  Rewards, advantages,
the Boolean mask, and the clipped importance weight are treated as
non-differentiated quantities.  Its outer reduction is
\texttt{seq-mean-segment-sum}: it sums token losses inside each
assistant turn, averages turns within a trajectory, and then averages the
$B_{\mathrm{valid}}$ valid trajectories.  Tool-observation tokens are excluded by the
response mask.  Consequently, neither raw trajectory length nor the number
of tool interactions alone determines a trajectory's outer weight, while a
long action retains its token-summed learning signal.  We use this recipe
for Pooled RL, Balanced RL, and category-expert RL; the implementation and
training settings are specified in Section~\ref{sec:experiments}.

\subsection{Refresh--Repair--Expand training of category experts}
\label{sec:rre}
\label{sec:experts}

We train one expert for each derived category $c\in\{A,B,C\}$. Category
separation defines each expert's scope, but does not by itself ensure strong
teachers for integration. The initial experts improve average training-instance
success rates, yet these gains coexist with instance-level regressions
(Section~\ref{sec:results_rre}). This motivates combining further RL with
explicit consolidation of successful trajectories and a refreshed assessment
of instance mastery.

A common
approach to long-horizon SWE post-training is to synthesize trajectories with a
larger teacher, distill them by SFT, and only then apply RL
\citep{song2026swemasterunleashingpotentialsoftware}.  Instead, all three
experts start from the same strong base policy $\pi_0$ and acquire their
category behavior directly through executable Agentic RL.  This removes an
external trajectory teacher from the expert-training loop.  Hereafter,
\emph{teacher} refers only to the resulting expert used by MOPD in
Section~\ref{sec:mopd}; no external model supplies its solution trajectories or
action targets. Teacher-free SWE-agent training is feasible at long horizon
\citep{golubev2025traininglongcontextmultiturnsoftware}. To support expert
development under changing instance mastery, we develop
\textbf{Refresh--Repair--Expand (RRE)}, a policy-adaptive loop that co-evolves
each expert and its training frontier.  The loop couples exploration of
category tasks with reuse of the expert's existing successful behavior:
the updated policy determines which instances receive repair supervision
and which tasks enter the next RL phase.
Algorithm~\ref{alg:rre_training} specifies this process.  Each round consists
of an RL phase and a Repair SFT phase, with a mastery refresh after each;
before another RL round, the repaired policy probes an expanded pool formed
by randomly adding previously omitted base-zero and base-one instances.

\begin{algorithm}[t]
\caption{Refresh--Repair--Expand training of category experts}
\label{alg:rre_training}
\small
\begin{algorithmic}[1]
\Require Shared base $\pi_0$; broader category pools $\{\mathcal D_c\}$; verifier $V$; rounds $R$
\Require Initial training sets $\{\mathcal S_c^0\}$; stored base mastery $\{p_c^0\}$ on the candidate pools
\Ensure Category experts $\{T_A,T_B,T_C\}$
\Statex \textit{Select}$(\Pi,c)=\arg\max_{\pi\in\Pi} S_c(\pi)$ selects within one training phase.
\For{each category $c\in\{A,B,C\}$}
  \State $\pi\gets\pi_0$; $\mathcal S\gets\mathcal S_c^0$; $p\gets p_c^0$ \Comment{Reuse stored base pass rates}
  \State $H\gets\{x:p_c^0(x)>0\}$
  \For{$r=1,\ldots,R$}
    \State $p_{\mathrm{pre}}\gets p|_{\mathcal S}$
    \Statex \hspace{\algorithmicindent}\hspace{\algorithmicindent}\textit{Explore and refresh on the current RL training set}
    \State $(\Pi_{\mathrm{RL}},\mathcal B^+)\gets\Call{AgenticRL}{\pi,\mathcal S,V}$
    \State $\pi_{\mathrm{RL}}\gets\textit{Select}(\Pi_{\mathrm{RL}},c)$
    \State $p_{\mathrm{RL}}\gets\Call{Refresh}{\pi_{\mathrm{RL}},\mathcal S,V}$
    \State $H\gets H\cup\{x:\mathcal B^+(x)\ne\varnothing\ \lor\ p_{\mathrm{RL}}(x)>0\}$
    \Statex \hspace{\algorithmicindent}\hspace{\algorithmicindent}\textit{Repair with successes from this RL phase, then refresh again}
    \State $q(x)\gets g_{\mathrm{rep}}(p_{\mathrm{RL}}(x))$ for $x$ with $\mathcal B^+(x)\ne\varnothing$
    \State $\mathcal D_{\mathrm{rep}}\gets\Call{SampleReplay}{\mathcal B^+,q}$
    \State $\Pi_{\mathrm{SFT}}\gets\Call{RepairSFT}{\pi_{\mathrm{RL}},\mathcal D_{\mathrm{rep}}}$
    \State $\pi\gets\textit{Select}(\Pi_{\mathrm{SFT}},c)$
    \State $p_{\mathrm{SFT}}\gets\Call{Refresh}{\pi,\mathcal S,V}$
    \State $H\gets H\cup\{x:p_{\mathrm{SFT}}(x)>0\}$
    \State Record mastery changes $p_{\mathrm{RL}}-p_{\mathrm{pre}}$ and $p_{\mathrm{SFT}}-p_{\mathrm{RL}}$
    \If{$r<R$}
      \Statex \hspace{\algorithmicindent}\hspace{\algorithmicindent}\hspace{\algorithmicindent}\textit{Expand beyond the preceding RL training set}
      \State $\mathcal E\gets\{x\in\mathcal D_c\setminus\mathcal S:p_c^0(x)\in\{0,1\}\}$
      \State $\mathcal U\gets\mathcal S\cup\Call{RandomSample}{\mathcal E}$
      \State $p\gets\Call{Refresh}{\pi,\mathcal U,V}$
      \State $H\gets H\cup\{x:p(x)>0\}$
      \State $\mathcal S\gets\Call{FrontierSelect}{\mathcal U,p,H}$
    \EndIf
  \EndFor
  \State $T_c\gets\pi$ \Comment{Best target-category model in the final training phase}
\EndFor
\State \Return $\{T_A,T_B,T_C\}$
\end{algorithmic}
\end{algorithm}

\begin{figure*}[t]
  \centering
  \includegraphics[width=\textwidth]{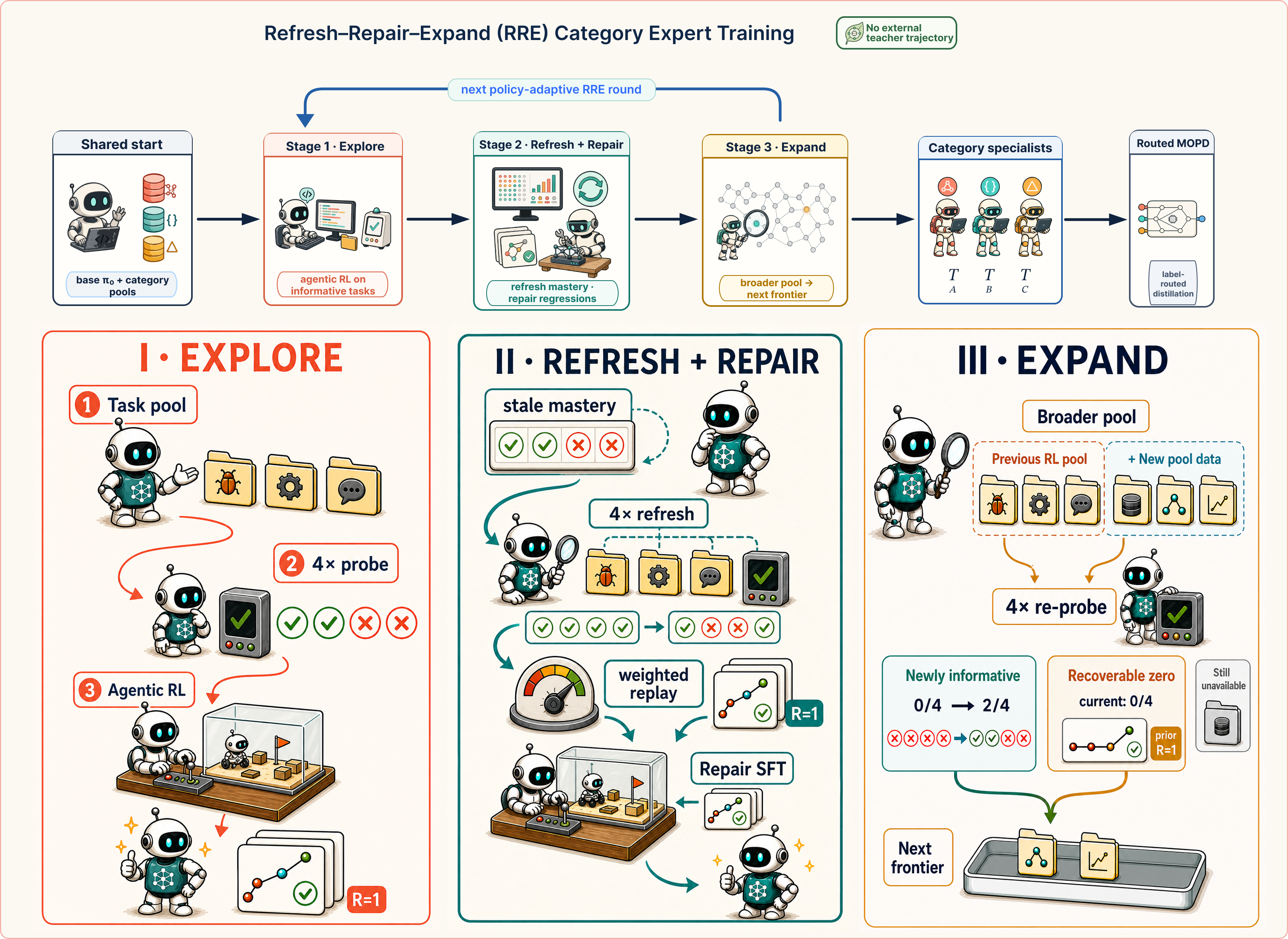}
  \caption{\textbf{Refresh--Repair--Expand training of category experts.}
  The same loop is instantiated independently for categories A, B, and C from
  a common strong base.  Initial selection reuses the base-model probes from
  corpus construction.  Agentic RL explores on an informative category
  frontier and records the expert's own verifier-approved trajectories.
  Mastery is refreshed on the preceding RL training set after both RL and
  Repair SFT.  The post-RL estimates guide successful-trajectory replay;
  the post-SFT estimates track the repaired policy's mastery.  Before the next
  RL phase, the selected SFT policy probes an expanded pool that augments the
  preceding RL set with randomly sampled base-zero and base-one tasks.
  Solid navy arrows
  track stage and policy progression, green arrows reuse self-generated
  verifier-approved trajectories, and dashed control arrows indicate
  mastery-weighted replay.  The blue return path closes the next RRE round.  No
  external trajectory teacher is used.}
  \label{fig:rre_training}
\end{figure*}

\paragraph{Signal-aware initial exploration.}
Let $\mathcal{D}_c$ be the candidate pool for category $c$.  We estimate the
mastery of instance $x$ under a policy $\pi$ with $M$
independent executable rollouts,
\begin{equation}
  \widehat p(x;\pi)=
  \frac{1}{M}\sum_{j=1}^{M}R(\tau_j),
  \qquad
  \tau_j\sim\pi(\cdot\mid x),
  \label{eq:rre_mastery}
\end{equation}
where $R(\tau)\in\{0,1\}$ is the verifier outcome.  Initial expert training
reuses the base-model mastery estimates computed during corpus construction:
the category training sets $\mathcal S_c^0$ inherit their pass rates from the
shared annotated corpus, with no additional initial refresh.  The RL selector
prioritizes non-extreme pass rates, for
which repeated rollouts expose both successes and failures, and admits a small
set of zero-pass-rate instances only when the model has produced a verified
success for that instance elsewhere.  Saturated instances and zero-pass-rate
instances without any success evidence consume little or no exploration
budget.  This follows the reward-signal motivation behind dynamic sampling
\citep{yu2025dapo}, but tracks per-instance mastery across training phases rather than
only filtering all-correct or all-incorrect groups inside an RL update.  The
selected set $\mathcal{S}_{c,\mathrm{RL}}^{(r)}$ is optimized with the Agentic
RL objective in Section~\ref{sec:minirl}; exact pass-rate strata and budgets are
specified in Section~\ref{sec:experiments}.

\paragraph{Coupling policy progress with data refresh and replay.}
The initial category pool is selected using executable pass-rate estimates from
the base policy.  These estimates characterize a \emph{policy--instance pair};
they are not immutable properties of the instance.  Once RL changes the
policy, the pass-rate landscape drifts and the original selection gradually
ceases to describe where useful training signal lies.  Selected instances can
become saturated and provide little rollout contrast, base-zero instances can
become learnable while remaining outside the fixed pool, and previously solved
instances can regress even as aggregate performance improves.  These
possibilities motivate revisiting both the instance selection and the
successful trajectories accumulated during RL.

Online resampling updates which tasks are explored next.  RRE couples that
choice with a separate supervised replay phase that directly trains on
successful behavior from the expert's earlier rollouts.  We therefore pause
after each RL phase to reevaluate the selected expert and allocate historical
successes to low-mastery instances.  We refresh again after Repair SFT because
supervised updates also change instance mastery.  Reassessing the preceding
RL set tracks these changes, while probing the broader candidate pool before
the next RL phase finds training signal outside the original selection.
These phase boundaries make the updated policy the basis for both repair
supervision and the next exploration distribution.

\paragraph{Refresh: synchronize the mastery map with the policy.}
Within each training phase, we select the model with the highest target-category
resolution $S_c$ among that phase's evaluated models, using the corresponding
Pro subset as specified in Section~\ref{sec:experiments}.  Denote the selected
post-RL and post-SFT policies in round $r$ by $\pi_{c,\mathrm{RL}}^{(r)}$ and
$\pi_{c,\mathrm{SFT}}^{(r)}$, respectively.  After RL, we reevaluate
$\mathcal S_{c,\mathrm{RL}}^{(r)}$ under $\pi_{c,\mathrm{RL}}^{(r)}$ to obtain
$\widehat p_{c,\mathrm{RL}}^{(r)}(x)$; let
$\widehat p_{c,\mathrm{pre}}^{(r)}(x)$ be the corresponding estimate under
the policy entering that RL phase.  The paired change
\begin{equation}
  \Delta_c^{(r)}(x)=
  \widehat p_{c,\mathrm{RL}}^{(r)}(x)-\widehat p_{c,\mathrm{pre}}^{(r)}(x)
  \label{eq:rre_delta}
\end{equation}
measures changes in estimated instance mastery: aggregate improvement does
not imply that every previously solved instance remains solved.  The
new $\widehat p_{c,\mathrm{RL}}^{(r)}$ also replaces the pre-RL mastery map.
The paired estimates locate observed regressions, while current mastery
guides replay allocation and future RL data selection.

\paragraph{Repair: consolidate self-generated successes.}
During the preceding RL phase, the expert has already accumulated a
trajectory buffer from its own on-policy rollouts,
\begin{equation}
  \mathcal{B}_c^{(r)}(x)=
  \left\{\tau:\tau\text{ was generated for $x$ in category-$c$ RL phase $r$},
  \ R(\tau)=1\right\}.
  \label{eq:rre_buffer}
\end{equation}
We replay only these verifier-approved trajectories.  An instance receives a
quota $q_c^{(r)}(x)=g_{\mathrm{rep}}(\widehat p_{c,\mathrm{RL}}^{(r)}(x))$, where
$g_{\mathrm{rep}}$ is non-increasing: low-mastery instances receive
more distinct successful trajectories, whereas currently mastered instances
receive fewer.  If fewer successful trajectories exist than the quota, they
are cycled only to meet the fixed repair budget.  The preceding delta is used
to diagnose which low-mastery cases are regressions; the actual replay quota is
determined by current mastery. A zero-success post-RL probe does not imply
an empty historical buffer; instances without any verified successful
trajectory in that buffer receive no repair examples. Starting from
$\pi_{c,\mathrm{RL}}^{(r)}$, Repair SFT minimizes
\begin{equation}
  \mathcal L_{c,\mathrm{rep}}^{(r)}(\pi)
  =-
  \sum_{x}\sum_{\tau\in
  \operatorname{Sample}(\mathcal{B}_c^{(r)}(x),q_c^{(r)}(x))}
  \log\pi(\tau\mid x).
  \label{eq:rre_repair}
\end{equation}
Here $\log\pi(\tau\mid x)$ abbreviates
$\sum_{t\in\mathcal I_{\mathrm{assistant}}(\tau)}
\log\pi(y_t\mid h_t)$, where $\mathcal I_{\mathrm{assistant}}(\tau)$
indexes only assistant-generated tokens; prompt and tool-observation tokens
are excluded from the loss. The $\operatorname{Sample}$ operator returns a
multiset, so repeated trajectories from cyclic replay contribute with their
sampling multiplicities.
The highest-scoring model on category $c$ within this SFT phase becomes
$\pi_{c,\mathrm{SFT}}^{(r)}$.  We then refresh mastery on the same RL
training set to obtain $\widehat p_{c,\mathrm{SFT}}^{(r)}$ and track its
change from $\widehat p_{c,\mathrm{RL}}^{(r)}$.  This refresh is performed
even after the final repair phase.
We call this step \emph{Repair SFT}: it is a reward-filtered
form of trajectory self-distillation, not a second source of expert
demonstrations.  It shares the teacher-free principle of recent self-distillation
methods \citep{hubotter2026sdpo}, while using complete, executable SWE
trajectories verified by the environment rather than feedback-conditioned
token distributions.

\paragraph{Expand: reopen the current learning frontier.}
Before the next RL phase, we augment $\mathcal S_{c,\mathrm{RL}}^{(r)}$
with a random sample of additional instances from $\mathcal D_c$ whose
base-model pass rates were zero or one.  Denote the resulting refresh pool by
$\mathcal U_c^{(r+1)}$.  Sampling these additional instances keeps reevaluation
within a bounded budget while revisiting tasks omitted by the initial
curriculum: previously unsolved tasks may now yield successes, and previously
saturated tasks may no longer be fully mastered.  We evaluate
$\pi_{c,\mathrm{SFT}}^{(r)}$ on $\mathcal U_c^{(r+1)}$ to obtain the next
pre-RL mastery map
$\widehat p_{c,\mathrm{pre}}^{(r+1)}$; it complements the post-SFT refresh on
the preceding training set.  Infrastructure-failed or incomplete refreshes
are rerun before an instance can enter the next training set.  Let
$H_c^{(r)}(x)=1$ indicate that $x$ has at
least one verified success in the base evaluation, an earlier expert
evaluation, or the expert's accumulated RL history.  RRE constructs the
next exploration set from
\begin{align}
  \mathcal{F}_c^{(r+1)}
    &=\{x\in\mathcal U_c^{(r+1)}:0<\widehat p_{c,\mathrm{pre}}^{(r+1)}(x)<1\},\\
  \mathcal{Z}_c^{(r+1)}
    &=\{x\in\mathcal U_c^{(r+1)}:\widehat p_{c,\mathrm{pre}}^{(r+1)}(x)=0\ \land\ H_c^{(r)}(x)=1\},\\
  \mathcal{S}_{c,\mathrm{RL}}^{(r+1)}
    &=\mathcal{F}_c^{(r+1)}\cup\mathcal{Z}_c^{(r+1)}.
  \label{eq:rre_expand}
\end{align}
The first set is the current informative frontier; the second retains
currently unsolved instances with evidence of historical success.  Pure
hard-zero and currently saturated instances are excluded from the main RL
budget.  Agentic RL then resumes from $\pi_{c,\mathrm{SFT}}^{(r)}$, closing
the refresh--repair--expand cycle shown in Figure~\ref{fig:rre_training}.
After the final round, $T_c=\pi_{c,\mathrm{SFT}}^{(R)}$: the expert is the
highest-scoring target-category model within the final training phase.

RRE is related to contemporary self-evolution frameworks that alternate
experience internalization and policy optimization.  In particular, SPEE
extracts transferable textual experience from successful and failed reasoning
trajectories, distills that experience, and then applies GRPO
\citep{ren2026spee}.  RRE advances this paradigm to long-horizon agentic SWE and
changes the object that evolves: instead of maintaining an abstract global
experience pool, it maintains a policy-adaptive, executable task frontier.
The refresh is grounded in repeated environment verification, the repair phase
reuses the same expert's successful action trajectories, and expansion is
conditioned on both current mastery and historical success.  Repeating this
loop independently yields $T_A$, $T_B$, and $T_C$, which share an origin and
agent interface and are optimized on different category pools.  Their
stage-wise performance and subsequent integration are evaluated in
Section~\ref{sec:results}.

\subsection{Label-routed multi-teacher on-policy distillation}
\label{sec:mopd}

Separate experts provide category-conditioned policies, while deployment
calls for a single model.  We use the same category map to assign each
training task to its corresponding expert during MOPD. The student is
initialized from the original base policy $\pi_0$ in the primary 27B setting and trained on a
category-balanced mixture of the three initial-RL task pools, formed by
upsampling the smaller pools to match the largest. The student
$\pi_\theta$ generates the complete trajectory, so teachers are
queried only on prefixes the student actually visits, following the on-policy
distillation setting of MOPD~\citep{ma2026mopdmultiteacheronpolicydistillation}.
For each category expert $T_k$, $k\in\{1,\dots,K\}$, let $m_k(x)\in\{0,1\}$
be the task-level route and let $q_t$ be the behavior log probability under
the rollout policy.  On sampled tokens we define two log ratios: the
student--teacher gap and the teacher--reference gap,
\begin{equation}
  d_{t,k}=q_t-\log\pi_{T_k}(y_t\mid h_t),
  \qquad
  e_{t,k}=\log\pi_{T_k}(y_t\mid h_t)-
  \log\pi_{\mathrm{ref}}(y_t\mid h_t).
\end{equation}
The routed multi-teacher advantage used in our runs includes a ReLU-gated
extrapolation term:
\begin{align}
  A_t^{\mathrm{MOPD}}
    &= \sum_{k=1}^{K}m_k(x)
       \bigl[-d_{t,k}+(\lambda_k-1)\max(e_{t,k},0)\bigr].
  \label{eq:mopd_advantage}
\end{align}
With $\lambda_k=1$, the second term vanishes and the advantage reduces exactly
to the MOPD imitation term; with $\lambda_k>1$, the second term extrapolates
beyond imitation on tokens assigned higher probability by the teacher than
by the reference.  MOPD
runs in pure distillation mode: each task is rolled out once, the environment
return does not enter the student loss, the group baseline and K1 reference
penalty are disabled, $\pi_{\mathrm{ref}}$ appears only as the extrapolation
anchor, and Equation~\ref{eq:mopd_advantage} is the entire advantage signal.
Response masks intersect teacher-route masks.  The A/B/C mapping is mutually
exclusive, so each routed task receives the signal of its corresponding
expert; an unrouted sample receives zero advantage.  The per-teacher imitation
coefficient is one in our configuration.

\paragraph{The student objective.}
The complete MOPD loss reuses the turn-aware reduction of
Section~\ref{sec:minirl}, with the routed advantage entering as a per-token
weight:
\begin{equation}
  \boxed{\;
  \mathcal{L}_{\mathrm{MOPD}}(\theta)
  =-\frac{1}{B_{\mathrm{valid}}}\sum_i\frac{1}{|\mathcal{S}_i|}
  \sum_{s\in\mathcal{S}_i}\sum_{t\in s}
  \tilde A_t^{\mathrm{MOPD}}\,
  \log\pi_\theta(y_t\mid h_t).\;}
  \label{eq:mopd_objective}
\end{equation}
Here $\tilde A_t^{\mathrm{MOPD}}=\operatorname{clip}\!\left(A_t^{\mathrm{MOPD}},
-c_a,c_a\right)$ is the bounded routed advantage, with finite clip bound
$c_a=5$. Teacher log-probabilities,
the behavior log-probability $q_t$, and the advantage itself are treated as
non-differentiated quantities, and only the student log-likelihood is
differentiated.  The gradient raises the student's probability of a sampled
token when its routed advantage is positive: the imitation component favors
tokens the routed teacher assigns more probability than the rollout policy
did, and the gated extrapolation component adds a nonnegative advantage
contribution on tokens favored by the teacher over the reference.  The
update is a vanilla policy gradient: unlike Agentic-miniRL, it applies no
proximal mask or importance-ratio correction, retaining only the finite
advantage clip so that no single token's student--teacher log-ratio can
dominate the token-summed turn reduction. Token losses are summed within each
assistant turn, and these turn-level sums are averaged within each trajectory
by the outer $1/|\mathcal{S}_i|$ factor.

\paragraph{Reward extrapolation and the reference policy.}
Standard on-policy distillation pulls the student toward the teacher.
We augment this imitation signal with reward extrapolation
(ExOPD)~\citep{yang2026learningbeyondteacher}: the
$(\lambda_k-1)\max(e_{t,k},0)$ term in Equation~\ref{eq:mopd_advantage} allows the
student update to include a reference-anchored extrapolation direction rather
than imitation alone.  We adopt this formulation from prior work on
cross-domain expert integration and apply it to intra-domain, label-routed
experts.

The coefficient $\lambda_k$ is set per teacher.  The audited configuration uses
$\lambda_k=1.25$.  For
$\pi_{\mathrm{ref}}$, we use the student's initial policy as the reference.
In the primary 27B setting, this is also the base from which category
specialization began. The 35B-A3B extension retains the same 27B teachers
and uses the 35B-A3B student's initial policy as its reference.
This choice costs one additional
frozen reference forward.  The integration hyperparameters
are reported in Section~\ref{sec:training_configuration}. 

\paragraph{Rectifying the extrapolation direction.}
The teacher--reference gap $e_{t,k}$ takes both signs across sampled tokens:
positive where the routed teacher assigns more probability than
$\pi_{\mathrm{ref}}$, negative when it assigns less.  Without gating, the
extrapolation term thus mixes positive and negative contributions, and in our
runs these components are comparable in magnitude, so the net extrapolation
magnitude remains well below that of the imitation term.  We therefore
rectify the extrapolation direction with a ReLU gate that retains the positive
teacher--reference log-probability gaps,
\begin{equation}
  \tilde e_{t,k}=\max(e_{t,k},\,0),
  \label{eq:mopd_rectifier}
\end{equation}
as already included in Equation~\ref{eq:mopd_advantage}, without changing
the routing, the imitation term, or the aggregation.  The gate zeroes only
the extrapolation contribution where the routed teacher assigns no higher
probability than the reference; the imitation signal remains unchanged.
With $\lambda_k\geq 1$, the retained extrapolation contribution is nonnegative
on every token and the term
acquires a systematic direction instead of acting as a signed residual that
cancels across tokens.  The ReLU-gated extrapolation is enabled in the
audited configuration.

\paragraph{Same-origin teachers and monitoring.}
The teachers are same-origin category experts, sharing tokenization,
prompt format, base checkpoint, and agent interface. In the primary 27B
setting the student shares this initialization as well. The 35B-A3B extension
tests reuse of the frozen teacher set with a different student architecture
(Appendix~\ref{app:expanded_evaluation}).
Section~\ref{sec:results_integration} reports training rollout scores and
student--teacher KL during integration. Extrapolation raises the number of frozen
teacher-side forwards from $K$ to $K{+}1$.

\section{Experimental setup}
\label{sec:experiments}

\subsection{Research questions}

Our central evaluation asks whether category-aware expert training and
integration produce a stronger single SWE policy.  Three research questions
connect Balanced RL and expert development to this final comparison.
\textbf{RQ1} examines whether Balanced RL alleviates the
uneven category gains observed under Pooled RL, alongside overall resolution.
\textbf{RQ2} examines the initial category experts, the instance-level learning
patterns that motivate further development, and how target-category performance
evolves through the RL and repair stages of RRE.
\textbf{RQ3} evaluates
whether label-routed MOPD can consolidate these experts into one policy
while retaining their category-specific gains and outperforming
Pooled RL and Balanced RL in overall and per-category resolution on Pro-618,
with additional overall comparisons on Pro V2 and Multilingual.
We use \emph{joint RL} as the umbrella term for training one shared
policy across categories. The two Agentic-miniRL baselines are named
\emph{Pooled RL} and \emph{Balanced RL}; figures may abbreviate them as
\emph{Pooled} and \emph{Balanced}.

\subsection{Base model and executable agent}

The principal 27B policies start from the same Qwen3.6-27B base
model~\citep{qwen2026qwen36}.  Training
and Pro-618 evaluation use the same R2E-Gym agent scaffold
\citep{jain2025r2egymproceduralenvironmentshybrid}: the model receives
the issue description and accumulated interaction history, invokes structured
repository and shell tools, and submits a patch that is scored by executable
fail-to-pass and pass-to-pass tests. A training rollout uses a
131{,}072-token context, at most 150 agent actions, and at most 12{,}288 newly
generated tokens per action; Agentic RL samples $G{=}8$ trajectories per task.

\subsection{Benchmarks and evaluation configurations}
\label{sec:benchmark_protocols}

Pro-618 is the fixed evaluation set for checkpoint selection, longitudinal
category analysis, and stage-wise expert comparisons. We additionally evaluate
the fixed final policies on SWE-bench Pro V2 and SWE-bench Multilingual. Table~\ref{tab:benchmark_protocols} distinguishes these
task populations and agent frameworks. The original 731-task Pro population,
a 175-task LiveCodeBench v6 set, and the 35B-A3B student extension are reported
in Appendix~\ref{app:expanded_evaluation}.

\begin{table}[htbp]
  \centering
  \caption{Evaluation sets and their roles. Pro-618 remains fixed for the
  original training-process analyses. Pro V2 and Multilingual evaluate the
  final policies on Pro V2 and Multilingual; Hard-51 inherits its V2 parent
  runs. Configured generation budgets are described in the text and
  Appendix~\ref{app:expanded_protocols}.}
  \label{tab:benchmark_protocols}
  \small
  \begin{tabularx}{\linewidth}{@{}lrlY@{}}
    \toprule
    Evaluation & Tasks & Agent framework & Role \\
    \midrule
    Pro-618 & 618 & R2E-Gym & Selection, category dynamics, expert stages \\
    Pro V2 & 642 & mini-swe-agent & Final-model evaluation \\
    Hard-51 & 51 & V2 parent runs & Difficult-subset analysis \\
    Multilingual & 300 & SWE-agent & Final-model evaluation \\
    \bottomrule
  \end{tabularx}
\end{table}

\paragraph{Pro-618.}
Our category-analysis set is an audited 618-instance subset of SWE-bench Pro
\citep{deng2025swebenchpro}, evaluated with the R2E-Gym agent scaffold.
The common category mapping yields 221, 201, and 196 instances in Pro-A,
Pro-B, and Pro-C. These scores support checkpoint selection, longitudinal
category analysis, stage-wise expert evaluation, and the original final
policy comparison. Base, Pooled RL, Balanced RL, MOPD, and expert endpoint
results report means and population standard deviations over three evaluation
rounds; checkpoint sweeps use the separately documented round selections.
The selection rules are specified below.

\paragraph{SWE-bench Pro V2 and Hard-51.}
The official V2 release contains 642 validated tasks with revised
instructions, verifiers, and images~\citep{scaleai2026swebenchprov2}.
It revises the original Pro task population and is distinct from our
audit-filtered Pro-618 set. We evaluate final policies with mini-swe-agent
2.4.6 and fresh-environment patch replay. Hard-51 is the official 51-task
subset of V2; its results are extracted from the same selected parent runs,
rather than obtained from a separate set of evaluations.

\paragraph{SWE-bench Multilingual.}
We evaluate all 300 instances using SWE-agent. The final MOPD policy, base model, Pooled RL, and Balanced RL
each have three selected evaluation rounds. Pooled RL and Balanced RL retain
their Pro-618-selected checkpoints. The existing category mapping assigns
72/25/176 instances to A/B/C and 27 to the unassigned group; the latter remain
in Full. Category results reported for the base and MOPD are recomputed from
the selected task-level outcomes.

\paragraph{Execution and scoring.}
Each rollout produces one candidate patch, scored by the benchmark's
executable verifier. The reported rate counts confirmed passes over the full
task denominator; unscored outcomes remain missing in the source records and
are not dropped from the denominator. Comparisons are made within a benchmark
and evaluation setting. Pro-618 uses common agent settings across the 27B
policies. Generation budgets and decoding requests for the final-model
benchmarks are detailed in Appendix~\ref{app:expanded_protocols}.
All base-model scores are obtained with our evaluation harnesses.

Pro-618 evaluation uses BF16 vLLM inference with a 200{,}000-token context,
temperature 1.0, top-$p$ 0.95, and top-$k$ 20, with one trajectory per task per
round. It permits 25{,}600 output tokens per call and 150 agent actions, with a
10{,}000-second environment timeout. Its verifier applies the model patch in a
fresh sandbox. The Multilingual setup permits up to 300 iterations
with a 7{,}200-second task limit. Its recorded generation budgets vary by model
(Appendix~\ref{app:expanded_protocols}).

\subsection{Pro-618 construction and audit}
\label{sec:pro618_construction}

SWE-bench Pro spans varied repository contexts, maintenance intents, and
modification scales~\citep{deng2025swebenchpro}. The cross-benchmark Labeler
profiles in Table~\ref{tab:labeler_benchmark_profiles} quantify this breadth:
47.3\% of Pro instances are bug fixes and 59.2\% require cross-module changes,
compared with 87.0\% bug fixes and 86.2\% single-file changes in SWE-bench
Verified. This breadth supports studying category-level learning within
repository-level SWE.

For repeated checkpoint and category comparisons, we fix an audit-filtered
subset of the original 731-task release. Released instance-level evidence
identifies specification/determinacy defects, underspecified grading choices,
reference-patch replay failures, and a prompt--test mismatch
\citep{kim2026determinacyaudit,openai2026signalnoise}.
Excluding these 113 tasks yields Pro-618, with 221/201/196 instances in
Pro-A/B/C. Its membership and evaluation setup remain fixed throughout the
longitudinal and stage-wise analyses. Appendix~\ref{app:pro618_audit} gives
the exclusion accounting.

\subsection{Training data and category-conditioned curricula}

The executable candidate pool described in Section~\ref{sec:data} contains
approximately 32K instances. Before training-corpus selection, \labeler{} uses
Qwen3.7-max to annotate this entire synthetic pool.  These labels are retained
for both initial category assignment and subsequent RRE pool expansion.
To construct the training corpus, we evaluate the base
model on these candidates and select instances with estimated pass rates
between 0 and 0.5, focusing training on tasks that the initial policy has not
yet reliably mastered.  This selection yields 3{,}048 synthetic instances.
We combine them with 3{,}675 tasks from a four-language slice of
SWE-rebench V2~\citep{badertdinov2026swerebenchv2languageagnosticswe}, forming
the 6{,}723-task corpus used by the Pooled RL and Balanced RL comparisons.
Every task has a
containerized repository, an issue-style specification, and executable
fail-to-pass and pass-to-pass tests.

The SWE-rebench V2 instances are also annotated by \labeler{} using
Qwen3.7-max.  The same Domain L1 rule in
Section~\ref{sec:labeler} maps eligible instances into A
(service/data/security), B (user-facing applications), or C
(systems/tooling/runtimes).  Of the 6{,}723 instances, 2{,}769 map to these
three categories: 601 to A, 516 to B, and 1{,}652 to C.  The remaining
instances fall outside the A/B/C mapping.  The Pooled RL see-saw experiment
trains one policy on all 6{,}723 tasks.  The stratified Balanced RL run samples 516 tasks from each
category, for 1{,}548 tasks in total.  Within this initial 6{,}723-task corpus,
category B contains 516 instances and therefore determines the common
per-category budget.

\begin{table}[ht]
  \centering
  \caption{\textbf{Training data used by the controlled experiments.}
  A, B, and C are shown in separate columns.  Candidate and repairable rows
  report the support available to each selection step; training and sampled
  rows report the materialized data.  Repair trajectories are sampled only
  from verifier-approved rollouts generated by the same expert in the
  immediately preceding RL phase. MOPD repeats A/B task records to match
  C's size; source-record counts and upsampled record counts are shown separately.}
  \label{tab:training_curricula}
  \scriptsize
  \setlength{\tabcolsep}{4pt}
  \renewcommand{\arraystretch}{1.05}
  \begin{tabularx}{\columnwidth}{@{}XXrrrr@{}}
    \toprule
    & & \multicolumn{3}{c}{Category} & \\
    \cmidrule(lr){3-5}
    Phase & Quantity & A & B & C & Total \\
    \midrule
    Pooled RL
      & training instances
      & --- & --- & --- & 6,723 \\
    Balanced RL
      & training instances
      & 516 & 516 & 516 & 1,548 \\
    \addlinespace[2pt]
    Initial expert RL
      & training records
      & 601 & 516 & 1,652 & 2,769 \\
    \addlinespace[2pt]
    \multirow[t]{2}{=}{First Repair SFT}
      & repairable instances
      & 571 & 501 & 1,498 & 2,570 \\
      & sampled preceding-RL trajectories
      & 1,623 & 1,481 & 4,384 & 7,488 \\
    \addlinespace[2pt]
    \multirow[t]{2}{=}{Expanded expert RL}
      & candidate instances
      & 1,000 & 1,000 & 2,282 & 4,282 \\
      & training instances
      & 566 & 502 & 1,435 & 2,503 \\
    \addlinespace[2pt]
    \multirow[t]{2}{=}{Second Repair SFT}
      & repairable rollout keys
      & 549 & 487 & 1,371 & 2,407 \\
      & sampled preceding-RL trajectories
      & 1,392 & 1,245 & 3,840 & 6,477 \\
    \addlinespace[2pt]
    \multirow[t]{2}{=}{MOPD integration}
      & training records from initial RL pools
      & 601 & 516 & 1,652 & 2,769 \\
      & category-balanced training records
      & 1,652 & 1,652 & 1,652 & 4,956 \\
    \bottomrule
  \end{tabularx}
\end{table}

Initial expert training uses these category-specific instances and reuses
their stored base-model pass rates without an additional initial mastery
refresh.  After each RL phase, we reevaluate the selected expert on that
phase's training instances using four fresh executable rollouts per instance.
These updated pass rates guide Repair SFT sampling from verifier-approved
successful trajectories generated by the same expert during that RL phase.
Table~\ref{tab:repair_sampling} allocates more successful
trajectories to instances whose current mastery is lower; a trajectory is
reused only when an instance has fewer distinct successful rollouts than its
target.  After Repair SFT, we refresh mastery on the same RL training
instances with four fresh rollouts per instance.  Before the next RL phase,
we additionally evaluate an expanded candidate pool with the selected
post-SFT expert to identify newly informative tasks.  The next RL set combines
the current frontier with historically supported recoverable-zero instances,
following Section~\ref{sec:rre}.

After the first repair, we augment each category's initial RL training set
with randomly sampled instances from the already labeled 32K synthetic pool
that had base-model pass rates of zero or one and were omitted from initial
RL.  The additions are assigned using their existing labels and the same
Domain L1 rule; no new labeling is required.  The resulting refresh pools contain
1{,}000, 1{,}000, and 2{,}282 instances for A/B/C, respectively.
The selected post-SFT experts reevaluate these pools to identify changes in
mastery and new training signal among previously excluded tasks.  Thus the
broader refresh uses sampled additions rather than rescoring the full 32K
synthetic candidate pool.  From these 4{,}282 candidates, the expanded-stage
selection materializes 566/502/1{,}435 training instances, as reported in
Table~\ref{tab:training_curricula}.
After this expanded RL phase, a second mastery refresh materializes
1{,}392/1{,}245/3{,}840 verifier-approved repair trajectories for A/B/C.

\begin{table}[ht]
  \centering
  \caption{\textbf{Mastery-adaptive sampling for Repair SFT.}
  Mastery is measured with four fresh executable rollouts.  Each selected
  trajectory must be a verifier-approved success previously generated by the
  same expert.}
  \label{tab:repair_sampling}
  \small
  \begin{tabular}{@{}ccp{0.48\columnwidth}@{}}
    \toprule
    Current successes & Target trajectories & Sampling emphasis \\
    \midrule
    0 or 1 of 4 & 4 & strongest repair pressure \\
    2 of 4      & 3 & unstable frontier \\
    3 of 4      & 2 & consolidation \\
    4 of 4      & 1 & retention \\
    \bottomrule
  \end{tabular}
\end{table}

For MOPD, we balance and combine the three experts' Stage~1 RL task pools,
retaining their A/B/C labels for teacher routing. The source pools contain
601/516/1{,}652 records. We upsample A and B by repeating task records to
match C's 1{,}652 records, yielding 4{,}956 training records with equal
category counts. This changes sampling frequency without adding new tasks;
the source pools contain 2{,}769 training records before upsampling.
The student starts from the original Qwen3.6-27B base rather than a
post-RRE expert. Consequently, the stored pre-Stage~1 base-model pass rates
also characterize the student's initial mastery, providing the same basis
for reusing this selected task pool without an additional initial refresh.
Training uses fresh student-generated trajectories on these tasks, not
the experts' stored Repair SFT trajectories.
\FloatBarrier

\subsection{Training infrastructure and optimization}
\label{sec:training_configuration}

\paragraph{Agentic RL.}
We implement all RL phases in ROLL, which separates distributed policy
optimization, model inference, environment interaction, and reward
computation~\citep{wang2025roll}.  We use the agent framework from R2E-Gym
\citep{jain2025r2egymproceduralenvironmentshybrid}, with ROLL executing each
trajectory in a ROCK sandbox provisioned from the task's repository image; the sandbox
isolates file-system changes and process failures while exposing the shell,
repository tools, and executable verifier to the agent
\citep{wang2025letitflow}.  Each sandbox receives 4 CPU cores and 16\,GiB of
memory, with a 9{,}600-second trajectory limit.

The initial and expanded RL phases use one common optimization configuration
for A, B, and C (Table~\ref{tab:training_hyperparameters}).  ROLL allocates 32
devices to policy optimization and 32 to vLLM rollout
generation.  Generation and optimization may overlap with an asynchrony bound
of one policy update.  Within an RL phase, the task manifest is fixed:
dynamic resampling is disabled and tasks are traversed without replacement
within each epoch.  The mastery refreshes and curriculum changes described in
Section~\ref{sec:rre} therefore occur between phases.

\paragraph{Repair SFT.}
After each RL phase, each selected expert is fully fine-tuned on its own
verifier-approved repair trajectories.  The loss is applied to assistant
response tokens while user, system, and tool-observation context remains
conditioning input.  We use the same 131{,}072-token limit as RL, Adam with a
cosine schedule, and bf16 training.  A/B/C use effective batches of
64/64/128, and each repair phase runs for approximately five corpus epochs.
All other SFT settings are shared
across categories and repair phases.

\begin{table}[h!]
  \centering
  \caption{\textbf{Core training configurations.}
  RL settings are shared by the initial and expanded phases and by all three
  category experts.  Category-valued SFT entries are ordered A/B/C.}
  \label{tab:training_hyperparameters}
  \scriptsize
  \setlength{\tabcolsep}{3.5pt}
  \begin{tabularx}{\columnwidth}{@{}lXX@{}}
    \toprule
    Setting & Agentic RL & Repair SFT \\
    \midrule
    Objective
      & \minirl{} with group-relative advantages
      & full-parameter, response-only cross-entropy \\
    Context
      & 131{,}072 tokens; 150 actions; 12{,}288 generated tokens/action
      & 131{,}072 tokens \\
    Batch
      & 16 tasks/update, $G{=}8$; 128 trajectories
      & micro-batch 1, accumulation 16; effective batch 64/64/128 \\
    Optimization
      & lr $3\!\times\!10^{-7}$; weight decay .01; cosine; 20 warmup steps
      & Adam, lr $5\!\times\!10^{-6}$; weight decay .01; cosine; 5\% warmup \\
    Stabilization
      & $(\epsilon_{\mathrm{low}},\epsilon_{\mathrm{high}})=(.20,.27)$;
        importance ratio $[0,5]$; KL coefficient .001; grad norm 20
      & bf16; grad norm 1 \\
    Generation
      & temperature 1.0; top-$p$ .95; top-$k$ 20
      & --- \\
    Parallelism
      & 32 train + 32 inference devices; train TP8/CP2; vLLM TP4
      & TP1/CP8/PP4; world size 128/128/256 \\
    Updates
      & one policy epoch per rollout batch
      & approximately five corpus epochs per repair phase \\
    \bottomrule
  \end{tabularx}
\end{table}

\paragraph{MOPD integration.}
The primary 27B student is trained by pure label-routed on-policy distillation with
the three final RRE experts frozen. Each rollout batch contains 128 tasks
with one trajectory per task, followed by one policy epoch, using learner
micro-batch size 1 and gradient accumulation of 64. We use a
learning rate of $3\times10^{-6}$, weight decay $0.01$, a cosine schedule
with no warmup, bf16 precision, and gradient-norm clipping at 20.
All three teachers use imitation coefficient 1 and extrapolation coefficient
$\lambda_k=1.25$ with the ReLU gate enabled; the token-level routed advantage
is clipped to $[-5,5]$.
The student uses the turn-aware vanilla policy-gradient objective in
Equation~\ref{eq:mopd_objective}, without an environment-reward term,
additional KL loss, or entropy bonus. Context, action, and per-action token
limits match expert RL, as do the decoding temperature, top-$p$, and top-$k$.
The upsampled task records are shuffled and traversed without replacement;
repeated copies of a task remain separate sampling entries. The training
seed is 42.
We allocate 32 devices each to student optimization and rollout generation,
16 to each teacher, and 16 to the frozen reference, with an asynchrony
bound of one policy update. The integration sandbox timeout is
7{,}200 seconds.
\FloatBarrier

\subsection{Reward integrity and anti-hacking protocol}
\label{sec:antihacking}

An executable success is meaningful only when the submitted source change,
rather than an unintended side effect, causes the required tests to pass.  We
distinguish two failure modes that are often conflated.  \emph{Verifier
exploitation} changes or bypasses the tests, harness, or execution state so
that an incorrect solution receives reward.  \emph{Solution leakage} recovers
the upstream fix from future Git history or an external repository.  The
latter may produce a functionally correct patch, but it does not measure the
agent's ability to solve the supplied task.  Both channels are realistic in
long-horizon coding RL~\citep{baker2025monitoringmisbehavior,
zhao2026specbench,rajan2026rewardhackability}.  Public SWE tasks also exhibit
memorization-sensitive performance, making it important to distinguish genuine
repository reasoning from recovered task artifacts
~\citep{liang2025swebenchillusion,
badertdinov2025swerebenchautomatedpipelinetask}.

An audit of unrestricted pilot trajectories exposed the characteristic
behaviors behind these risks: agents queried \texttt{git log --all}, read
future objects with \texttt{git show}, added or fetched upstream remotes, and
searched public hosting services for the corresponding fix.  We consequently
apply the defense-in-depth protocol in Table~\ref{tab:reward_integrity}.  The
controls operate at different boundaries: task construction removes textual
answer leakage; repository sanitization restricts the rollout state; deferred
tests and fresh-sandbox replay protect the verifier; and trajectory auditing
catches external answer retrieval that a repository-local control cannot see.

\begin{table}[htbp]
  \centering
  \caption{Defense-in-depth protocol for preserving reward integrity.  Each
  control closes a different path from an executable pass to a misleading
  training or evaluation signal.}
  \label{tab:reward_integrity}
  \small
  \setlength{\tabcolsep}{5pt}
  \renewcommand{\arraystretch}{1.08}
  \begin{tabularx}{\textwidth}{>{\raggedright\arraybackslash}p{0.19\textwidth}Y Y}
    \toprule
    Risk surface & Control & Integrity condition \\
    \midrule
    Problem-statement leakage &
    Separate fix and test patches; reject generated statements that reveal
    source/test paths or patch-added code. &
    The agent receives the issue-style specification and base repository, not
    the reference implementation or held-out test patch. \\
    Future repository state &
    Reset to the base revision; remove remotes, non-current branches,
    post-base tags, reflogs, and unreachable objects.  Pro-618 removes all
    tags in the rollout sandbox. &
    Future fixes cannot be recovered through local \texttt{git log},
    \texttt{git show}, tags, or dangling objects. \\
    Verifier and environment tampering &
    Defer held-out test injection; export only a sanitized source patch and
    replay it in a fresh instance sandbox. &
    Reward depends on the replayed patch under evaluator-controlled tests, not
    on rollout-side files, processes, packages, or test edits. \\
    External answer retrieval &
    Inspect the full action trajectory for target-repository search,
    clone/fetch, patch/raw downloads, and exact fix-SHA access. &
    Confirmed retrieval is an integrity failure, receives a score of zero,
    and remains in the fixed evaluation denominator. \\
    \bottomrule
  \end{tabularx}
\end{table}

For ordinary training instances, repository sanitization resets the worktree
to the declared base revision, removes upstream refs and non-current branches,
deletes tags not ancestral to that revision, expires reflogs, and prunes
unreachable objects.  Retaining ancestral tags avoids breaking projects whose
build systems derive version metadata from Git.  The stricter Pro-618 policy
removes all remotes, non-current branches, tags, reflogs, and unreachable
objects in the agent sandbox, while preserving image-built dependencies and
runtime files.  A failed sanitization invalidates the episode.

The evaluator never scores the mutable rollout workspace.  It extracts the
agent's source patch, removes unsupported binary hunks and known image-runtime
artifacts, starts a fresh sandbox from the same instance image, resets it to
the base revision, applies only the sanitized patch, and then injects the
held-out test patch and executes the verifier.  Thus edits to visible tests,
shell state, installed packages, or long-lived processes cannot carry into the
scoring environment.  Complete trajectories are additionally scanned for
target-repository search, clone/fetch, raw patch download, and exact fix-SHA
access.  Confirmed external answer retrieval is reported as an integrity
failure, assigned a score of zero, and retained in the evaluation denominator.

\FloatBarrier

\subsection{Baselines and comparison settings}

\paragraph{Pooled RL and the category see-saw.}
To answer RQ1, we evaluate the common base and every scheduled checkpoint of
the Pooled RL 6{,}723-task run on Pro-618.  We report the overall curve together with
Pro-A/B/C, rather than selecting only the best aggregate checkpoint.  The
1{,}548-task stratified Balanced RL run repeats the same analysis with 516
training instances per category.

For the Pooled RL endpoint in Table~\ref{tab:main_results}, we select the
checkpoint with the highest mean Pro-618 Full resolution among checkpoints
with three valid evaluation rounds.  We report Full and Pro-A/B/C means and
standard deviations from those same three rounds of that checkpoint.
For Balanced RL, we select one of the checkpoints tied for the highest Full
resolution in the single-round sweep, fix that checkpoint, and report its
mean and standard deviation across three evaluation rounds, including the
initial screening round.

\paragraph{Expert formation and RRE.}
To answer RQ2, A, B, and C experts share the same initialization and
training algorithm but receive their route-specific curricula.  We evaluate
each candidate checkpoint once on the corresponding Pro-A, Pro-B, or Pro-C
subset within each RL or Repair SFT phase, and select the checkpoint with
the highest target-category resolution to enter the next phase.  We then
fix the selected checkpoint and conduct repeated evaluations; the stage
table reports the mean and standard deviation over three evaluation rounds.
The final expert is selected within the last training phase by the same rule. These category
scores guide model selection and performance reporting; training-instance
pass rates are estimated separately to guide repair-trajectory sampling and
the next RL curriculum. Table~\ref{tab:rre_stage_results} reports each expert's
target-category resolution after initial RL, the first Repair SFT, expanded
RL, and the second Repair SFT.

\paragraph{Training-instance diagnostics.}
We align base-model, post-RL, and post-Repair SFT success-rate estimates on matched
records from the initial experts' training sets: 601 for A, 516 for B,
and 1,652 for C. The post-RL checkpoints are the selected initial experts.
The post-SFT estimates come from the
mastery refresh before expanded RL; newly probed tasks are excluded from
this matched-instance comparison. Actual repair-training membership identifies
which previously regressed instances received SFT examples.
Each instance's rate uses its actual
number of sampled attempts, typically four, and category means give each
training record equal weight. Repeated records share their instance-level
success-rate estimate. These training-set estimates diagnose observed
changes in instance mastery, separately from Pro-618 checkpoint selection
and performance reporting. Appendix~\ref{app:instance_dynamics} details
duplicate handling and the paired-snapshot protocol.

\paragraph{Expert integration.}
To answer RQ3, the MOPD student is compared with the common base,
Pooled RL and Balanced RL, and
category-routed experts.
The routed expert system selects the corresponding expert for each category.
The MOPD comparison
therefore tests how much of the expert improvement is
recovered in one policy.  The primary 27B student is initialized from the common base,
the same checkpoint from which the category experts began.

\subsection{Metrics, uncertainty, and failure accounting}

Final policy comparisons report overall and category-wise resolution.
\emph{Full} denotes the overall benchmark score in tables and figures;
\emph{Full gain} and \emph{overall gain} denote the same score difference.
On Pro-618, Full gain, minimum category gain $G_{\mathrm{sim}}$, and
\ssg{} summarize improvement over the common base, following
Equations~\ref{eq:simultaneous_gain} and~\ref{eq:ssg}.
Table~\ref{tab:pooled_balanced_metrics} averages these metrics over the
50 displayed checkpoints per method and reports final Full gain separately.
The process metrics are computed from per-checkpoint two-round mean scores
in Figure~\ref{fig:pooled_balanced}. Its compressed evaluation-index axis
preserves checkpoint order within each run, rather than aligning elapsed
training steps across runs.
Table~\ref{tab:final_policy_gains} compares the final single policies using
their three-round mean scores. Category-wise trajectories show the opposing
movements that characterize the temporal see-saw. Endpoint integration is
also evaluated by the aggregate and category lifts defined below.

For a candidate policy $\pi$, we measure its category lift over the
Pooled RL policy $\pi_{\mathrm{pooled}}$ as
\begin{equation}
  d_c(\pi)=S_c(\pi)-S_c(\pi_{\mathrm{pooled}}),
  \qquad
  d_{\min}(\pi)=\min_{c\in\{A,B,C\}}d_c(\pi).
  \label{eq:min_category_lift}
\end{equation}
We call $d_{\min}$ the \emph{minimum category lift over Pooled RL}.  Positive
$d_c(\pi_{\mathrm{MOPD}})$ for all three categories constitutes a
category-wise Pareto improvement over Pooled RL; $d_{\min}$ reports the weakest
of these three improvements as a single, directly interpretable number.  We
report the corresponding overall lift separately using the official Pro-618
aggregate.

We also compute the three category lifts and their minimum relative to
Balanced RL by replacing the comparator in
Equation~\ref{eq:min_category_lift}.

Expert formation is summarized by stage-wise target-category scores,
training rollout curves, and matched instance-level mastery changes.
Integration is summarized by category-expert gain recovery.  For target
expert $T_c$,
\begin{equation}
  \mathrm{Recovery}_c=
  \frac{S_c(\pi_{\mathrm{MOPD}})-S_c(\pi_{\mathrm{base}})}
       {S_c(\pi_{T_c})-S_c(\pi_{\mathrm{base}})}.
  \label{eq:recovery}
\end{equation}
When $T_c$ does not improve over the base, Recovery is undefined rather than
using a non-positive denominator.

Checkpoint sweeps are longitudinal diagnostics. Final endpoint comparisons
use repeated executable rollout rounds, retain paired task identities, and
report mean resolution and population standard deviation across three rounds,
with the task denominators stated. For the SWE-bench Multilingual
comparison between MOPD and the base, we additionally report a paired
instance-bootstrap 95\% interval. Each model's confirmed-pass indicators
for an instance are averaged across its three selected rounds; the 300 paired
instances are resampled with replacement 10{,}000 times (random seed 42),
and the 2.5th and 97.5th percentiles give the interval. These are evaluation
repeats, not independent training seeds or pass@3. Unscored outcomes retain
their missing status in the source records while remaining in the fixed
denominator. Integrity violations contribute no confirmed pass.

\FloatBarrier
\section{Results and analysis}
\label{sec:results}

The main outcome is the performance of the integrated single policy relative
to Pooled RL and Balanced RL.  To interpret that comparison, we first
examine the effect of balanced sampling on category-level learning.
We next trace the evolution of the three RRE experts and evaluate how much of
their category-specific improvement is retained by a single MOPD policy.

\subsection{Category dynamics under Pooled RL and Balanced RL}
\label{sec:results_balanced}

One way to use task categories is to balance the data composition of a shared
training stream.  To answer RQ1, we compare Pooled RL with
training on a category-balanced subset of 1{,}548 instances (516 per category),
stratified by base-model pass rate. The comparison extends the observations in
Section~\ref{sec:problem} by examining whether a more balanced training
mixture alleviates the observed category see-saw.
Guided by Table~\ref{tab:gain_interpretation}, we examine whether Balanced RL
yields higher Full gain, higher minimum category gain
$G_{\mathrm{sim}}$, and lower \ssg{} than Pooled RL.

\begin{figure}[htbp]
  \centering
  \includegraphics[width=\linewidth]{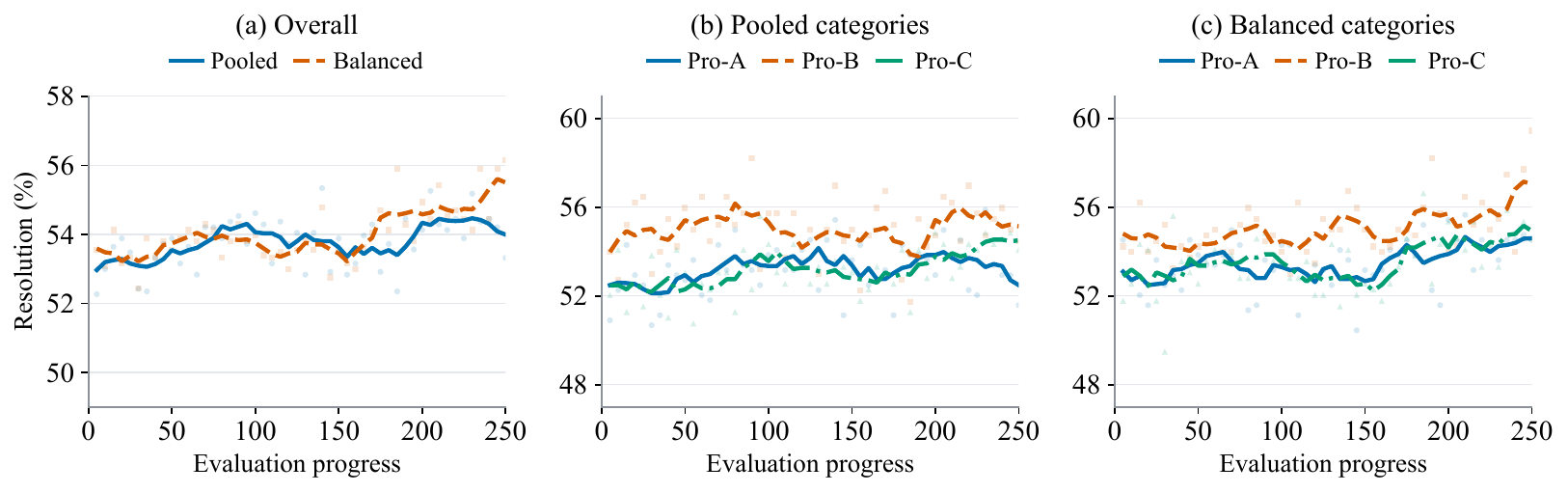}
  \caption{\textbf{Category dynamics under Pooled RL and Balanced RL.}
  Panel (a) compares Pro-618 Full resolution; panels (b) and (c) show the
  corresponding Pro-A/B/C trajectories on identical category-score axes.
  Faint points show two-round mean evaluation scores at each checkpoint.
  Lines show five-point centered moving averages.
  Category scores continue to exhibit
  opposing movements under Balanced RL.}
  \label{fig:pooled_balanced}
\end{figure}

Figure~\ref{fig:pooled_balanced}(a) shows similar overall performance across
the two training runs, despite Balanced RL using only 1{,}548 distinct
training instances compared with 6{,}723 for Pooled RL.  The endpoint results
in Table~\ref{tab:main_results} likewise show close Full resolution
(55.34\% versus 55.50\%).  Balanced RL also shows a local advantage on
Pro-B during parts of training, and its final model has a slightly
higher three-round mean Pro-B resolution (57.05\% versus 56.72\%).
Thus, in this setting, a smaller training set focused on the target categories
can retain competitive performance.

Table~\ref{tab:pooled_balanced_metrics} distinguishes final-model
performance from training-process summaries. Relative to the same base-model
Full score of 52.64\%, the final Pooled RL model gains
2.86 percentage points and the final Balanced RL model gains
2.70 points, using their three-round evaluation means. Both training recipes
therefore produce substantial overall improvement. The lower training-process
Full gains, 1.13 and 1.37 points respectively, average across the 50
displayed checkpoints per method rather than measuring final-model improvement.

The process summaries assess whether gains are sustained across categories.
Balanced RL has higher average Full gain (1.37 versus 1.13 points)
and minimum category gain $G_{\mathrm{sim}}$ (0.19 versus 0.08 points)
than Pooled RL. Its average \ssg{} is also higher
(1.18 versus 1.05 points).
The minimum-gain averages refer to the weakest category gain at each
evaluated checkpoint, not to the category scores of the final models.
Thus, balancing improves both average overall and minimum category gains,
but does not reduce the aggregate-to-minimum gap; it does not jointly improve all three
quantities sought in Table~\ref{tab:gain_interpretation}.
Thus, the process-level limitation concerns the consistency of category gains,
not an absence of overall learning in the final models.

\begin{table}[htbp]
  \centering
  \caption{\textbf{Category-gain summaries under Pooled RL and Balanced RL.}
  Training-process columns average per-checkpoint metrics over the
  50 displayed checkpoints per method, using two-round mean scores as in
  Figure~\ref{fig:pooled_balanced}.
  The last column reports the final models'
  three-round mean Full gain, using the same models as
  Table~\ref{tab:main_results}. All gains use the same three-round base-model
  reference; values are in percentage points.}
  \label{tab:pooled_balanced_metrics}
  \small
  \setlength{\tabcolsep}{5pt}
  \begin{tabular*}{\linewidth}{@{\extracolsep{\fill}}lrrrr@{}}
    \toprule
    & \multicolumn{3}{c}{Training-process mean} & \\
    \cmidrule(lr){2-4}
    Training mixture & Full gain $\uparrow$ & $G_{\mathrm{sim}}$ $\uparrow$ & \ssg{} $\downarrow$ & Final Full gain $\uparrow$ \\
    \midrule
    Pooled RL & 1.13 & 0.08 & 1.05 & \textbf{+2.86} \\
    Balanced RL & 1.37 & 0.19 & 1.18 & \textbf{+2.70} \\
    \bottomrule
  \end{tabular*}
\end{table}

The trajectories reveal a related temporal pattern.
Figure~\ref{fig:pooled_balanced}(c) still contains opposing category
movements: an improvement in one category can coincide with a regression
in another. Balanced RL can improve all three categories at individual
points, but their gains are not consistently sustained together over training.
Taken together, these results answer RQ1: Balanced RL maintains close
overall performance with fewer distinct training instances and raises
average minimum category gain, but does not reduce the average gap or
eliminate opposing category movements. This motivates examining a second way to use the category
structure: developing experts in separate training streams before integrating
them into one policy.

\subsection{Category-expert evolution through RRE}
\label{sec:results_rre}

To answer RQ2, we first assess the initial category experts and diagnose
their instance-level learning. We then examine how RRE develops these
experts before their subsequent integration by MOPD.

\paragraph{Does initial category-specific RL suffice?}
The initial experts improve their target-category Pro-618 scores over the
base model, reaching 53.54\%, 55.06\%, and 52.55\% on A, B, and C,
respectively (Table~\ref{tab:rre_stage_results}). These means remain below
the Pooled RL endpoint's 54.75\%, 56.72\%, and 55.10\%. To investigate why
initial category-specific RL yields only modest target-category gains,
we examine how success rates change across individual training instances.

\paragraph{Average progress coexists with instance-level regressions.}
Table~\ref{tab:stage1_instance_dynamics} follows the same matched training
records from the base model through initial RL and Repair SFT.
During initial RL, mean success rates increase
by 7.26, 6.76, and 4.60 points for A, B, and C. Yet across 2,769 training
records, observed rates increase on 1,108, decrease on 839 (30.3\%), and
remain unchanged on 822 (Table~\ref{tab:training_gain_loss}, Total rows).
Thus, average progress within a category does not
imply uniform improvement across its training instances. These are changes
between two sampled estimates, not continuous training trajectories or
per-instance tests of forgetting.

\begin{table}[htbp]
  \centering
  \caption{Training-record success rates through initial RL and first Repair SFT.
  Scores are equal-weight means over the same matched training records (\%).
  The final two columns report SFT gains in percentage points.}
  \label{tab:stage1_instance_dynamics}
  \small
  \begin{tabular*}{\linewidth}{@{\extracolsep{\fill}}lrrrrrr@{}}
    \toprule
    Expert & $N$ & Base & Initial RL & Repair SFT & $\Delta_{\mathrm{SFT-RL}}$ & $\Delta_{\mathrm{SFT-Base}}$ \\
    \midrule
    A & 601 & 38.79 & 46.05 & \textbf{56.28} & $+10.23$ & $+17.49$ \\
    B & 516 & 38.78 & 45.54 & \textbf{59.59} & $+14.05$ & $+20.81$ \\
    C & 1,652 & 38.68 & 43.28 & \textbf{52.75} & $+9.47$ & $+14.07$ \\
    \bottomrule
  \end{tabular*}
\end{table}

This uneven learning motivates the consolidation and refresh steps of RRE.
Repair SFT reuses verifier-approved successes accumulated during the preceding
RL phase, allocating more replay to currently low-mastery instances with
available trajectories. A low post-RL estimate can therefore coexist with
historical successes available for replay.

\paragraph{Recovery and gain retention after repair.}
The post-SFT refresh shows higher mean success rates in all three training
categories: 56.28\%, 59.59\%, and 52.75\% for A, B, and C, respectively.
These exceed the initial-RL rates by 10.23, 14.05, and 9.47 points and the
base rates by 17.49, 20.81, and 14.07 points
(Table~\ref{tab:stage1_instance_dynamics}). Of the 839 records whose
observed rates decreased during initial RL, 751 were included in repair
training: 179 for A, 141 for B, and 431 for C. Of these, 125 (69.8\%),
103 (73.0\%), and 264 (61.3\%) recover to or above their respective base
rates; another 7, 5, and 30 partially recover (Table~\ref{tab:repair_recovery}).
Thus, the aggregate improvement is accompanied by recovery on many of the
previously regressed instances actually included in SFT.

\begin{table}[htbp]
  \centering
  \caption{Training-record gains and losses relative to the base model.
  Both stages use the same matched training records within each category.
  Higher/equal/lower count observed pass rates relative to Base,
  not changes from the immediately preceding stage. Parentheses in the
  Repair SFT rows show the change in each count relative to Initial RL.}
  \label{tab:training_gain_loss}
  \small
  \begin{tabular*}{\linewidth}{@{\extracolsep{\fill}}llrrrr@{}}
    \toprule
    Expert & Stage & $N$ & Higher & Equal & Lower \\
    \midrule
    A & Initial RL & 601 & 259 & 151 & 191 \\
    A & Repair SFT & 601 & 336 ($+77$) & 161 ($+10$) & 104 ($-87$) \\
    \addlinespace[3pt]
    B & Initial RL & 516 & 216 & 154 & 146 \\
    B & Repair SFT & 516 & 314 ($+98$) & 121 ($-33$) & 81 ($-65$) \\
    \addlinespace[3pt]
    C & Initial RL & 1,652 & 633 & 517 & 502 \\
    C & Repair SFT & 1,652 & 877 ($+244$) & 398 ($-119$) & 377 ($-125$) \\
    \midrule
    Total & Initial RL & 2,769 & 1,108 & 822 & 839 \\[3pt]
    Total & Repair SFT & 2,769 & 1,527 ($+419$) & 680 ($-142$) & 562 ($-277$) \\[3pt]
    \bottomrule
  \end{tabular*}
\end{table}

Relative to the same base reference, the number of higher-scoring records
increases from 259 to 336 for A, 216 to 314 for B, and 633 to 877 for C;
the corresponding lower-scoring counts decrease from 191 to 104, 146 to 81,
and 502 to 377 (Table~\ref{tab:training_gain_loss}). Thus, all three categories
show both broader gains and fewer observed losses after repair. The magnitude
decomposition in Appendix~\ref{app:instance_dynamics} likewise shows larger
positive contributions and smaller negative contributions relative to Base.

Recovery coexists with further changes in instance mastery. Across all
2,769 training records, 1,211 rates increase after SFT, 978 remain unchanged, and
580 decrease. Of the 1,108 records that initially improved over Base,
404 lose some of their post-RL gains; 135 still exceed Base and 183 return
to the base rate, leaving 86 below it. Hence, 78.7\% of this 404-record
cohort remain at or above Base despite their decline from the RL endpoint.
By category, these proportions are 81.6\% (84/103) for A, 80.0\% (60/75)
for B, and 77.0\% (174/226) for C.
These observations show substantial recovery and positive net gains without
uniform improvement on every instance. They also motivate refreshing mastery
after SFT: the updated estimates, together with probes of the expanded
candidate pool, guide the next RL curriculum. To assess whether the benefits
extend beyond the training instances, we relate these recovery results to
the corresponding target-category Pro-618 evaluations below.

\begin{figure}[htbp]
  \centering
  \includegraphics[width=\linewidth]{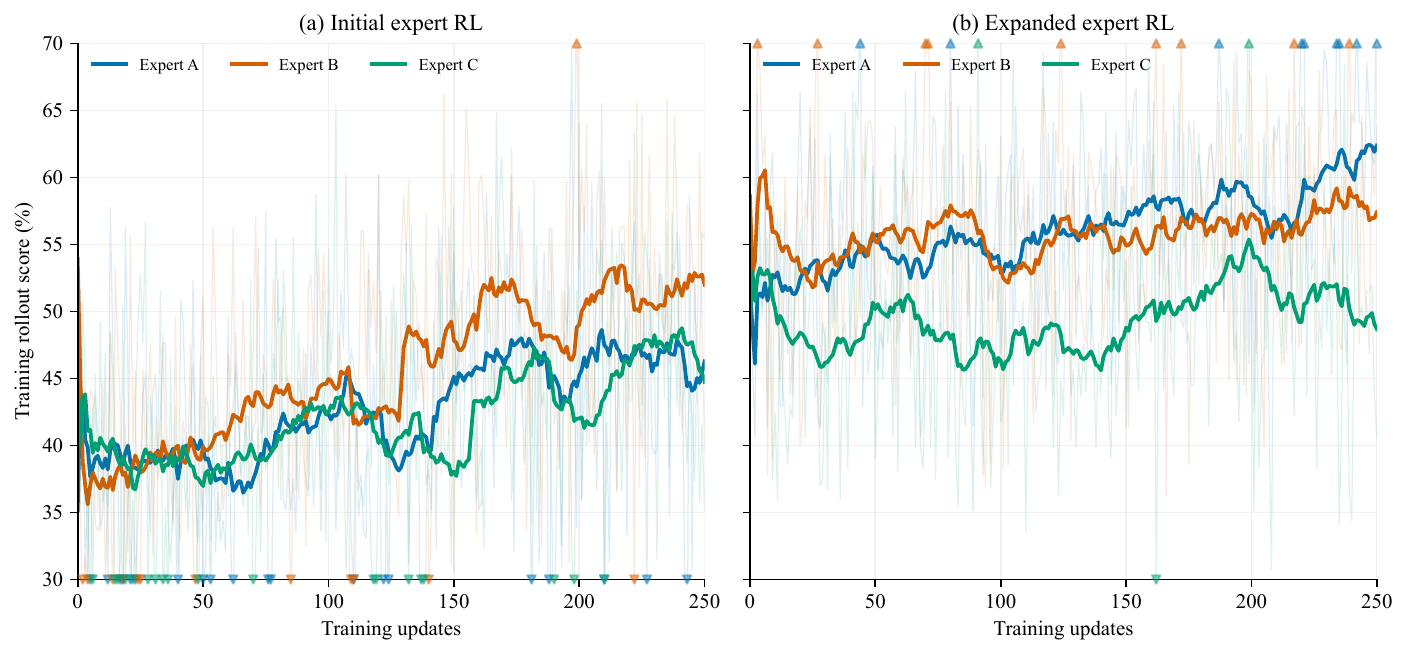}
  \caption{\textbf{Category-expert training dynamics in the two RL stages.}
  Panels (a) and (b) show training rollout scores for initial RL (Stage 1)
  and expanded RL (Stage 3), respectively. Faint traces show raw scores;
  bold traces show 20-update trailing means, with shorter windows at the
  start. Both panels use the same 30--70\% vertical range; boundary
  triangles mark raw scores outside this range. Each expert is trained on
  its own category pool, which is refreshed between stages.}
  \label{fig:rre_training_dynamics}
\end{figure}

\paragraph{Learning within the RL stages.}
Figure~\ref{fig:rre_training_dynamics} shows the training process in the initial
and expanded RL stages. Comparing mean raw rollout scores in the early
and late portions of the displayed trajectories, initial RL improves A/B/C by
7.76/12.74/6.58 percentage points; expanded RL improves them by
9.53/3.65/2.15 points. All three
experts therefore make progress within each stage over the shared window,
although the trajectories fluctuate and include local regressions.
These within-stage changes are measured on each expert's own training
stream; the two stages use different task pools.

\paragraph{Stage-wise target-category gains.}
Table~\ref{tab:rre_stage_results} follows each expert through the four stages.
Each column evaluates an expert on its corresponding category, using the
stage-selected checkpoints described in Section~\ref{sec:experiments}.
The base-model evaluation provides the reference row.

\begin{table}[htbp]
  \centering
  \caption{Stage-wise RRE expert results. Entries report target-category
  resolution (\%) as mean $\pm$ population standard deviation across three
  evaluation rounds. Parentheses below each entry give the mean gain over
  the base model in percentage points.}
  \label{tab:rre_stage_results}
  \small
  \setlength{\tabcolsep}{5pt}
  \begin{tabular*}{\linewidth}{@{\extracolsep{\fill}}lrrr@{}}
    \toprule
    Training stage & Expert A / Pro-A & Expert B / Pro-B & Expert C / Pro-C \\
    \midrule
    Base & $51.73 \pm 0.43$ & $54.39 \pm 1.17$ & $51.87 \pm 1.27$ \\[3pt]
    Initial RL & \shortstack[r]{$53.54 \pm 0.77$ \\[-1pt] {\scriptsize $(+1.81)$}} & \shortstack[r]{$55.06 \pm 0.62$ \\[-1pt] {\scriptsize $(+0.66)$}} & \shortstack[r]{$52.55 \pm 1.10$ \\[-1pt] {\scriptsize $(+0.68)$}} \\[3pt]
    First Repair SFT & \shortstack[r]{$57.32 \pm 0.43$ \\[-1pt] {\scriptsize $(+5.58)$}} & \shortstack[r]{$56.55 \pm 0.62$ \\[-1pt] {\scriptsize $(+2.16)$}} & \shortstack[r]{$55.44 \pm 0.96$ \\[-1pt] {\scriptsize $(+3.57)$}} \\[3pt]
    Expanded RL & \shortstack[r]{$58.67 \pm 1.19$ \\[-1pt] {\scriptsize $(+6.94)$}} & \shortstack[r]{$58.37 \pm 2.00$ \\[-1pt] {\scriptsize $(+3.98)$}} & \shortstack[r]{$56.97 \pm 0.96$ \\[-1pt] {\scriptsize $(+5.10)$}} \\[3pt]
    Second Repair SFT & \shortstack[r]{$59.58 \pm 1.40$ \\[-1pt] {\scriptsize $(+7.84)$}} & \shortstack[r]{$58.87 \pm 1.31$ \\[-1pt] {\scriptsize $(+4.48)$}} & \shortstack[r]{$57.48 \pm 0.64$ \\[-1pt] {\scriptsize $(+5.61)$}} \\[3pt]
    \bottomrule
  \end{tabular*}
\end{table}

The recovery observed on training instances is accompanied by improved
performance on the corresponding Pro-618 categories. Following the first
repair phase, mean target-category resolution increases over initial RL by
3.77, 1.50, and 2.89 percentage points for A, B, and C, respectively.
Together with the higher training success rates and fewer below-base
instances in Tables~\ref{tab:stage1_instance_dynamics}
and~\ref{tab:training_gain_loss}, these results support the role of repair
in expert development: recovery on training tasks is accompanied by
category-level gains beyond those tasks.

The selected checkpoints show increasing mean target-category resolution
across the four stages for all three experts.
Expanded RL yields further gains, followed by smaller mean improvements
after the second repair. By the end of the chain, the experts exceed the
base model on their corresponding categories by 7.84, 4.48, and 5.61 points.
The gains also extend beyond the first RRE round: compared with the first
repair endpoint, the second repair endpoint improves mean resolution by
2.26, 2.32, and 2.04 points for A, B, and C, respectively.
The first repair is followed by the largest adjacent-stage gain, with
further improvements observed after expanded RL and the second repair.
After expanded RL, all three target-category means exceed the Pooled RL
endpoint; the final experts' margins over that reference are 4.83, 2.16,
and 2.38 points on A, B, and C, respectively. The advantage thus emerges
through expert development rather than from the initial category RL phase alone.

Together, the instance-level diagnostics, within-stage learning curves,
and stage-wise Pro-618 results answer RQ2. Initial category-specific RL
leaves uneven progress across training instances; repair recovers many
observed losses while raising mean target-category resolution, and
refreshed curricula and expanded RL continue expert development.
The next section asks whether these expert gains can be retained in one
integrated policy, rather than only in a category-routed expert system.

\subsection{Integrating expert gains into a single policy}
\label{sec:results_integration}

To answer RQ3, the final comparison asks whether the improvements developed by separate
experts can be retained in a single deployed model.

\begin{figure}[htbp]
  \centering
  \includegraphics[width=\linewidth]{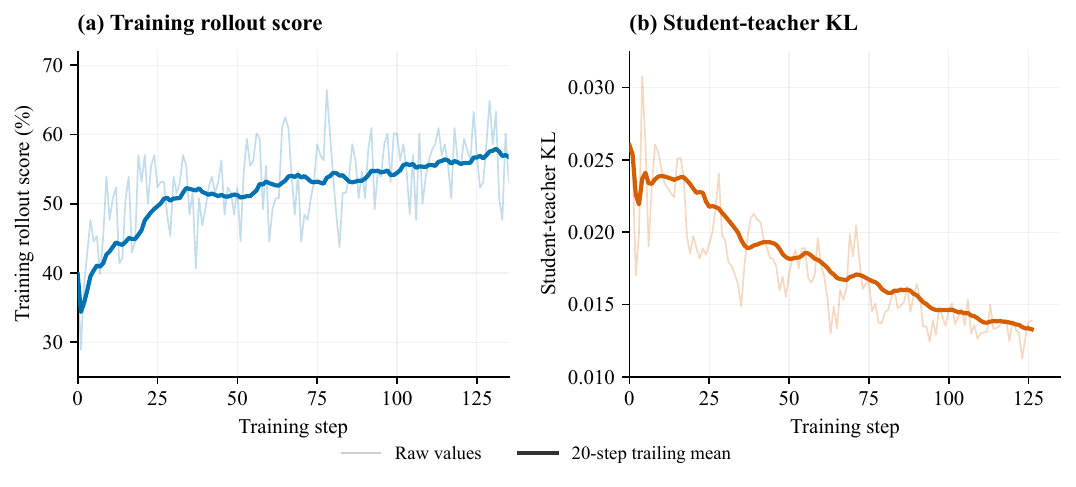}
  \caption{\textbf{Learning during multi-teacher on-policy distillation.}
  Panel (a) shows the student's training rollout score; panel (b) shows
  student--teacher KL. Faint traces show raw values and bold traces show
  20-step trailing means, with shorter windows at the start.
  Teacher alignment improves alongside the student's training task performance.}
  \label{fig:mopd_training_dynamics}
\end{figure}

\paragraph{Learning during multi-teacher integration.}
Figure~\ref{fig:mopd_training_dynamics} tracks the student's training rollout
score and student--teacher KL during MOPD. Comparing the mean raw values
in the early and late portions of training, rollout score increases from 45.51\% to
56.91\%, a gain of 11.41 percentage points, while mean KL decreases by
43.0\%. The declining KL indicates
closer alignment with the routed teachers on student-generated training
contexts, accompanied by stronger task performance on the category-balanced
training stream.  These concurrent trends connect teacher alignment to
the student's improving training performance; we next evaluate the
integrated policy on the benchmarks.

\paragraph{Overall and category-wise benchmark performance.}
Table~\ref{tab:main_results} compares MOPD with the base model,
Pooled RL and Balanced RL, and the category-routed experts.
\textbf{Category-routed experts} denotes a three-model reference that uses
Expert A on Pro-A, Expert B on Pro-B, and Expert C on Pro-C.
This row reports the experts before single-model integration, whereas
\method{} reports the integrated single policy.

\begin{table}[htbp]
  \centering
  \caption{Overall and category-wise resolution on Pro-618 (\%). Base,
  Pooled RL/Balanced RL, and MOPD scores, and category-wise entries of the routed expert
  reference, report mean $\pm$ population standard deviation across three
  evaluation rounds. Parentheses below each non-base score give the change
  from the corresponding base-model mean in percentage points, computed
  before rounding. The category-routed experts report
  the corresponding expert's result in each category before single-model integration.}
  \label{tab:main_results}
  \small
  \newcommand{\scoregain}[2]{\shortstack[r]{#1\\[-1pt]{\scriptsize (#2)}}}
  \begin{tabular*}{\linewidth}{@{\extracolsep{\fill}}lrrrr@{}}
    \toprule
    Policy & Full \up & Pro-A \up & Pro-B \up & Pro-C \up \\
    \midrule
    Base model & \BasePro & \BaseProA & \BaseProB & \BaseProC \\
    Pooled RL & \scoregain{\UnifiedPro}{+2.86} & \scoregain{\UnifiedProA}{+3.02} & \scoregain{\UnifiedProB}{+2.32} & \scoregain{\UnifiedProC}{+3.23} \\[3pt]
    Balanced RL & \scoregain{\BalancedPro}{+2.70} & \scoregain{\BalancedProA}{+2.71} & \scoregain{\BalancedProB}{+2.65} & \scoregain{\BalancedProC}{+2.72} \\
    Category-routed experts (3 models) & --- & \scoregain{\OracleProA}{+7.84} & \scoregain{\OracleProB}{+4.48} & \scoregain{\OracleProC}{+5.61} \\
    \midrule
    \method & \scoregain{\CategoryMOPDPro}{+5.39} & \scoregain{\CategoryMOPDProA}{+6.33} & \scoregain{\CategoryMOPDProB}{+4.98} & \scoregain{\CategoryMOPDProC}{+4.76} \\
    \bottomrule
  \end{tabular*}
\end{table}

The final MOPD policy achieves 58.04\% mean Full resolution across three
evaluation rounds, improving over the base model by 5.39 percentage points.
Its Pro-A/B/C scores are 58.07\%, 59.37\%, and 56.63\%, corresponding to
gains of 6.33, 4.98, and 4.76 points over Base. Compared with Pooled RL,
MOPD improves Full resolution by 2.54 points and A/B/C by 3.32, 2.65,
and 1.53 points. Compared with Balanced RL, the corresponding improvements
are 2.70 points overall and 3.62, 2.32, and 2.04 points by category.
Thus, the higher aggregate score is accompanied by higher mean resolution
in every category, with minimum category lifts of 1.53 and 2.04 points over
Pooled RL and Balanced RL, respectively (Equation~\ref{eq:min_category_lift}).

\begin{figure}[htbp]
  \centering
  \includegraphics[width=\linewidth]{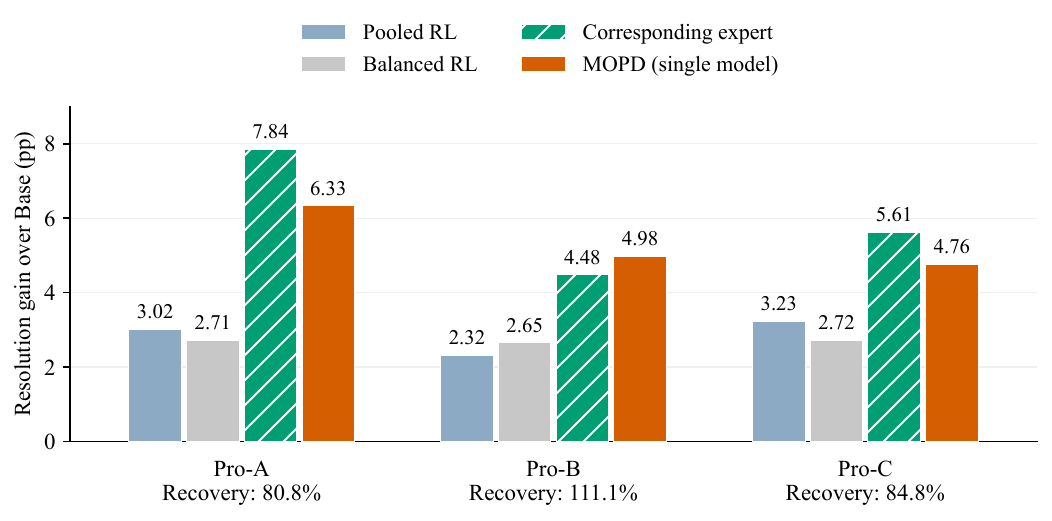}
  \caption{\textbf{Category gains before and after single-model integration.}
  Bars show three-round mean resolution gains over the same base model.
  Each expert is evaluated on its corresponding category; MOPD uses one
  integrated model for all categories. Recovery below each group is the
  fraction of the corresponding expert's gain retained by MOPD
  (Equation~\ref{eq:recovery}). Round standard deviations are reported in
  Table~\ref{tab:main_results}.}
  \label{fig:mopd_category_gains}
\end{figure}

\paragraph{Which expert gains survive integration?}
Figure~\ref{fig:mopd_category_gains} compares the final models' category gains
relative to the same base policy. MOPD retains 80.8\%, 111.1\%, and 84.8\%
of the corresponding expert gains on A, B, and C, respectively, using
Equation~\ref{eq:recovery}. Recovery above 100\% on B means that the student
exceeds its corresponding expert's mean score, by 0.50 points. On A and C,
the student remains 1.51 and 0.85 points below the corresponding experts, while still
outperforming both joint-RL baselines. Integration therefore preserves
enough of the developed expert gains to improve all three categories over
both joint-RL baselines, with the lowest proportional recovery on A.

\paragraph{Overall gain, minimum category gain, and gain balance.}
Table~\ref{tab:final_policy_gains} compares these three aspects for the final
single policies, using the same base reference. Unlike the training-process
averages in Table~\ref{tab:pooled_balanced_metrics}, these metrics are computed
from each final model's three-round mean benchmark scores.

\begin{table}[htbp]
  \centering
  \caption{\textbf{Final-policy gains on Pro-618.}
  Metrics are computed from three-round mean scores relative to the same base
  model, in percentage points. $G_{\mathrm{sim}}$ is the minimum A/B/C gain;
  \ssg{} is Full gain minus $G_{\mathrm{sim}}$. Bold marks the best value
  among the three single policies for each metric.}
  \label{tab:final_policy_gains}
  \small
  \begin{tabular*}{\linewidth}{@{\extracolsep{\fill}}lrrr@{}}
    \toprule
    Policy & Full gain $\uparrow$ & $G_{\mathrm{sim}}$ $\uparrow$ & \ssg{} $\downarrow$ \\
    \midrule
    Pooled RL & 2.86 & 2.32 & 0.54 \\
    Balanced RL & 2.70 & 2.65 & \textbf{0.04} \\
    \method & \textbf{5.39} & \textbf{4.76} & 0.63 \\
    \bottomrule
  \end{tabular*}
\end{table}

MOPD achieves the largest Full gain (5.39 points) and minimum category gain
$G_{\mathrm{sim}}$ (4.76 points). Relative to Pooled RL and Balanced RL,
the minimum category gain increases by 2.44 and 2.11 points, respectively.
Together with the positive A/B/C differences in Table~\ref{tab:main_results},
this shows that the aggregate improvement extends to every category,
including the category with the smallest gain over Base.

Relative to Pooled RL, MOPD nearly doubles both the overall and minimum
category gains, while \ssg{} increases only slightly, from 0.54 to 0.63
percentage points. Compared with Balanced RL, MOPD achieves higher overall
and minimum category gains, although the gains are less uniform
(\ssg{}: 0.63 versus 0.04 points). Under the interpretation in
Table~\ref{tab:gain_interpretation}, this is shared improvement with a wider
gap, rather than aggregate improvement accompanied by a stagnant or declining
minimum gain. The A/B/C scores in Table~\ref{tab:main_results} further confirm
that MOPD improves every category over both baselines. These results
distinguish a higher level of improvement across categories from a more
uniform distribution of gains.

\begin{figure}[htbp]
  \centering
  \includegraphics[width=\linewidth]{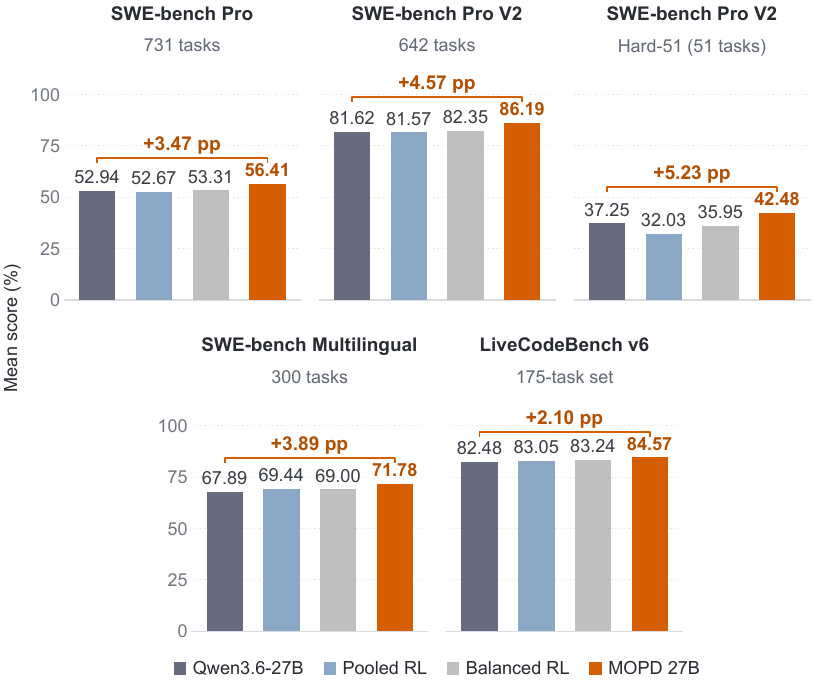}
  \caption{\textbf{Final 27B policy performance on four evaluated task sets
  and the Hard-51 subset.} Bars show three-round mean scores;
  brackets mark MOPD's gains over Qwen3.6-27B in percentage
  points. Hard-51 uses the same runs as its SWE-bench Pro V2 parent set.
  Full scores and round variability are given in
  Table~\ref{tab:expanded_eval_results}.}
  \label{fig:updated_benchmarks}
\end{figure}

\paragraph{Final-model benchmark performance.}
Figure~\ref{fig:updated_benchmarks} summarizes the fixed final 27B policies
on the 731-task SWE-bench Pro set, SWE-bench Pro V2 and its Hard-51 subset,
SWE-bench Multilingual, and the 175-task LiveCodeBench v6 set.
On Pro V2, MOPD reaches \ExpMOPDProVTwoMean\% mean resolution, compared with
\ExpBaseProVTwoMean\% for Qwen3.6-27B, an increase of
\ExpMOPDProVTwoGainBase{} percentage points. Its mean exceeds Pooled RL and
Balanced RL by \ExpMOPDProVTwoGainPooled{} and
\ExpMOPDProVTwoGainBalanced{} points, respectively. On Hard-51, MOPD reaches
\ExpMOPDHardMean\%, improving over the base by \ExpMOPDHardGainBase{} points
and over the stronger joint-RL baseline, Balanced RL, by
\ExpMOPDHardGainBalanced{} points.

\FloatBarrier

\paragraph{SWE-bench Multilingual results.}
MOPD achieves \ExpMOPDMultiMean\% mean resolution, compared with
\ExpBaseMultiMean\% for the base model, an increase of
\ExpMOPDMultiGainBase{} percentage points. Pooled RL and Balanced RL reach
\ExpPooledMultiMean\% and \ExpBalancedMultiMean\%, respectively; MOPD's
corresponding margins are \ExpMOPDMultiGainPooled{} and
\ExpMOPDMultiGainBalanced{} points. Relative to the base, the recomputed
A/B/C category means increase by 3.24, 12.00, and 3.79 points, while the
unassigned group decreases by 1.23 points
(Table~\ref{tab:multilingual_results}).

\paragraph{Extended benchmark coverage.}
Appendix~\ref{app:expanded_evaluation} reports complete six-model results.
The 27B MOPD policy reaches \ExpMOPDProFullMean\% on the 731-task Pro
execution setup and \ExpMOPDLCBMean\% on the 175-task LiveCodeBench v6 set.

\paragraph{Reusing experts across student architectures.}
We also use the same three frozen 27B category experts to train a
Qwen3.6-35B-A3B student, a mixture-of-experts (MoE) model. This holds the
teacher set fixed while changing the student architecture and initialization.
The 35B-A3B student improves over its own base by
\ExpMOPDMoEProVTwoGainBase{} points on Pro V2,
\ExpMOPDMoEMultiGainBase{} points on Multilingual, and
\ExpMOPDMoEHardGainBase{} points on Hard-51, with positive mean gains on
Pro 731 and LiveCodeBench as well (Table~\ref{tab:expanded_eval_results}).
One set of RRE-developed experts therefore supports both the dense 27B and
MoE 35B-A3B students, making expert development reusable across these two
student integrations.

\paragraph{Summary of single-policy integration.}
Taken together, the results answer RQ3 and connect expert development to the
final deployed policy. Initial category-specific RL yields modest
target-category gains; RRE develops stronger experts, and MOPD retains much
of their improvement in one model. On Pro-618, the resulting policy achieves
higher mean resolution than both joint-RL baselines overall and in all three
categories. Pro V2 and Multilingual evaluations also show higher
overall means. Expert-gain recovery on Pro-618 remains uneven:
A retains the smallest fraction of its expert's gain (80.8\%), identifying
further A-category gain retention as a concrete opportunity to improve
integration. This gap coexists with positive A-category lifts over both
joint-RL baselines.

\section{Discussion}
\label{sec:discussion}

\paragraph{When should one domain be decomposed?}
A shared task domain need not imply uniform learning demands. Within
repository-level software engineering, tasks differ in their engineering
contexts, tool interactions, and verification requirements. Decomposition
becomes worth exploring when these differences form interpretable task groups
and pooled training exhibits persistent differences in their learning
trajectories that aggregate performance obscures. In our experiments,
category gains move in opposing directions during parts of training, and
balancing the training mixture does not clearly remove this pattern. These
observations motivate investigating category-specific learning as an
alternative to changing mixture proportions alone. They frame decomposition
as a hypothesis to evaluate, rather than an assumed advantage of specialization.

\paragraph{Developing experts beyond category separation.}
Category separation changes which tasks a policy learns from, but does not
by itself make those tasks fully learnable or preserve previously acquired
behavior. Our initial-RL diagnostics show that average training improvement
can coexist with declining success rates on individual instances. This
suggests that expert development must address not only differences across
categories, but also changing mastery within each category. RRE responds
to this second problem by revisiting the expert's own successful trajectories
and refreshing task selection as the policy changes. The observed
training-instance recovery and successive target-category improvements
support treating specialization as an iterative development process, rather
than a one-time partition of the training pool.

\paragraph{Higher performance does not imply uniformly retained gains.}
Single-model integration should be assessed both by the gains it delivers
over the joint-RL baselines and by how much of each expert's improvement it retains.
These perspectives reveal different aspects of the outcome. On Pro-618, MOPD
improves overall resolution and every category over both joint-RL baselines, while
achieving a higher minimum category gain. Yet its gains remain uneven, and
the fraction of expert improvement retained differs across categories.
In particular, C has the smallest gain over Base, whereas A has the lowest
expert-gain recovery. Thus, the category with the smallest final improvement
is not necessarily the one with the largest proportional integration loss.
This distinction helps identify whether further progress should focus on
developing stronger experts or transferring their existing capabilities
more effectively.

\section{Limitations}
\label{sec:limitations}

SWE tasks can span multiple categories, whereas our hard-routing configuration
assigns each task to one expert category.  This simplifies overlapping task
structure for category-conditioned training.  Training multiple experts and running teacher forwards adds
substantial compute, while the best $K$ may vary with data, base model, and
budget.  Same-origin teachers may be too similar to add signal or, after heavy
specialization, too distant for stable integration.  Benchmark pass rates do
not fully measure patch quality, and test-driven rewards can invite evaluator
hacking or reflect brittle task specifications~\citep{baker2025monitoringmisbehavior,
zhao2026specbench,rajan2026rewardhackability}.  Repository sanitization and
fresh-sandbox replay prevent local-history leakage and rollout-state
tampering, while trajectory audits screen external retrieval; these controls
do not eliminate pretraining memorization or every weak-verifier failure.  Any result on
Pro-618 remains conditional on its audit filtering and checkpoint-selection
role. The expanded evaluation covers the original 731-task population and
Pro V2, but scores across different task versions and agent settings are not
directly interchangeable. Recorded generation budgets and sampling requests
also vary in the expanded results, limiting attribution of their differences
to training alone. Teacher reuse is evaluated with one additional MoE student.
Finally,
the experiments evaluate the complete training framework and its stage-wise
outcomes; they do not isolate the contributions of each optimization component,
replay choice, or routing granularity.  Single-run category trajectories also
include evaluation variability, and broader comparisons of sampling and
routing strategies remain open.

\paragraph{Societal and release considerations.}
More capable coding agents can improve developer productivity but can also
produce insecure changes, automate vulnerability discovery, or create patches
whose tests pass without satisfying maintainability requirements.  Release
decisions should include provenance, licenses, benchmark contamination checks,
unsafe-code handling, and safeguards proportionate to model capability.

\section{Conclusion}
\label{sec:conclusion}

We develop a category-aware expert-training and policy-integration framework
for heterogeneous repository-level SWE tasks.  SWE Labeler provides
evidence-grounded task categories; Agentic-miniRL and RRE develop same-origin
experts through executable-reward learning, mastery refresh, and replay of
their own successful trajectories; label-routed MOPD consolidates the experts
into one deployed policy.  The evaluation connects Pooled RL and Balanced RL, expert development, and
single-model integration through overall and per-category resolution on a
fixed Pro-618 analysis set. The final policy achieves 58.04\% mean resolution
on Pro-618, with higher mean scores than both joint-RL baselines overall
and in each category. Final-model evaluations also yield
\ExpMOPDProVTwoMean\% on Pro V2 and \ExpMOPDMultiMean\% on SWE-bench
Multilingual, exceeding the base by \ExpMOPDProVTwoGainBase{} and
\ExpMOPDMultiGainBase{} percentage points, respectively.
The same three 27B experts also improve a 35B-A3B student across the reported
sets. Expert-gain recovery on Pro-618 remains uneven, identifying room to
improve single-model integration. This framework treats task categories as
units for organizing post-training, with the practical goal of retaining
expert gains in a single SWE agent.

\bibliographystyle{plainnat}
\bibliography{references}

\appendix
\section{Evaluation benchmark construction and audit}
\label{app:pro618_audit}

Pro-618 is an audit-filtered subset of the 731 SWE-bench Pro instances.
This construction defines the fixed original analysis population; expanded
final-model evaluations are described in Appendix~\ref{app:expanded_evaluation}.
Benchmark audits have shown that specification gaps, incorrect reference
solutions, and incomplete grading logic can change measured performance and
even model rankings~\citep{kim2026determinacyaudit,openai2026signalnoise}.  We
therefore remove only cases backed by released instance-level evidence.

\begin{table}[h]
  \centering
  \caption{Construction of the Pro-618 evaluation subset.  Counts are mutually
  exclusive at the final exclusion stage.}
  \label{tab:pro618_construction}
  \small
  \begin{tabularx}{\columnwidth}{@{}Xr@{}}
    \toprule
    Construction stage & Instances \\
    \midrule
    Full SWE-bench Pro & 731 \\
    Public audit: mechanically evidenced specification/determinacy issue & $-83$ \\
    Public audit: two-model-family underspecification finding & $-26$ \\
    Public audit: reference patch fails isolated grader replay & $-3$ \\
    Official OpenAI audit: prompt--test mismatch & $-1$ \\
    \midrule
    Pro-618 & 618 \\
    \bottomrule
  \end{tabularx}
\end{table}

The first three exclusion groups follow the pinned public determinacy-audit
artifact~\citep{kim2026determinacyaudit}; the final case follows the official
OpenAI audit~\citep{openai2026signalnoise}.  We report the source audit and
excluded-instance count for each exclusion group.  The accompanying manifest
records task membership and exclusion reasons.
Applying the common Domain L1 routing rule after filtering gives 221 Pro-A, 201
Pro-B, and 196 Pro-C instances.

\section{SWE Labeler and category construction}
\label{app:labeler}

\subsection{Label inventory}
\label{app:taxonomy}

This appendix records the final label axes used in our experiments.  The
experiment artifact exports the exact machine-readable definitions, active
prompts, and configuration used to assign data; each dataset manifest
additionally records the labeling model and source export.

\paragraph{Task Type L1 labels (26).}
\texttt{bug-fix}, \texttt{performance-fix},
\texttt{compatibility-fix}, \texttt{feature}, \texttt{enhancement},
\texttt{backward-compatibility}, \texttt{from\_scratch}, \texttt{test-add},
\texttt{test-fix}, \texttt{coverage}, \texttt{debug-support},
\texttt{refactor}, \texttt{style}, \texttt{deprecation},
\texttt{build-fix}, \texttt{ci-fix}, \texttt{deployment},
\texttt{packaging}, \texttt{config}, \texttt{user-docs},
\texttt{design-spec}, \texttt{diagram}, \texttt{code-explanation},
\texttt{security-fix}, \texttt{auth}, and \texttt{privacy}.  The current
definition file contains 119 Task Type L2 labels.

\paragraph{Domain L1 labels (21).}
\texttt{web\_frontend}, \texttt{mobile\_dev}, \texttt{desktop\_gui},
\texttt{web\_backend}, \texttt{database\_storage},
\texttt{data\_science}, \texttt{data\_engineering}, \texttt{os\_system},
\texttt{iot\_embedded}, \texttt{devops\_infra},
\texttt{media\_graphics}, \texttt{game\_dev},
\texttt{devtools\_test}, \texttt{lang\_runtime},
\texttt{pkg\_manager\_cli}, \texttt{cms\_ecommerce},
\texttt{blockchain\_web3}, \texttt{security\_auth},
\texttt{docs\_knowledge}, \texttt{non\_coding}, and \texttt{other}.
The current definition file contains 108 Domain L2 labels.

Every L2 entry records its name, source attribution, source-derived definition,
operational definition, observable signals, and boundary notes.  Some L1
labels intentionally have no L2 children; the runtime returns
\texttt{unspecified} instead of inventing one.

\subsection{Taxonomy grounding and structural audit}
\label{app:taxonomy_grounding}

\paragraph{Source selection by role.}
Table~\ref{tab:taxonomy_source_roles} summarizes how external material is used.
Normative standards stabilize established software-engineering distinctions;
curated catalogs provide named defect or transformation concepts; empirical
developer metadata suggests coarse domain neighborhoods; and official framework
documentation supplies ecosystem-specific meaning.  None of these sources is
treated as a ready-made universal taxonomy for SWE-agent training.

\begin{center}
  \begin{minipage}{\textwidth}
  \refstepcounter{table}\label{tab:taxonomy_source_roles}
  \centering
  \textbf{Table~\thetable: Source families used to ground the label inventory.}
  \par\medskip
  \small
  \begin{tabularx}{\textwidth}{p{0.16\textwidth}p{0.30\textwidth}Y}
    \toprule
    Axis / family & Representative sources & Role in \labeler \\
    \midrule
    Task Type: defect and security
      & MITRE CWE entries and NIST guidance~\citep{mitre_cwe682,mitre_cwe697,grassi2017digitalidentity}
      & Name recurrent failure mechanisms and separate neighboring mechanisms. \\
    Task Type: maintenance and change
      & ISO/IEC~25010, ISO/IEC/IEEE~14764, Semantic Versioning, and Conventional Commits~\citep{iso25010_2023,iso14764_2022,semver2000,conventionalcommits100}
      & Anchor quality, maintenance, compatibility, feature, build, CI, and change-intent families. \\
    Task Type: test and refactor
      & ISTQB testing terminology and Fowler's refactoring catalog~\citep{istqb2023ctfl,fowler2018refactoring}
      & Supply established test activities and behavior-preserving transformation concepts. \\
    Repository Domain
      & Stack Overflow tag co-occurrence clusters plus standards and official ecosystem documentation~\citep{hoffa2019stacktags}
      & Seed coarse technology neighborhoods, then define repository-purpose and framework-specific boundaries. \\
    Scope, complexity, and time
      & Repository topology, Epoch AI's SWE-bench analysis, and cognitive-load theory~\citep{brand2025swebenchskills,sweller1988cognitiveload}
      & Define a project-specific scope rule and adapt empirical brackets or theoretical inspiration into SWE decision rules. \\
    \bottomrule
  \end{tabularx}
  \end{minipage}
\end{center}

\paragraph{Representative evidence chains.}
The following examples instantiate Equation~\ref{eq:label_evidence}.  In each
case, a source concept is translated into evidence available to the offline
annotator and separated from its nearest alternatives.

\paragraph{Logic error.}
The \texttt{bug-fix.logic\_error} label uses CWE-682 and CWE-697 to anchor
incorrect calculations and comparisons.  The SWE
rule accepts an incorrect condition, arithmetic rule, comparator, or
algorithmic result.  Array/loop boundary mistakes instead map to
\texttt{off\_by\_one}, while unintended mutation maps to
\texttt{state\_corruption}.

\paragraph{API-signature refactoring.}
For \texttt{refactor.api\_signature\_refactor}, Fowler's API, algorithm, and
data-structure refactorings provide the source
concept~\citep{fowler2018refactoring}.  The operational rule covers a changed
signature, parameter object, equivalent algorithm, or representation while
preserving behavior.  Merely creating a helper is \texttt{extract\_method}; an
identifier-only change is \texttt{rename\_symbol}.

\paragraph{Scientific and numerical software.}
The \texttt{data\_science.scientific\_numerical} label uses ACM CCS
mathematical software~\citep{acmccs2012} and official scientific-
library documentation anchor numerical, symbolic, array, geometry, and
domain-science library internals.  Training or serving a learned model uses an
ML label; feature preparation on top of these libraries is
\texttt{data\_preprocessing}.

\paragraph{Identity lifecycle.}
For \texttt{security\_auth.identity\_lifecycle}, NIST digital-identity guidance
supplies the authenticator-lifecycle anchor
\citep{grassi2017digitalidentity}.  The project rule covers enrollment,
renewal, revocation, session, credential, and device-trust lifecycle work.
Token-issuance protocols map to \texttt{oauth\_oidc}, authorization policy to
\texttt{rbac\_acl}, and primitive implementation to \texttt{crypto\_impl}.

\paragraph{Cross-module scope.}
The \texttt{scope.cross\_module} label is a project-defined
repository-topology rule, with benchmark task-size
analysis as motivational context~\citep{brand2025swebenchskills}.  It requires
changes across top-level packages or architectural layers within one
repository.  Several files in one module remain
\texttt{multi\_file\_same\_module}; independent repositories are
\texttt{cross\_repo}.

\paragraph{Global reasoning.}
For \texttt{complexity.global\_reasoning}, high element interactivity is
theoretical inspiration
\citep{sweller1988cognitiveload}; the SWE rule requires tracing data flow, call
chains, or state across components.  Choosing among valid architectures is
instead \texttt{design\_decision}.  This is an adaptation, not a claim that
cognitive-load theory directly specifies SWE labels.  Likewise, the Stack
Overflow clustering is an empirical seed for domain organization rather than
a gold annotation set.

\paragraph{Final admission and boundary protocol.}
Every label in the frozen inventory satisfies four requirements: (i) it is not
already covered by an existing operational definition; (ii) it represents one
coherent concept rather than a narrow implementation variant; (iii) it has an
attributable source, observable signals, and an explicit non-overlap boundary;
and (iv) sparse or heterogeneous residuals can abstain through
\texttt{unspecified}.  Consequently, related API transformations share
\texttt{api\_signature\_refactor}, and enrollment, renewal, revocation, session,
credential, and device-trust operations share \texttt{identity\_lifecycle}.
The inventory does not include broad catch-alls such as a generic business-app
backend label or a separate label for every individual refactoring operation.

\paragraph{Schema-completeness audit.}
In the final inventory, all 239 decision records---119 Task Type L2 labels, 108
Domain L2 labels, and 12 orthogonal levels---contain the six schema fields
described around Equation~\ref{eq:label_evidence}.  All 26 Task Type L1 families,
21 Domain L1 families, and three orthogonal dimensions record a source
framework.  This audit verifies the completeness of label definitions and
their traceability to source concepts.

\paragraph{Orthogonal axes.}
Modification scope has four levels: \texttt{single\_file},
\texttt{multi\_file\_same\_module}, \texttt{cross\_module}, and
\texttt{cross\_repo}. Cognitive complexity has four levels:
\texttt{mechanical}, \texttt{local\_reasoning},
\texttt{global\_reasoning}, and \texttt{design\_decision}. Estimated time has
four levels: \texttt{trivial\_15min}, \texttt{small\_1h},
\texttt{medium\_4h}, and \texttt{large\_4h\_plus}.

\paragraph{Auxiliary Phase-1 metadata.}
The implementation records three fields outside the semantic label vector.
\texttt{code\_language} chooses one of 21 normalized values (including
\texttt{other} and \texttt{unknown}); \texttt{task\_spec\_type} is
\texttt{fuzzy} or \texttt{specific}; and \texttt{dependency\_context} is one
of \texttt{codebase}, \texttt{env}, \texttt{codebase+env}, or \texttt{none}.
They describe the evidence needed to interpret an instance and support audit or
stratification.  They are not additional task categories and do not enter the
main expert mapping $g_\psi$.

\paragraph{Not an implemented label axis.}
Fault localization, code navigation, test generation, and long-horizon planning
are not separately annotated axes in the current label schema.

\paragraph{Frozen structural audit.}
Table~\ref{tab:labeler_audit} summarizes the final structural snapshot on the
1,531-task collection from SWE-bench Verified, SWE-bench Pro, and
SWE-bench Multilingual. Because this collection was used to finalize
the inventory, these values measure coverage and internal consistency rather
than held-out semantic accuracy.

\begin{table}[h]
  \centering
  \caption{Structural audit of the frozen SWE Labeler. Repository consistency
  is computed from nonempty Domain L1 outputs.}
  \label{tab:labeler_audit}
  \small
  \begin{tabular}{lr}
    \toprule
    Audit item & Final value \\
    \midrule
    Task Type L1 / L2 inventory & 26 / 119 \\
    Domain L1 / L2 inventory & 21 / 108 \\
    Orthogonal dimensions / levels & 3 / 12 \\
    Task-L2 \emph{unspecified} & 70 (4.6\%) \\
    Domain-L2 \emph{unspecified} & 35 (2.3\%) \\
    Consistent nonempty Domain L1 by repository & 58/64 (90.6\%) \\
    Persisted illegal concrete L1--L2 assignments & 0 \\
    \bottomrule
  \end{tabular}
\end{table}

The audit contains one empty Domain L1 output, and six repositories retain more than
one nonempty Domain L1 label.  These include genuinely ambiguous products such
as a music server that can be viewed as either a web backend or a media system.
For reproducibility, the released artifact records the labeling model, active
prompt, definition files, raw export, and grouping manifest together; the
current JSONL schema does not independently encode all of these artifact
identifiers in each record.

\begin{figure*}[t]
  \centering
  \includegraphics[width=\textwidth]{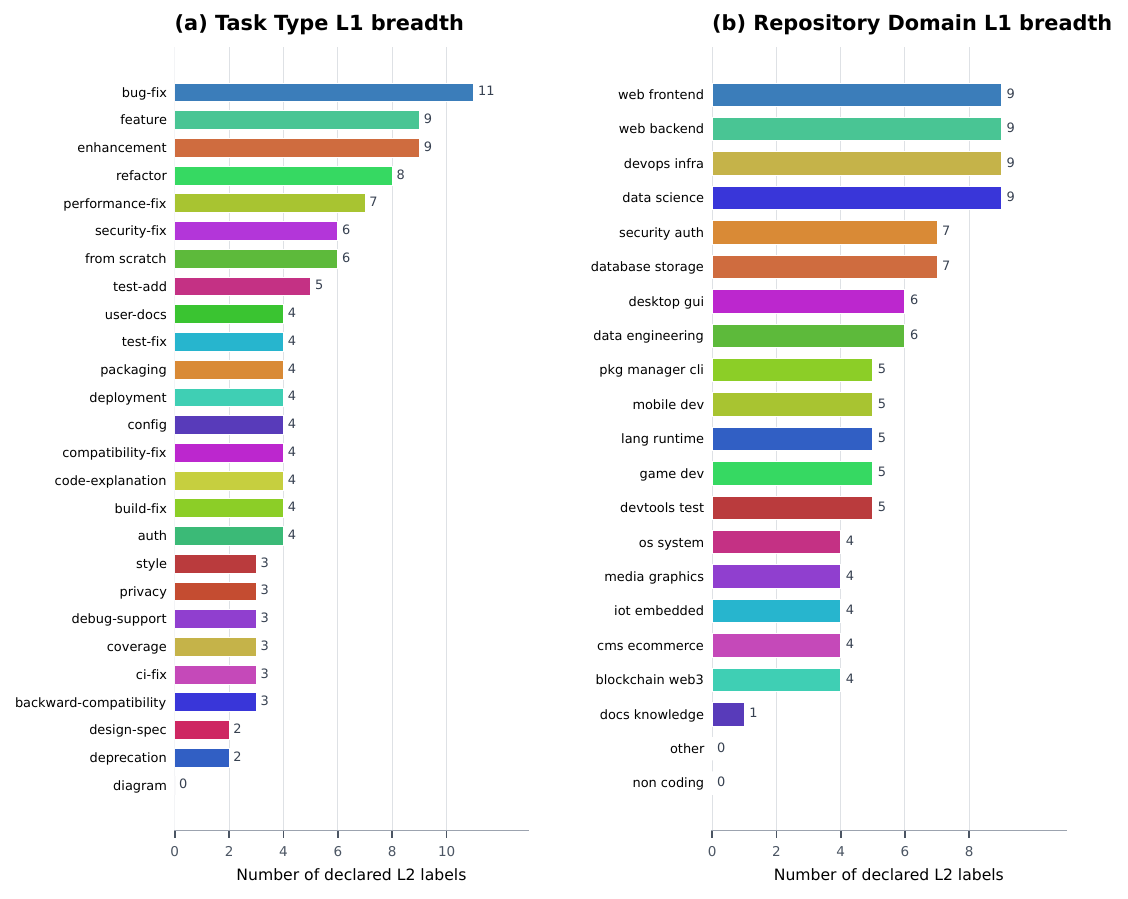}
  \caption{\textbf{Exact L2 breadth of every L1 family.}
  Bars report the number of declared L2 children, sorted within each semantic
  axis.  Zero-child families intentionally fall back to \texttt{unspecified};
  the values describe taxonomy structure, not empirical sample prevalence.}
  \label{fig:taxonomy_breadth}
\end{figure*}

\subsection{Annotation and validation}
\label{app:trajectory_labeling}

Instance annotation uses the issue description, repository metadata, gold fix,
and test patch as offline evidence. The first phase assigns L1 labels, and the
second selects L2 labels from the corresponding L1 families. These annotation
inputs are separate from the repair agent's evaluation input, which does not
include the gold fix or test patch.

For full-trajectory annotation, compact evidence additionally retains user
requests, code edits, tool actions, and test outcomes. Valid existing L1 labels
can be preserved before L2 refinement. Outputs are checked against the declared
label inventory and hierarchy: concrete L2 labels must belong to the resolved
L1 family, and \texttt{unspecified} is used alone. Invalid outputs are excluded
from the labeled export.

\subsection{Three-route construction and routing specification}
\label{app:routing}

\begin{figure*}[h]
  \centering
  \includegraphics[width=\textwidth]{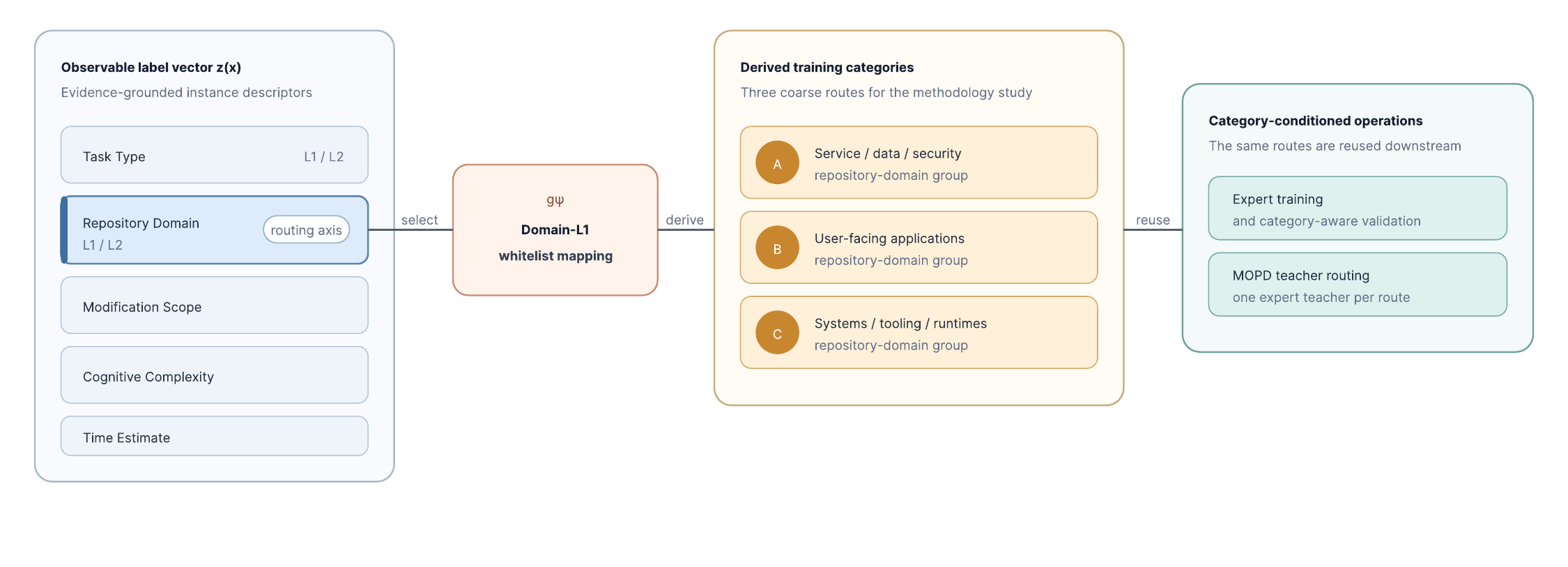}
  \caption{\textbf{From SWE labels to category-conditioned training and routing.}
  Repository Domain L1 is mapped by a fixed whitelist into three derived
  training categories.  The resulting A/B/C assignment is reused for expert
  training, category-aware validation, and MOPD teacher routing.}
  \label{fig:label_to_category}
\end{figure*}

The three-route map operates solely on Repository Domain L1.  It is
deterministic and disjoint: every listed Domain L1 value maps to exactly one
route, and all other values map to $\varnothing$.  Task Type, L2 labels, and
the scale axes are retained for stratification and analysis rather than used as
additional routing predicates.

\begin{table*}[h]
  \centering
  \caption{Repository Domain L1 mapping for the three-route expert
  configuration.}
  \label{tab:routing_registry}
  \normalsize
  \renewcommand{\arraystretch}{1.12}
  \begin{tabularx}{\textwidth}{@{}l l Y@{}}
    \toprule
    Route & Engineering context & Domain L1 whitelist \\
    \midrule
    A & Service/data/security & \texttt{web\_backend}, \texttt{database\_storage}, \texttt{data\_engineering}, \texttt{data\_science}, \texttt{cms\_ecommerce}, \texttt{security\_auth}, \texttt{blockchain\_web3} \\
    B & User-facing applications & \texttt{web\_frontend}, \texttt{desktop\_gui}, \texttt{mobile\_dev}, \texttt{media\_graphics}, \texttt{game\_dev} \\
    C & Systems/tooling/runtimes & \texttt{devops\_infra}, \texttt{devtools\_test}, \texttt{pkg\_manager\_cli}, \texttt{os\_system}, \texttt{iot\_embedded}, \texttt{lang\_runtime} \\
    \bottomrule
  \end{tabularx}
\end{table*}

\section{Training implementation details}
\label{app:algorithm}

\subsection{Agentic-miniRL update order}

For each Agentic-miniRL batch, the implementation follows this order:
\begin{enumerate}
  \item Generate multi-turn trajectories with the rollout policy and
  record inference log probabilities and response-segment masks.
  \item Execute the environment and apply RLOO to the response-level returns as
  in Equation~\ref{eq:rloo}.
  \item Expand the centered return to the terminal token, subtract the K1
  reward, and compute token-level return-to-go as in
  Equation~\ref{eq:k1_return}.
  \item Recompute current and old trainer log probabilities, form the
  behavior correction and sign-aware mask in
  Equation~\ref{eq:minirl_ratios_mask}, and optimize the complete turn-aware
  loss in Equation~\ref{eq:minirl_objective}.
\end{enumerate}

\section{Additional training-instance analyses}
\label{app:instance_dynamics}

The training-instance diagnostic aligns the base model, selected Stage-1
checkpoints, and the post-Stage-2 SFT
policies on the same matched training
records. There are 601, 516, and 1,652 training records in A, B, and C.
Each source record contributes equally to the summaries. Repeated records
share their instance-level rate estimates and do not constitute independent
probes. The repeated C task has four base attempts and eight post-RL attempts,
with a post-RL rate of $5/8$.

Each estimate is the observed success count divided by the actual attempt
count. In the base data, A and B each contain one eight-attempt instance;
C contains nine eight-attempt and three three-attempt instances. All remaining
base estimates use four attempts. Post-RL estimates use four attempts except
for the duplicated C instance above. Post-SFT estimates use four attempts
except for one A instance and seven C instances with eight attempts.
The post-SFT refresh probes a broader pool, but only the original matched
instances enter this comparison. Failures, including infrastructure
failures, remain in the supplied attempt denominators. Category means weight
training records equally rather than pooling episodes with different sample
counts. Up, same, and down compare the exact success-count fractions before
rounding.

\begin{table}[htbp]
  \centering
  \caption{Training-record success rates before and after initial expert RL.
  Base and post-RL rates are equal-weight means over matched
  training records (\%); $\Delta$ is in percentage points.
  Up/same/down count records with higher/equal/lower observed pass rates.
  These are paired empirical estimates, typically from four rollouts per instance.}
  \label{tab:stage1_instance_transitions}
  \small
  \begin{tabular*}{\linewidth}{@{\extracolsep{\fill}}lrrrrrrr@{}}
    \toprule
    Expert & $N$ & Base & Post-RL & $\Delta$ & Up & Same & Down \\
    \midrule
    A & 601 & 38.79 & 46.05 & $+7.26$ & 259 & 151 & 191 \\
    B & 516 & 38.78 & 45.54 & $+6.76$ & 216 & 154 & 146 \\
    C & 1,652 & 38.68 & 43.28 & $+4.60$ & 633 & 517 & 502 \\
    \bottomrule
  \end{tabular*}
\end{table}

Repair recovery is evaluated on training records whose initial-RL rates fall below
Base and whose membership in the actual SFT data is confirmed. There are
179, 141, and 431 such records for A, B, and C (751 total). Recovery to
Base means $p_{\mathrm{SFT}}\geq p_{\mathrm{Base}}$; partial recovery means
$p_{\mathrm{RL}}<p_{\mathrm{SFT}}<p_{\mathrm{Base}}$; no improvement means
$p_{\mathrm{SFT}}\leq p_{\mathrm{RL}}$. The separate gain-retention cohort
requires $p_{\mathrm{RL}}>p_{\mathrm{Base}}$ and
$p_{\mathrm{SFT}}<p_{\mathrm{RL}}$ (404 records). Its above/equal/below-Base
outcomes use that cohort's denominator, not all 580 post-SFT decreases.
All cohort boundaries use integer-count fractions before rounding.

\begin{table}[htbp]
  \centering
  \caption{Observed recovery and gain retention after first Repair SFT.
  Entries give training-record counts with percentages in parentheses.
  Each percentage uses the category-specific denominator shown in its panel.
  Regression means an observed initial-RL rate below Base.
  The second cohort is a subset of all training records whose rates decrease after SFT.}
  \label{tab:repair_recovery}
  \small
  \begin{tabular*}{\linewidth}{@{\extracolsep{\fill}}lrrr@{}}
    \toprule
    \multicolumn{4}{@{}l}{\textit{Previously regressed training records included in Repair SFT}} \\
    Outcome after Repair SFT & A ($N=179$) & B ($N=141$) & C ($N=431$) \\
    \midrule
    At or above Base & 125 (69.8\%) & 103 (73.0\%) & 264 (61.3\%) \\
    Above RL but below Base & 7 (3.9\%) & 5 (3.5\%) & 30 (7.0\%) \\
    At or below RL & 47 (26.3\%) & 33 (23.4\%) & 137 (31.8\%) \\
    \midrule
    \multicolumn{4}{@{}l}{\textit{Initial-RL gains followed by a decline after SFT}} \\
    Outcome after Repair SFT & A ($N=103$) & B ($N=75$) & C ($N=226$) \\
    \midrule
    Still above Base & 36 (35.0\%) & 28 (37.3\%) & 71 (31.4\%) \\
    Equal to Base & 48 (46.6\%) & 32 (42.7\%) & 103 (45.6\%) \\
    Below Base & 19 (18.4\%) & 15 (20.0\%) & 52 (23.0\%) \\
    \bottomrule
  \end{tabular*}
\end{table}

To distinguish the magnitudes of gains and losses, let
$d_i=p_i^{\mathrm{stage}}-p_i^{\mathrm{Base}}$ for each of the $N$ matched
records in a category. The positive contribution is
$100N^{-1}\sum_i\max(d_i,0)$ and the negative contribution is
$100N^{-1}\sum_i\min(d_i,0)$. Their sum equals the stage's mean gain over
Base in percentage points. Both contributions use all matched records as
the denominator, rather than separate means over improving and declining
subsets. Table~\ref{tab:training_gain_contributions} reports this decomposition.

\begin{table}[htbp]
  \centering
  \caption{Positive and negative contributions to mean training-record gains
  relative to Base (percentage points). Each contribution is normalized
  by all matched training records in that category; positive plus negative equals net gain.}
  \label{tab:training_gain_contributions}
  \small
  \begin{tabular*}{\linewidth}{@{\extracolsep{\fill}}llrrr@{}}
    \toprule
    Expert & Stage & Positive & Negative & Net \\
    \midrule
    A & Initial RL & $+16.85$ & $-9.59$ & $+7.26$ \\
    A & Repair SFT & $+22.46$ & $-4.97$ & $+17.49$ \\
    \addlinespace[3pt]
    B & Initial RL & $+15.12$ & $-8.36$ & $+6.76$ \\
    B & Repair SFT & $+25.61$ & $-4.80$ & $+20.81$ \\
    \addlinespace[3pt]
    C & Initial RL & $+13.68$ & $-9.08$ & $+4.60$ \\
    C & Repair SFT & $+20.71$ & $-6.64$ & $+14.07$ \\
    \bottomrule
  \end{tabular*}
\end{table}

These matched snapshots describe observed changes within the selected training
set. They are not a longitudinal test of persistent forgetting, and finite
rollout sampling contributes to the measured changes. Moreover, the initial
pool was selected using base-model mastery estimates, so the diagnostic is
conditional on that training-set selection. Selecting instances with low or
high observed RL rates also makes regression to the mean relevant to recovery
and retention comparisons. These before/after observations do not isolate
SFT's causal effect from a no-SFT control. Actual SFT membership establishes
training-data exposure; a zero-success post-RL probe does not establish an
empty historical success buffer.

\FloatBarrier
\section{Extended evaluation of fixed final policies}
\label{app:expanded_evaluation}

\subsection{Complete results and round variability}

Table~\ref{tab:expanded_eval_results} reports all six models on the four
evaluated task sets and the nested Pro V2 Hard-51 subset. Each mean counts
confirmed passes across three selected rounds over three times the full task
count. The same selected parent rounds define both Pro V2 and Hard-51 results.
Round standard deviations describe evaluation variability and do not measure variability
across independent training seeds.

\begin{table}[htbp]
  \centering
  \caption{Complete final-model evaluation (\%). Entries report mean $\pm$ population standard deviation over three selected evaluation rounds. Hard-51 is nested within Pro V2. Pro 731 denotes the revised execution setup over the original 731-task population; LCB denotes LiveCodeBench (v6, 175 tasks). Bold marks the highest mean within each student-family block.}
  \label{tab:expanded_eval_results}
  \small
  \setlength{\tabcolsep}{3pt}
  \begin{tabular*}{\linewidth}{@{\extracolsep{\fill}}lccccc@{}}
    \toprule
    Policy & Pro V2 & Hard-51 & Multilingual & Pro 731 & LCB v6 \\
    Tasks per round & 642 & 51 & 300 & 731 & 175 \\
    \midrule
    Qwen3.6-27B & $81.62\pm0.89$ & $37.25\pm1.60$ & $67.89\pm0.42$ & $52.94\pm0.59$ & $82.48\pm0.54$ \\
    Pooled RL & $81.57\pm1.11$ & $32.03\pm0.92$ & $69.44\pm1.40$ & $52.67\pm0.19$ & $83.05\pm0.27$ \\
    Balanced RL & $82.35\pm1.20$ & $35.95\pm7.57$ & $69.00\pm0.00$ & $53.31\pm0.56$ & $83.24\pm0.27$ \\
    MOPD 27B & $\mathbf{86.19\pm0.65}$ & $\mathbf{42.48\pm0.92}$ & $\mathbf{71.78\pm0.87}$ & $\mathbf{56.41\pm0.23}$ & $\mathbf{84.57\pm0.47}$ \\
    \midrule
    Qwen3.6-35B-A3B & $73.31\pm0.81$ & $28.10\pm4.03$ & $63.44\pm0.83$ & $48.75\pm0.56$ & $76.95\pm0.27$ \\
    MOPD 35B-A3B & $\mathbf{77.00\pm0.90}$ & $\mathbf{33.99\pm3.33}$ & $\mathbf{66.44\pm0.42}$ & $\mathbf{51.12\pm0.13}$ & $\mathbf{78.10\pm0.54}$ \\
    \bottomrule
  \end{tabular*}
\end{table}

The revised 731-task execution setup retains the original Pro task population
while addressing parts of the execution pipeline. This coverage evaluation
does not establish that every specification or grading issue underlying the
Pro-618 exclusions has been resolved. Official Pro V2 additionally revises
task definitions and verifiers~\citep{scaleai2026swebenchprov2}; its membership
and setup differ from the original Pro-618 analysis. The 731-task population
also includes tasks used during Pro-618 checkpoint selection, so this expanded
coverage is not an independent selection-free test.

The LiveCodeBench experiment uses a 175-task set associated with v6 and a
single logical code-generation trial per task per round, followed by isolated
tests~\citep{jain2024livecodebench}. It is a code-generation evaluation, not
a multi-turn repository-agent rollout. We report this specific evaluated set
rather than treating it as the entire cumulative LiveCodeBench release.

\subsection{Multilingual category comparison}

Table~\ref{tab:multilingual_results} recomputes the base--MOPD category
comparison from the selected Multilingual outcomes and the existing
300-instance mapping. A/B/C contain 72/25/176 tasks; the remaining 27 are
unassigned and retained in Full. MOPD improves all three routed category
means relative to the base, while the unassigned group decreases. The paired
instance-bootstrap 95\% interval for the overall difference from the base is
[\ExpMultiBaseCILow, \ExpMultiBaseCIHigh] percentage points. It uses these same
selected task identities, with each model's three confirmed-pass indicators
averaged before resampling.

\begin{table}[htbp]
  \centering
  \caption{SWE-bench Multilingual category comparison of Qwen3.6-27B and MOPD (\%). Full reports three-round mean $\pm$ population standard deviation; category entries are three-round means recomputed from the selected task-level outcomes. A/B/C/U contain 72/25/176/27 instances, respectively. U denotes unassigned tasks and remains in Full.}
  \label{tab:multilingual_results}
  \small
  \begin{tabular*}{\linewidth}{@{\extracolsep{\fill}}lrrrrr@{}}
    \toprule
    Policy & Full $\uparrow$ & A $\uparrow$ & B $\uparrow$ & C $\uparrow$ & U $\uparrow$ \\
    \midrule
    Qwen3.6-27B & $67.89\pm0.42$ & 66.20 & 62.67 & 69.89 & 64.20 \\
    MOPD 27B & $71.78\pm0.87$ & 69.44 & 74.67 & 73.67 & 62.96 \\
    \bottomrule
  \end{tabular*}
\end{table}

\subsection{Execution protocols and score accounting}
\label{app:expanded_protocols}

Table~\ref{tab:expanded_eval_protocols} summarizes the recorded execution
settings. The expanded evaluations use model-specific requests rather than
one uniform generation budget. On Multilingual, the base and MOPD27 request
the same 65{,}536-token output limit; Pooled/Balanced request 81{,}920 tokens
and presence penalty 1.5, while MOPD35 requests 131{,}072 tokens and presence
penalty 1.5. Presence penalty is not explicit in the inspected base/MOPD27
requests. Recent runs also use native polling, whereas historical runs use
synchronous or streaming calls.

\begin{table}[htbp]
  \centering
  \caption{Recorded execution settings for the expanded evaluations.
  Generation values are request settings from configuration readbacks;
  omitted fields do not establish the model server's effective defaults.}
  \label{tab:expanded_eval_protocols}
  \small
  \begin{tabularx}{\linewidth}{@{}>{\raggedright\arraybackslash}p{0.17\linewidth}>{\raggedright\arraybackslash}p{0.22\linewidth}Y@{}}
    \toprule
    Evaluation & Execution scaffold & Recorded limits and requests \\
    \midrule
    Pro V2 / Hard-51 & mini-swe-agent 2.4.6; fresh-environment patch replay &
    14{,}400\,s task limit; approximately 2{,}940\,s in-agent budget.
    Temperature, top-$p$, top-$k$, and maximum output were not explicit in
    the inspected requests. Hard-51 uses the same parent runs. \\
    Multilingual & SWE-agent, no-hack preset &
    Temperature 1, top-$p$ 0.95; 300 iterations; 7{,}200\,s task limit.
    Maximum output is 65{,}536 tokens for the bases and MOPD27,
    81{,}920 for Pooled/Balanced, and 131{,}072 for MOPD35. \\
    Pro 731 & SWE-agent, no-hack preset &
    Temperature 1; 32{,}768 output tokens; 300 iterations; 7{,}200\,s task
    limit. Recorded batches include top-$p$/top-$k$ of 0.8/1 and 0.95/20. \\
    LCB v6, 175 tasks & Single logical code-generation trial and isolated judge &
    Top-$p$ 0.95; 81{,}920 or 131{,}072 output tokens by source run;
    requests include temperature 0.2 or 1. Judge limit 1{,}800\,s,
    with 6\,s per test. \\
    \bottomrule
  \end{tabularx}
\end{table}

Pro 731 batches and LCB source runs differ in some sampling requests and
output budgets. Model services can also apply deployment defaults in place
of request overrides, so request equality alone does not prove effective
sampling equality. LCB provenance retains the distinction between raw full
rounds and historical offline retry merges; a merge is not an additional
independent full round. These differences limit attribution of the reported
score gaps to the training method alone. Reported means use the full task
denominator. Unscored task-round outcomes contribute no confirmed pass and
remain missing in the source records; Qwen3.6-27B has three such outcomes on
Multilingual.

\subsection{Reusing 27B teachers with a 35B-A3B student}

\begin{figure}[htbp]
  \centering
  \includegraphics[width=\linewidth]{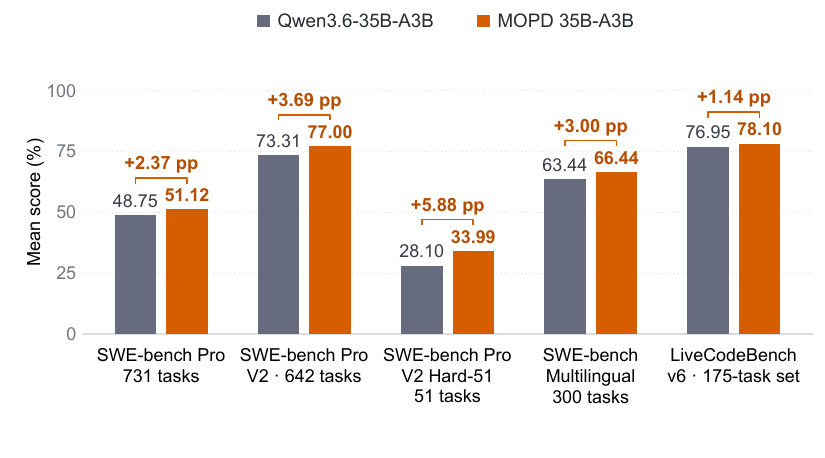}
  \caption{\textbf{Reusing the same three 27B experts with a 35B-A3B student.}
  Bars compare Qwen3.6-35B-A3B with the MOPD student trained from it, using
  the same final RRE experts as the 27B integration. Values are three-round
  mean scores; brackets mark gains over the base in percentage points.
  Hard-51 uses the same parent runs as SWE-bench Pro V2. Round variability
  is reported in Table~\ref{tab:expanded_eval_results}.}
  \label{fig:teacher_reuse_comparison}
\end{figure}

The extension initializes the student from Qwen3.6-35B-A3B and reuses the
same three final 27B RRE experts. The frozen extrapolation reference is the
35B-A3B student's initial policy. Thus the teacher set remains same-origin,
while the student has a different architecture and initialization from the
27B student. Figure~\ref{fig:teacher_reuse_comparison} compares the resulting
student with its 35B-A3B base across the four evaluated task sets and Hard-51;
complete scores are reported in Table~\ref{tab:expanded_eval_results}.

\FloatBarrier

On Hard-51, the mean rises from \ExpBaseMoEHardMean\% to
\ExpMOPDMoEHardMean\%, a gain of \ExpMOPDMoEHardGainBase{} percentage points.
The corresponding gains are \ExpMOPDMoEProVTwoGainBase{} points on Pro V2,
\ExpMOPDMoEMultiGainBase{} on Multilingual, \ExpMOPDMoEProFullGainBase{} on
Pro 731, and \ExpMOPDMoELCBGainBase{} on LiveCodeBench. Together with the
27B results, this experiment uses one RRE-developed teacher set for both a
dense student and an MoE student. The category specialization learned during
expert development is therefore reusable across these two student
initializations, while the student and its frozen reference are configured
for each integration run.

\end{document}